\documentclass[aps,prl,reprint,superscriptaddress,nofootinbib,nobibnotes]{revtex4-2}

\usepackage{amsmath,amssymb,mathtools}
\usepackage{graphicx}
\usepackage{xcolor}
\definecolor{blueviolet}{rgb}{0.2, 0.2, 0.6}
\definecolor{webgreen}{rgb}{0,.5,0}
\definecolor{webbrown}{rgb}{.6,0,0}
\definecolor{revisiongreen}{rgb}{0,0.55,0}
\definecolor{revisionred}{rgb}{0.80,0,0}
\usepackage[normalem]{ulem}
\usepackage[activate={true,nocompatibility},final,tracking=true,kerning=true,spacing=true
]{microtype}
\microtypecontext{spacing=nonfrench}
\usepackage[pdftex,
	bookmarks=false,
	colorlinks=true, 
	urlcolor=webbrown, 
	linkcolor=blueviolet, 
	citecolor=webgreen,
	pdfstartpage=1,
	pdfstartview={FitH},  
	bookmarksopen=false
	]{hyperref}
\usepackage{enumitem}

\setlist[itemize]{leftmargin=*,itemsep=2pt,topsep=3pt}
\setlist[enumerate]{leftmargin=*,itemsep=2pt,topsep=3pt}

\newcommand{\TU}{T_{\rm U}}
\newcommand{\TDB}{T_{\rm DB}}
\newcommand{\facc}{{f_{\mathrm{acc.}}}}
\newcommand{\fU}{{f_{\mathrm{U,acc.}}}}
\newcommand{\fminacc}{{f_{\min,\mathrm{acc.}}}}
\newcommand{\nulab}{{\delta\nu_{\mathrm{lab}}}}

\graphicspath{{figures/}{./}}

\begin{document}

\title{Measuring the Acceleration-Dependent Temperature of the Minkowski Vacuum}

\author{Daine L. Danielson}
\email{daine@mit.edu}
\affiliation{Center for Theoretical Physics --- a Leinweber Institute, Massachusetts Institute of Technology, Cambridge, MA 02139, USA}
\affiliation{Black Hole Initiative, Harvard University, Cambridge, MA 02138, USA}
\author{Netta Engelhardt}
\email{engeln@mit.edu}
\affiliation{Center for Theoretical Physics --- a Leinweber Institute, Massachusetts Institute of Technology, Cambridge, MA 02139, USA}
\affiliation{Black Hole Initiative, Harvard University, Cambridge, MA 02138, USA}
\author{Lindley A. Winslow}
\email{lindley.winslow@mpp.mpg.de}
\affiliation{Laboratory for Nuclear Science, Massachusetts Institute of Technology, Cambridge, MA 02139, USA}
\affiliation{Max-Planck-Institut f\"ur Physik, 85748 Garching bei M\"unchen, Germany}

\date{\today}

\begin{abstract}
\noindent Relativistic quantum field theory is the structure on which the Standard Model and beyond-Standard Model physics are built. One of its most striking predictions is the Unruh effect: an observer undergoing uniform linear acceleration measures the Minkowski vacuum as a thermal bath at a fixed temperature. Despite decades of theoretical interest, this effect has never been directly observed. We propose to use the high acceleration gradients provided by wakefield accelerators, coupled with a forward {laboratory-frequency $\sim$THz detector array}, to measure the resulting forward-beamed radiation, and outline the unique signature provided by quantum detailed balance to extract the signal from backgrounds. While challenging, such a measurement appears within reach of present-day accelerator and detector technologies.
\end{abstract}

\maketitle


\section{Introduction}
One of the basic predictions of relativistic quantum field theory is that an observer undergoing uniform linear acceleration measures the standard vacuum state of Minkowski space to be a thermal bath at fixed temperature. This is the celebrated prediction of the Fulling-Davies-Unruh effect~\cite{Ful73,Dav75,Unr76} (``Unruh effect'' for brevity), and it is remarkable in its universality: it is independent of spin \cite{Takagi86,Crispino08}, coupling \cite{BisWic75,BisWic76}, or any of the theory-specific details of the Lagrangian besides Lorentz invariance and locality---and therefore applies to the Standard Model and all such extensions. A local form is also expected in gravitating theories, where it underlies Hawking radiation from black holes \cite{Haw74,Haw75,Crispino08}.

Failure of the Unruh effect where relativistic quantum field theory otherwise succeeds would challenge the framework underlying particle physics.\footnote{Our prior is of course that the Unruh effect indeed manifests in nature as expected; this is precisely why it is essential to experimentally test it.} Because it could not be repaired species by species, and because the same structure underlies Hawking radiation, such a result would also force a rethinking of the black hole information paradox. 

{Complementary experimental programs have sought signatures in storage rings, atomic and cavity quantum electrodynamics, ultraintense laser interactions, relativistic quantum information, positron spectra, and classical descriptions of accelerated-charge radiation \cite{BellLeinaas83,BellLeinaas87,ChenTajima99,Scully03,Schutzhold06,MartinMartinez11,Cozzella18, Lynch21}; see also \cite{Earman11,PenaSudarsky14}.  The proposal of \cite{Cozzella17,Cozzella18} is close in spirit to ours below.  Its proposed observable, however, is reproduced exactly by classical electrodynamics \cite{HiguchiMatsasSudarsky92a,HiguchiMatsasSudarsky92b,HiguchiMatsas93,Pena2005,LandulfoFullingMatsas19}.  {Here $\facc>0$ denotes ordinary frequency conjugate to accelerating-frame proper time.}  Our proposed experiment described below instead targets the ordered $+\facc/-\facc$ covariance ratio: as shown in the Appendix, neither a deterministic classical field nor a reciprocal stochastic classical field produces the particular asymmetry associated with the Unruh effect.}
\begin{figure}
    \centering
\includegraphics[width=\columnwidth]{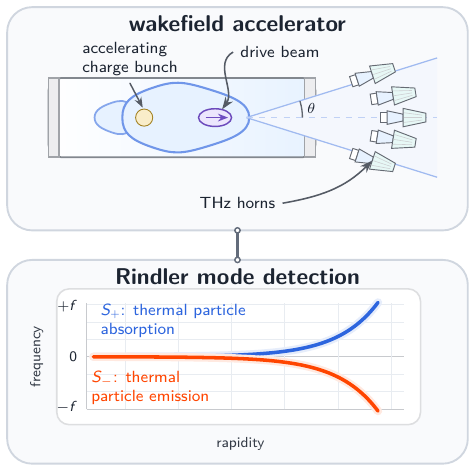}
\caption{Schematic experimental protocol.  Beam diagnostics determine the
accelerated trajectory and hence the map between retarded lab time and proper
time.  Polarization selection, photon counting, and dual homodyne reconstruct the ordered $+\facc$ and $-\facc$ Rindler-mode responses.}
\label{fig:protocol}
\end{figure}

We propose a test with current-generation wakefield accelerators, whose particle or laser drivers excite high-gradient plasma waves that accelerate electrons \cite{TajimaDawson79,Esarey09}.  Tens-of-GV/m gradients and multi-GeV beams have been demonstrated in laser-, electron-, and proton-driven programs, including BELLA, FACET-II, FLASHForward, and AWAKE \cite{Geddes:2004tb,Leeman:2006,Gonsalves19,Picksley:2024,Yakimenko19,natureZhang,Lindstrom21,AWAKE18}.  We measure the charge bunch's forward-beamed thermalization radiation (Fig.~\ref{fig:protocol}) through inertial chirped modes corresponding to finite-energy Rindler photons. We propose using the unique signature provided by the so-called ``quantum detailed balance'' ratio to extract the corresponding Unruh temperature of the  vacuum.

\section{Theory: Unruh Effect}

\begin{figure}
    \centering
    \includegraphics[width=0.6\linewidth]{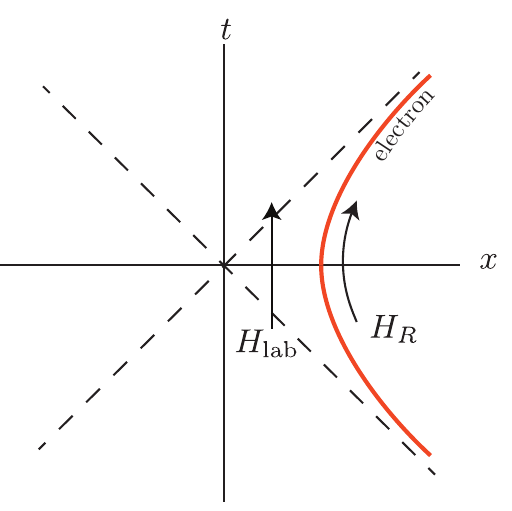}
    \caption{The Rindler decomposition of Minkowski space: the right Rindler wedge is the $x>0$ region outside of the origin lightcone. A uniformly accelerated trajectory is illustrated in orange. The boosts corresponding to the Rindler Hamiltonian $H_{R}$ are denoted along the trajectory, as are the inertial time coordinate translations corresponding to the standard Hamiltonian $H_{\rm lab}$.  }
\label{fig:RindlerDecomp}
\end{figure}

We begin with a conceptual overview of the Unruh effect itself. For simplicity we will illustrate the effect in 1+1-dimensional Minkowski space. In Cartesian inertial coordinates, the Hamiltonian $H$ generates time translations in the standard $t$-direction. Consider now a detector in Minkowski space undergoing uniform acceleration. This detector is confined to the right spacelike quadrant, known as the Rindler wedge {\cite{Rindler66,Crispino08}}. The future lightcone of the origin constitutes a horizon for the detector: no timelike or lightlike signals from behind this lightsheet can reach the detector. This is illustrated in 
Figure~\ref{fig:RindlerDecomp}.

In the non-inertial rest frame of the detector, time translations are boosts along the detector's worldline; the Hamiltonian of the Rindler wedge $H_{R}$ generating translations in the detector's proper time is therefore proportional to the boost generator, rather than to the inertial Hamiltonian. These define two different notions of energy, and it thus stands to reason that the vacuum state of $H$ is not identical to the vacuum state of $H_{R}$. The former is the vacuum under time translations whereas the latter is the vacuum under boosts. Observe, however, that specifying the eigenstates of $H_{R}$ leaves the full state on Minkowski space unspecified due to the left spacelike Rindler quadrant. The restriction of the Minkowski vacuum to the right Rindler wedge is therefore mixed. Indeed, {the landmark analyses \cite{BisWic75,BisWic76,Unr76,Sewell82,Crispino08} show} that a uniformly accelerated detector with proper acceleration $a$ responds to the Minkowski vacuum as if immersed in a thermal bath at temperature
        \begin{equation}
            \TU=\frac{\hbar a}{2\pi k_B c}.
        \end{equation}
 This is the Unruh effect: the Minkowski vacuum restricted to a Rindler wedge is a thermal state according to the Hamiltonian $H_{R}$. Thermal states are characterized by the Kubo-Martin-Schwinger (KMS) condition {\cite{Kubo57,MartinSchwinger59,Sewell82}} for the behavior of two-point functions in the complex plane; in the case of the Unruh effect, this is:
\begin{equation}
\langle \hat{O}_{1}(\tau) \hat{O}_{2}(\tau')\rangle_{T_\mathrm{U}} = \langle \hat{O}_{2}(\tau') \hat{O}_{1}(\tau +2\pi ic/a)\rangle_{T_\mathrm{U}}
\end{equation}
where $\hat{O}_{1}$ and $\hat{O}_{2}$ are operators, $\tau$ and $\tau'$ are proper times along the detector's trajectory, and $\beta$ is the inverse temperature. {Equivalently, the two Wightman orderings are related by the same imaginary-proper-time shift.}  It is worth noting that non-inertial accelerations will in general perceive the Minkowski vacuum to be some excited state (e.g. a detector undergoing uniform circular acceleration {\cite{LetawPfautsch80,BellLeinaas83,BellLeinaas87,DaviesDrayManogue96}}). However, the features of a smooth horizon, an exact thermal spectrum, and a clean mapping between symmetries of Minkowski space are all unique to the uniformly accelerating detector---i.e., the classic Unruh effect.

If the effect is so essential, then, why has it not yet been observed? As we describe quantitatively below, a direct detection that is robust against systematic noise requires a challenging combination of frequency range, signal duration, and intensity. A direct-spectrum measurement would try to isolate the positive-frequency response of an accelerated detector and test the Planckian occupation
\begin{equation}
n_U(\facc)\!=\!\left(e^{\facc/\fU}-1\right)^{-1},\;
\fU\!\equiv\!\frac{k_BT_U}{h}
\end{equation}
where $\fU$ is the {accelerated-frame Unruh frequency}. In the accelerating frame, the detector (or any charged body) absorbs and emits thermal ``Rindler'' photons, with proper-time frequency rates denoted by $S_\tau(+\facc)$ and $S_\tau(-\facc)$, respectively.  Their ratio is fixed by quantum detailed balance {\cite{DeWitt79,UnrWal84,Crispino08}}. A rough estimate of the measured power is
\begin{equation}    \label{eq:direct-spectrum-main}
S_{\tau}(+\facc)={\cal J}(\facc)n_U(\facc)+B_+(\facc),
\end{equation}
where ${\cal J}(\facc)$ is the response/collection transfer function and $B_+(\facc)$ denotes additive backgrounds (we discuss intrinsic backgrounds, but other backgrounds can be treated similarly). The inferred temperature therefore depends on absolute calibration of both the detector response and the background subtraction.

Here we propose a smoking-gun signal of the Unruh effect
for which common multiplicative calibration errors are less important.
Inertial detectors reconstruct the positive- and negative-Rindler-frequency
responses from two chirped laboratory sidebands associated with photon
emission and absorption by collective modes of an accelerating charge
bunch. Their ratio measures the asymmetry while canceling the transfer
function ${\cal J}(\facc)$ to leading order.

The resulting inferred temperature is
        \begin{equation} \label{eq:thermal-detailed-balance-main}
            \TDB(\facc)=\frac{h \facc}{k_B\ln[S_\tau(-\facc)/S_\tau(+\facc)]}
        \end{equation}
        and the Unruh prediction is $\TDB(\facc)=\TU$ across the accessible band, with slope fixed by the measured acceleration.  The ratio suppresses common multiplicative calibration errors, but not asymmetric classical leakage into the same modes; the controls below target that contamination.

\section{Theory: quantum detailed balance}\label{sec:QDB}
Acceleration radiation by itself is not a distinctive signature of the Unruh effect.  Both classical and quantum charges radiate when accelerated, and ordinary bremsstrahlung is already accounted for by classical electrodynamics.  The distinct prediction of Fulling, Davies, and Unruh is that, in the co-accelerating frame, a charged body interacting with the quantized photon field \textit{thermalizes} with the Minkowski vacuum.  In that frame the body absorbs and emits thermal ``Rindler'' photons, with proper-time frequency rates denoted by $S_\tau(+\facc)$ and $S_\tau(-\facc)$, respectively.  Their ratio is fixed by quantum detailed balance. Though observers may disagree whether a field transition corresponds to photon emission or absorption \cite{UnrWal84}, the occurrence of the transition is frame invariant.  {An inertial observer resolves a finite Rindler mode into two chirped, positive-Minkowski-frequency partner packets; their occupations and anomalous covariance reconstruct the same ordered response.}

{The relevant signature is therefore in the ordered fluctuation spectrum of the electromagnetic field, not in the mean bremsstrahlung power \cite{Clerk10,Weinstein14}.}  Let $\hat{\mathcal E}(\tau)$ denote the component of the electromagnetic readout observable, after removal of the reproducible coherent classical displacement, sampled along an approximately uniformly accelerated trajectory.  The two non-symmetrized proper-time spectra are
\begin{equation}\begin{split}
 S^{>}_{\tau}(\facc)=\int d\Delta\tau\,e^{i2\pi \facc\Delta\tau}
 \left\langle \hat{\mathcal E}(\tau+\Delta\tau)\hat{\mathcal E}(\tau)\right\rangle,\\
 S^{<}_{\tau}(\facc)=\int d\Delta\tau\,e^{i2\pi \facc\Delta\tau}
 \left\langle \hat{\mathcal E}(\tau)\hat{\mathcal E}(\tau+\Delta\tau)\right\rangle.
\end{split}\end{equation}
For $\facc>0$, define $S_\tau(+\facc)\equiv S^{>}_{\tau}(-\facc)$ and $S_\tau(-\facc)\equiv S^{>}_{\tau}(+\facc)$.  These orderings coincide for a classical field but differ for a quantum field by the familiar $n$ versus $n+1$ asymmetry.

{Following  Unruh and Wald~\cite{UnrWal84}, we label the two positive-Minkowski-frequency partner modes $F_{\mathrm I,f}$ and $F_{\mathrm{II},f}$, with annihilation operators $\hat a_{\mathrm I}$ and $\hat a_{\mathrm{II}}$, and define $\delta\hat a_J\equiv\hat a_J-\langle\hat a_J\rangle$ for $J=\mathrm I,\mathrm{II}$.  In the ideal single-pair limit, a Rindler mode can be decomposed in terms of these modes as annihilated by
\begin{equation*}
 \delta\hat b_f=u_f\,\delta\hat a_{\mathrm I}
 +v_f \delta\hat a_{\mathrm{II}}^\dagger,
 \qquad |u_f|^2-|v_f|^2=1,
\end{equation*}
with
\begin{equation*}
 u_f=\frac{1}{\sqrt{1-e^{-\facc/\fU}}},\qquad
 v_f=\frac{1}{\sqrt{e^{\facc/\fU}-1}}.
\end{equation*}
Thus $|v_f|^2=n_U(\facc)$ and $|u_f|^2=n_U(\facc)+1$.

Define the centered partner occupations and anomalous covariance by
\begin{equation*}
 N_J\equiv\langle\delta\hat a_J^\dagger\delta\hat a_J\rangle
 \quad(J=\mathrm I,\mathrm{II}),
 \qquad
 M_{\mathrm{I,II}}\equiv
 \langle\delta\hat a_{\mathrm I}\delta\hat a_{\mathrm{II}}\rangle.
\end{equation*}
The two ordered Rindler-mode counts,
$n_+\equiv\langle\delta\hat b_f^\dagger\delta\hat b_f\rangle$ and
$n_-\equiv\langle\delta\hat b_f\delta\hat b_f^\dagger\rangle$, are
\begin{align*}
 n_\pm={}&|u_f|^2\left(N_{\mathrm I}+\frac12\right)
 +|v_f|^2\left(N_{\mathrm{II}}+\frac12\right)\\
 &+2\,\mathrm{Re}\!\left[u_f^*v_fM_{\mathrm{I,II}}^*\right]
 \mp\frac12.
\end{align*}
Consequently,
\begin{equation*}
 n_--n_+=\langle[\delta\hat b_f,\delta\hat b_f^\dagger]\rangle=1.
\end{equation*}
For reciprocal source matrix elements, $S_\tau(+\facc)\propto n_+$ and $S_\tau(-\facc)\propto n_-$.  It is this exact commutator difference---absent for commuting deterministic or stochastic classical fields---that supplies the nonclassical signature emphasized in the Appendix.}

The power spectra of a Wightman function satisfying the KMS condition give the quantum detailed-balance temperature~\eqref{eq:thermal-detailed-balance-main} {\cite{Kubo57,MartinSchwinger59,Clerk10}}.  For uniform acceleration, $T=\TU$, so
\begin{equation}
    \ln\frac{S_\tau(-\facc)}{S_\tau(+\facc)}
    =\frac{\facc}{\fU}.
    \label{eq:unruh-log-slope-main}
\end{equation}
Thus the proposed measurement is a slope measurement in proper-time frequency,
with no fitted temperature once the acceleration is independently known.  
{The positive-Rindler-frequency spectrum~\eqref{eq:direct-spectrum-main} is a useful baseline, but measures occupation rather than the excitation--de-excitation asymmetry and requires absolute calibration of ${\cal J}$ and $B_+$ across the band.  Detailed balance uses the same weak branch while replacing this with relative $+\facc/-\facc$ calibration; it is therefore our primary observable.}

It remains to connect the proper-time spectra to what an inertial detector
records.  Write the rapidity of the accelerated electron as
\begin{equation}
 \eta=\frac{a\tau}{c},\qquad
 \gamma=\cosh\eta,\qquad
 \beta=\tanh\eta .
\end{equation}
For a far-field detector at angle $\theta$ relative to the instantaneous
velocity, the retarded lab time is
\begin{equation}
 u(\eta)=\frac{c}{a}\left[\sinh\eta-
\cos\theta\left(\cosh\eta-1\right)\right].
 \label{eq:retarded-time-main}
\end{equation}
{Using the known map $u\leftrightarrow\tau$, two transducer settings select the Minkowski-partner waveforms
\begin{align}
 z_{\mathrm I}(u;\facc)&={\cal A}_{\mathrm I}{\cal M}(\eta)W(\eta)
 \exp\!\left[-i2\pi\facc\frac{c}{a}\eta(u)\right],\nonumber\\
 z_{\mathrm{II}}(u;\facc)&={\cal A}_{\mathrm{II}}{\cal M}(\eta)W(\eta)
 \exp\!\left[+i2\pi\facc\frac{c}{a}\eta(u)\right],
 \label{eq:chirp-template-main}
\end{align}
where ${\cal A}_{\mathrm I,\mathrm{II}}$ are calibrated complex normalizations, $W$ is the finite acceleration window, and ${\cal M}$ includes propagation, collection, and polarization dependence.  The mode operators selected by $z_{\mathrm I}$ and $z_{\mathrm{II}}$ are precisely $\hat a_{\mathrm I}$ and $\hat a_{\mathrm{II}}$ defined above.}  For the estimates below, $W$ is taken to be common and real, independent of $\facc$ over the fit band.  The instantaneous laboratory sideband offset is
\begin{equation}
 \nulab(\eta,\theta)=
 \frac{\facc}{\cosh\eta-\cos\theta\sinh\eta}
 =\frac{\facc}{\gamma(1-\beta\cos\theta)} .
 \label{eq:lab-chirp-frequency-main}
\end{equation}
{Including the transducer carrier $\nu_c$, the positive-frequency outputs are ${\cal E}_J(u)=e^{-i2\pi\nu_c u}z_J(u)$ for $J=\mathrm I,\mathrm{II}$; with physical laboratory frequencies $\nu=\nu_c\pm\nulab>0$.

The collection geometry matters because the radiated intensity is not
isotropic, with an angular power distribution proportional to
\begin{equation}
    \frac{dP_\mathrm{cl.}}{d\Omega}\propto
    \frac{\sin^2\theta}{(1-\beta\cos\theta)^5} \quad\text{and}\quad \frac{dP_\mathrm{sig.}}{d\Omega}\propto
    \frac{\sin^2\phi}{(1-\beta\cos\theta)^3}
\label{eq:longitudinal-angular-pattern-main}
\end{equation}
for the classical Li\'enard field, and for the azimuthally polarized emission from a transverse dipole bunch mode, respectively.
\section{Experimental protocol and feasibility}

The angular pattern in Eq. (12) makes the experiment
intrinsically angle resolved: at large rapidity the power is concentrated in
a narrow forward cone. 
A polarized receiver resolves the azimuthally polarized light of transverse dipole transitions of the bunch, thus rejecting much of the coherent classical background. Table~\ref{tab:balanced-wide-angle-inputs} includes the selected-source acceptance and receiver loss.  In this regime the fixed proper-time modes of Eq.~\eqref{eq:chirp-template-main} appear to inertial detectors as broadband chirps, with
 laboratory sideband offsets
extending from mm-wave through the THz band to the mid-infrared; Fig.~\ref{fig:signal-classical-spectrograms} compares their lab-frame support with the classical spectrum.  The polarization-resolved response, angular fractions, and laboratory mapping are given in the Supplemental Material.
\begin{table}[t]
{
\caption{Reference source, receiver, and rejection benchmark. $N_{\rm eff}$ and $N_{\rm count}$ are effective sample sizes for the homodyne and counting source--reference differences.}
\label{tab:balanced-wide-angle-inputs}
\squeezetable
\renewcommand{\arraystretch}{0.97}
\begin{ruledtabular}
\begin{tabular}{@{}p{0.66\columnwidth}p{0.25\columnwidth}@{}}
\textbf{Quantity} & {\raggedright\textbf{Reference value}\par}\\
\multicolumn{2}{c}{\textit{Trajectory and source}}\\
Effective accelerating field $E$ & $1~\mathrm{GV/m}$\\
Bunch charge $Q_b$ & $0.65~\mathrm{nC}$\\
Rapidity endpoints $(\eta_i,\eta_f)$ & $(1,8)$\\
Rapidity span $\Delta\eta\equiv\eta_f-\eta_i$ & $7$\\
Collection region ${\cal C}$ & $0\leq\theta\leq0.2~\mathrm{rad}$\\
Selected bunch response ${\cal R}_{\rm src}$ & $0.16$\\
\multicolumn{2}{c}{\textit{Readout, fit, and rejection}}\\
Reference fit point $\kappa\equiv\facc/\fminacc$ & $1.2$\\
Partner-counting efficiency $\eta_{\rm count}^{(N)}$ & $0.018$\\
Two-mode quadrature efficiency $\eta_{\rm tom}$ & $0.09$\\
Residual field-amplitude error $\sigma_A$ & $10^{-3}$\\
Liénard field leakage $L_{\rm tom}$ & $2\times10^{-16}$\\
Homodyne effective samples $N_{\rm eff}$ & $3\times10^9$\\
Counting effective samples $N_{\rm count}$ & $1.5\times10^9$\\
\multicolumn{2}{c}{\textit{Derived benchmark figures}}\\
Unruh frequency $\fU$ & $15~\mathrm{GHz}$\\
Minimum frequency $\fminacc$ & $84~\mathrm{GHz}$\\
Reference Rindler frequency $\facc$ & $100~\mathrm{GHz}$\\
Weak mode-II power $\overline P_{\mathrm{II},\rm det}^{\rm arrival}$ & $4.7\times10^{-15}~\mathrm W$\\
Absorption probability $P_{\rm abs}^{R}$ & $6.7\times10^{-5}$\\
Raw energy dynamic range $D_{\rm cl}$ & $6.3\times10^{21}$\\
Statistical contamination $\sigma_{\Xi}^{\rm stat}$ & $2.3\times10^{-5}$\\
Classical contamination target $\Xi_z^{\rm tot}$ & $\lesssim10^{-2}$\\
\end{tabular}
\end{ruledtabular}
}
\end{table}

\begin{figure*}[t]
\centering
\begin{minipage}[t]{0.481\textwidth}
    \centering
    \includegraphics[width=\linewidth]{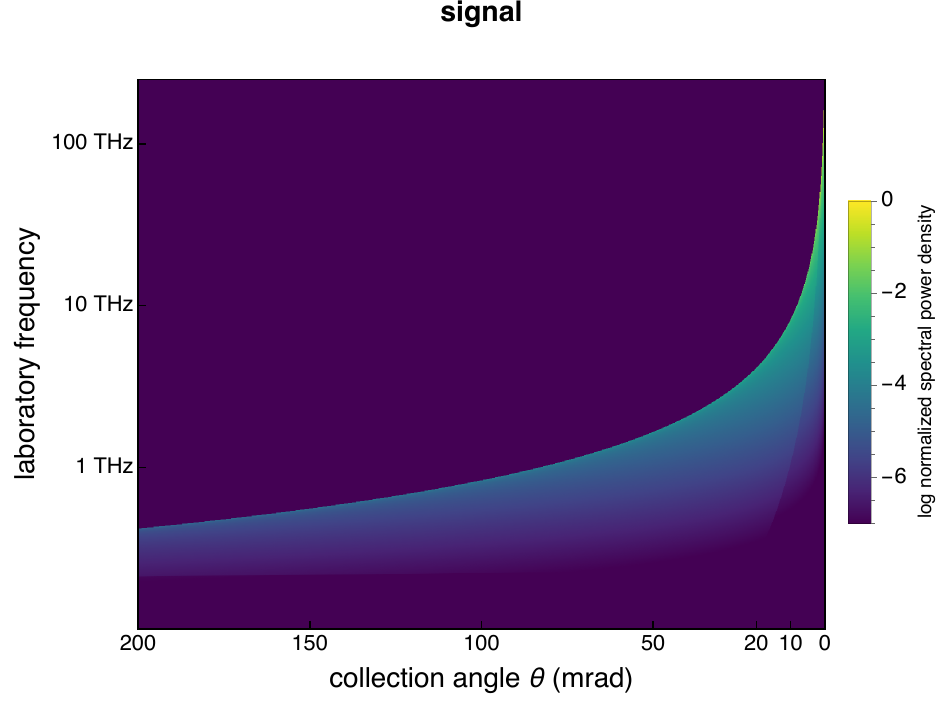}
    \vspace{0.2ex}
    {\small \textbf{(a)} Weak Minkowski mode support.}
\end{minipage}
\hfill
\begin{minipage}[t]{0.481\textwidth}
    \centering
    \includegraphics[width=\linewidth]{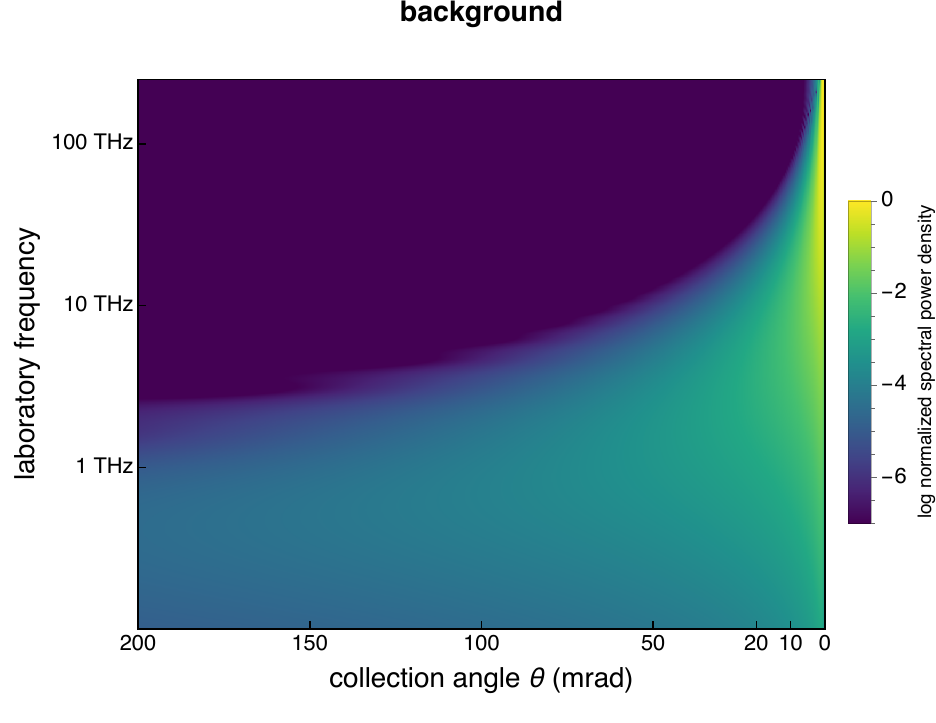}
    \vspace{0.2ex}
    {\small \textbf{(b)} Classical spectrum.}
\end{minipage}
\caption{Angle--laboratory-sideband-offset spectrograms for the $E=1~\mathrm{GV/m}$ trajectory, $\eta:1\to8$.  Panel (a) shows the selected transverse-dipole mode support at $\facc=\fminacc$; panel (b) shows the raw longitudinal bremsstrahlung spectrum before polarization selection.  In both panels $\theta=0$ lies at the right edge.  Each spectral power density is normalized independently to its panel maximum.}
\label{fig:signal-classical-spectrograms}
\end{figure*}

Because the fit samples the thermal tail, the finite-duration infrared cutoff is a central design parameter.  We therefore operate below the maximum wakefield gradient: lower $E$ lengthens the interaction and lowers $\fminacc$, whereas higher $E$ raises $\fU$ but also raises the cutoff.  At fixed $\Delta\eta$, their ratio is unchanged, so the chosen range balances interaction duration and stage length against the {laboratory sideband-offset band} and detector response.

{The receiver must preserve the chirped phase~\cite{WangTHz19,SalaminTHz19,BeneaChelmus19}.  Mode-selective conversion sorts the $z_{\mathrm I}$ and $z_{\mathrm{II}}$ partner packets into a common optical band~\cite{Ansari17,Serino25,Vollmer14}; photon counting measures $N_{\mathrm I}$ and $N_{\mathrm{II}}$~\cite{Huang21MIR,Vishnuradhan26}, while dual balanced homodyne measures $M_{\mathrm{I,II}}$~\cite{Babichev04,Tasker24}.  These reconstruct the ordered counts $n_\pm$ defined above~\cite{Clerk10,Weinstein14}. Subscripts $0$, $g$, and $e$ label the transition-detuned reference, lower-state, and upper-state preparations. The leading-order mode changes are
\begin{equation*}
 P_{\rm abs}^{R}=n_{+,0}-n_{+,g},\qquad
 P_{\rm em}^{R}=n_{+,e}-n_{+,0},
\end{equation*}
which are proportional to $S_\tau(+\facc)$ and $S_\tau(-\facc)$, respectively.  We denote the weak-partner counting efficiency by $\eta_{\rm count}^{(N)}$ and the covariance measurement efficiency by $\eta_{\rm tom}$; the loss model and calibration are discussed in the Supplemental Material.}  {For a proper-time acceleration duration $\Delta\tau$, finite acceleration restricts the resolvable band to}
\begin{equation}
    \fminacc\simeq\frac{1}{\Delta\tau},
    \qquad
    \frac{\fminacc}{\fU}=\frac{4\pi^2}{\Delta\eta},
\end{equation}
Naively extrapolating $P_{\rm abs}^{R}/\Delta\tau$ as a constant proper-time rate for uninterrupted acceleration from rest gives a median laboratory waiting time for one Rindler absorption exceeding $10^{30000}~\mathrm{s}$, since $t=(c/a)\sinh(a\tau/c)$. This being rather impractical for an experimental program, we instead repeat finite accelerations of fresh bunches, thereby avoiding this cumulative time dilation.

{Let $\nu_{\mathrm{II},\rm out}(E,\kappa)$ denote the converted optical frequency of the weak Minkowski partner II.}  The absolute weak-branch scale is, for $\kappa=\facc/\fminacc$ and $\kappa_*=1.2$,
\begin{equation*}
\begin{split}
\overline P&_{\mathrm{II},\rm det}^{\rm arrival}(\kappa,E)
\approx4.7\times10^{-15}~\mathrm W
\left(\frac{Q_b}{0.65~\mathrm{nC}}\right)
\left(\frac{{\cal R}_{\rm src}}{0.16}\right)\\
&\quad\times
\left(\frac{\eta_{\rm count}^{(N)}}{0.018}\right)
\left(\frac{E}{1~\mathrm{GV/m}}\right)^2
\left(\frac{\kappa}{1.2}\right)^2\\
&\quad\times
\frac{\nu_{\mathrm{II},\rm out}(E,\kappa)}{300~\mathrm{THz}} 
\exp\!\left[-\frac{4\pi^2}{\Delta\eta}(\kappa-\kappa_*)\right].
\end{split}
\end{equation*}
{The Supplemental Material derives the source response and efficiencies. Balanced acquisition requires roughly 5.7 years at the target usable rate of 100~Hz before scans.}

We use
$D_{\rm cl}\equiv E_{\rm cl}^{\rm shot}/(P_{\rm abs}^{R}h\bar\nu_{\mathrm{II},\rm lab})$
for the classical-to-weak energy ratio. Here $E_{\rm cl}^{\rm shot}$ is the
all-angle radiation energy per shot
from a coherently accelerated bunch, before receiver rejection.
The mean lab frequency of partner~II in the forward cone is
$\bar\nu_{\mathrm{II},\rm lab}\simeq\kappa(22~\mathrm{THz})(E/[1~\mathrm{GV/m}])$. Rejection of the large classical background $D_{\rm cl}\simeq6.3\times10^{21}$ relies on precise prediction of the coherent classical waveform using beam diagnostics. We target
$L_{\rm tom}=\rho_{\rm pol}L_{\rm rest}=2\times10^{-16}$, combining
polarization leakage $\rho_{\rm pol}=2\times10^{-10}$ with subsequent mode
leakage $L_{\rm rest}=10^{-6}$, calibrated on the transmitted waveform.
Raw-energy-normalized mean bias $\Lambda^{\rm bias}$ and bounded
systematics $\Lambda^{\rm other}$ give the total allowance
$\Xi_z^{\rm tot}=D_{\rm cl}(|\Lambda^{\rm bias}|+\Lambda^{\rm other})
+z\sigma_{\Xi}^{\rm stat}$, where $z$ sets the statistical allowance.
The modeled relative classical uncertainty after mean correction is
\begin{equation*}
\begin{aligned}
 \sigma_{\Xi}^{\rm stat}
 &\simeq D_{\rm cl}\frac{\sigma_A^2L_{\rm tom}}{\sqrt{N_{\rm eff}}}\\
 &\simeq 2.3\times10^{-5}
 \left(\frac{Q_b}{0.65~\mathrm{nC}}\right)
 \left(\frac{0.16}{{\cal R}_{\rm src}}\right)
 \left(\frac{1~\mathrm{GV/m}}{E}\right)\\
 &\quad\times
 \left(\frac{1.2}{\kappa}\right)^3
 \frac{\exp[4\pi^2\kappa/\Delta\eta]-1}
      {\exp[4\pi^2(1.2)/\Delta\eta]-1}\\
 &\quad\times
 \left(\frac{\sigma_A}{10^{-3}}\right)^2
 \left(\frac{L_{\rm tom}}{2\times10^{-16}}\right)
 \left(\frac{3\times10^9}{N_{\rm eff}}\right)^{1/2}.
\end{aligned}
\end{equation*}
It rises to $4.5\times10^{-4}$ at $\kappa=2$, where
requiring other classical contributions below $0.005$ gives $\Xi_5^{\rm tot}\lesssim0.0073$. The Supplemental Material develops
the full readout sensitivity and control requirements.}
\section{Discussion and outlook}
This Letter has proposed an experimental protocol for testing a prediction of the thermal nature of the Minkowski vacuum according to an accelerating probe: the Unruh detailed balance ratio. This prediction is a direct consequence of locality and Lorentz covariance of quantum field theory. A positive detection of the detailed balance ratio would strengthen confidence in the framework of relativistic quantum field theory as a non-gravitational description of nature. A clean observation of the ratio would furthermore allow quantitative bounds on departures from KMS behavior and hence on possible deviations from local quantum field theory. 

Conversely, a well-controlled null result would challenge a central pillar of local quantum field theory, which underlies all of particle physics as well as perturbative quantum gravity. The implications for the latter are significant: the same near-horizon quantum-field structure underlies Hawking radiation from black holes \cite{Haw74,Haw75,Crispino08}. Observers must accelerate to hover outside of the black hole horizon: in the absence of a violent breakdown of effective field theory at the event horizon, such as the proposed firewall scenario \cite{AMPS}, this signals that a stationary external observer perceives the black hole to be in an approximately thermal state and thus radiating. A violation of the Unruh effect in a non-gravitational setting would suggest modifications may be necessary to the current understanding of Hawking radiation and black hole evaporation. 

{The experiment could be repeated at wakefield facilities worldwide, using distinct accelerator architectures, beam environments, and detector implementations as a cross-check on facility-specific systematics.  Reproducing detailed-balance slopes across sites would provide a particularly strong validation of the signal.} The next stages of the project require detailed simulations of the machine integrated with the radiation generation process to inform the design of the detection system. Co-optimizing the running of the accelerator with the frequency range and power of the signal, along with minimizing environmental backgrounds, provides an exciting experimental challenge to match the important theoretical implications of the discovery.

\section{Acknowledgments}
We thank Cameron Geddes and Brian Naranjo for information on the current performance of wakefield accelerators as well as Elliott Gesteau, Dolev Bluvstein, and William Unruh for valuable comments and discussion. D.L.D. acknowledges support as a Black Hole Initiative Fellow, and has benefited from the activities of COST Action CA23115: Relativistic Quantum Information, funded  by COST (European Cooperation in Science and Technology). His contribution to this publication is funded in part by the Gordon and Betty Moore Foundation (Grant \#13526). It was also made possible through the support of a grant from the John Templeton Foundation (Grant \#63445). The opinions expressed in this publication are those of the author(s) and do not necessarily reflect the views of these Foundations. The work of N.E. is supported in part by the Department of Energy under Early Career Award DE-SC0021886 and the HEP-QIS program under grant number DE-SC0025937, by the Heising-Simons Foundation under grant no. 2023-4430, by the Moore Foundation via the Black Hole Initiative, and by the MIT Physics Department. L.W. was supported by NSF award 2411650, DOE Office of Nuclear Physics award DE-SC0011091 and DOE QuantISED 2.0 award DE-SC0026237 and is now supported by the Max Planck Society.


\section{Appendix: Classical-field response of the mode-matched detector}

\label{app:classicalPrediction}

\vspace{-0.4cm}

{This Appendix analyzes the response of a mode-matched detector to the classical field sourced by an accelerating charge, including possible stochastic fluctuations of the field. We show that a deterministic classical field produces a coherent
mean waveform, while a stochastic classical field produces an apparatus- and
beam-dependent covariance.  Neither case produces the ordered KMS detailed
balance relation  in the proposed measurement ~\cite{Boyer80,HiguchiMatsas93,PenaSudarsky14}.}

\noindent \textbf{Definite classical field--} {Let the reciprocal bunch transition couple to the same real field through the two conjugate proper-time phases.}  For a horn at polar angle \(\theta\), the retarded time of
radiation emitted by a charge with rapidity \(\eta=a\tau/c\) is
\begin{equation*}
 u(\eta,\theta)=\frac{c}{a}
 \left[\sinh\eta-\cos\theta\,(\cosh\eta-1)\right].
 \label{eq:app-classical-u}
\end{equation*}
The far-zone classical radiation field of a longitudinally accelerated charge
has the schematic form {\cite{Jackson99,Boulware80}}
\begin{equation*}
Z_{\rm cl}(\eta,\theta)
 \propto
 \frac{\sin\theta}{\gamma^3(1-\beta\cos\theta)^3}.
\end{equation*}
Here $\gamma=\cosh\eta$ and $\beta=\tanh\eta$; polarization leakage enters the receiver response.  The important point is
that \(Z_{\rm cl}\) is a reproducible \(c\)-number waveform fixed by the
classical trajectory.  The matched classical drive is
{
\begin{equation*}
 a^{\rm cl}_{\sigma}(\facc,\theta)
 =
 {\cal N}
 \int_{\eta_i}^{\eta_f} d\eta\,
 W(\eta,\theta) Z_{\rm cl}(\eta,\theta)
 e^{i\sigma 2\pi \facc c\eta/a},
 \label{eq:app-classical-mode-amplitude}
\end{equation*}
}
where $\sigma=\pm$, ${\cal N}$ is a common normalization, and \(W\) includes polarization leakage, the receiver window, and the Jacobian \(du/d\eta\).  In this first approximation the same real, frequency-independent $W$ is used in both settings, and the real classical waveform obeys
\begin{equation}
 a^{\rm cl}_{-}(\facc,\theta)
 =
 \left[a^{\rm cl}_{+}(\facc,\theta)\right]^*,
 \qquad
 |a^{\rm cl}_{-}|^2=|a^{\rm cl}_{+}|^2 .
 \label{eq:app-classical-conjugate}
\end{equation}
{Thus a deterministic classical radiation field drives the reciprocal induced-transition probabilities equally. By contrast, Eq.~\eqref{eq:unruh-log-slope-main} gives $S_\tau(+\facc)/S_\tau(-\facc)=1.15\times10^{-3}$ at the reference point $\facc=1.2\fminacc$.}

{A simplified  bunch model uses ground-state excitation for the $+$ channel and excited-state de-excitation for the $-$ channel.  In proper time it couples to the collected classical field through}
\begin{equation}
 H_{\rm int}(\tau)=
 -\chi(\tau)\,\hat{\bf d}(\tau)\cdot{\bf E}_{\rm cl}(\tau),
 \label{eq:app-classical-hint}
\end{equation}
where {\(\chi\) is the same real finite switching function, taken independent of $\facc$ over the fit band,} and
\begin{equation}
 \hat{\bf d}(\tau)
 =
 {\bf d}_{eg}\sigma_+ e^{i2\pi \facc\tau}
 +
 {\bf d}_{eg}^*\sigma_- e^{-i2\pi \facc\tau} .
\end{equation}
{Here ${\bf d}_{eg}$ is the transition dipole matrix element and $\sigma_\pm$ are the corresponding raising and lowering operators.}
To first order in the coupling, the excitation and de-excitation probabilities
are, in units where $\hbar=1$,
\begin{align}
 P_{g\to e}^{\rm cl}
 &=
 \left|
 {\bf d}_{eg}\cdot
 \int d\tau\,\chi(\tau){\bf E}_{\rm cl}(\tau)e^{i2\pi \facc\tau}
 \right|^2, \\
 P_{e\to g}^{\rm cl}
 &=
 \left|
 {\bf d}_{eg}^*\cdot
 \int d\tau\,\chi(\tau){\bf E}_{\rm cl}(\tau)e^{-i2\pi \facc\tau}
 \right|^2 .
 \label{eq:app-classical-antenna-probs}
\end{align}
For a real field and real switching function the two Fourier components are
complex conjugates, so \(P_{g\to e}^{\rm cl}=P_{e\to g}^{\rm cl}\) for the
same antenna polarization. A classical wave can drive the antenna, but it
stimulates the upward and downward transitions symmetrically.  It supplies a
first moment.  After subtracting the reproducible coherent waveform, the
connected matched-mode covariance is zero in the ideal deterministic classical
model.

\noindent \textbf{Stochastic classical field--} 
A more generous classical counterfactual is to allow the field to fluctuate
from shot to shot {\cite{Boyer80,HiguchiMatsas93}}.
\begin{equation}
 C^{\rm cl}_{ij}(\tau,\tau')
 = C^{\rm cl}_{ji}(\tau',\tau)=
 \left\langle
 \delta E_i(\tau)\delta E_j(\tau')
 \right\rangle_{\rm cl}.
 \label{eq:app-classical-covariance}
\end{equation}
Here $\delta E_i$ is the fluctuation about the ensemble mean.  Averaging the antenna probabilities over this classical ensemble gives
\begin{align*}
 \left\langle P_{g\to e}^{\rm cl}\right\rangle
 &=d^i_{eg}d^{j*}_{eg}
 \int d\tau d\tau'\,\chi(\tau)\chi(\tau')\\[-1mm]
 &\quad\times e^{i2\pi \facc(\tau-\tau')}
 C^{\rm cl}_{ij}(\tau,\tau'),\\
 \left\langle P_{e\to g}^{\rm cl}\right\rangle
 &=d^{i*}_{eg}d^{j}_{eg}
 \int d\tau d\tau'\,\chi(\tau)\chi(\tau')\\[-1mm]
 &\quad\times e^{-i2\pi \facc(\tau-\tau')}
 C^{\rm cl}_{ij}(\tau,\tau').
\end{align*}
Relabeling \(\tau\leftrightarrow\tau'\) and \(i\leftrightarrow j\) in the
second expression, and using Eq.~\eqref{eq:app-classical-covariance}, shows
that a classical stochastic field under the same two conjugate coupling settings gives equal upward and
downward response.  Equivalently, the same ideal ordering-sensitive transducer gives equal ensemble-averaged responses in its two hardware settings.  There is no operator-ordering distinction and no
classical analogue of the quantum \(n\) versus \(n+1\) asymmetry.

Classical accelerator jitter does produce a residual covariance after mean
subtraction.  If \(\lambda_i\) denote shot-to-shot beam, plasma, or detector
nuisance parameters, then to leading order
\begin{align*}
 \delta a^{\rm cl}_{\sigma}(\facc)
&=
 \sum_i
 \frac{\partial a^{\rm cl}_{\sigma}(\facc)}{\partial\lambda_i}
 \delta\lambda_i,\\
C^{\rm cl}_{\sigma\sigma'}(\facc)
 &=
 \sum_{ij}
 \frac{\partial a^{\rm cl}_{\sigma}}{\partial\lambda_i}
 \Sigma_{ij}
 \frac{\partial a^{\rm cl *}_{\sigma'}}{\partial\lambda_j},
 \qquad
 \Sigma_{ij}=\langle\delta\lambda_i\delta\lambda_j\rangle .
\end{align*}
Here $\sigma,\sigma'=\pm$ label the two response channels.
This is precisely the residual classical contamination budgeted in the main
text.  Its size and spectral structure are set by the accelerator covariance
matrix \(\Sigma_{ij}\), detector transfer functions, and derivatives of
the classical waveform.  They are not fixed by Lorentz invariance or
proper acceleration, and they do not imply
\(T_{\rm DB}(\facc)=\hbar a/(2\pi k_B c)\) over an accelerated-frame frequency band.

Thus an arbitrary classical two-mode covariance may mimic the mean partner occupations. The anomalous covariance, however, violates the classical bound $|M_{\mathrm I,\mathrm{II}}|^2\leq N_{\mathrm I}N_{\mathrm{II}}$, but does not in itself witness the Unruh effect. The complete test combines all three to determine the unequal reciprocal bunch-transition rates.  A spurious
asymmetry can arise only from nonreciprocal source–detector
coupling, channel-dependent gain or added noise, conjugate-
mode leakage, or a setting-dependent active source. These 
instrumental systematics must be bounded by wrong-chirp,
synthetic-injection, and acceleration-scaling controls.
\bibliographystyle{apsrev4-2-author-truncate}
\bibliography{all}
\end{document}


\title{Supplemental Material for ``Measuring the Acceleration-Dependent Temperature of the Minkowski Vacuum''}

\author{Daine L. Danielson}
\email{daine@mit.edu}
\affiliation{Center for Theoretical Physics --- a Leinweber Institute, Massachusetts Institute of Technology, Cambridge, MA 02139, USA}
\affiliation{Black Hole Initiative, Harvard University, Cambridge, MA 02138, USA}
\author{Netta Engelhardt}
\email{engeln@mit.edu}
\affiliation{Center for Theoretical Physics --- a Leinweber Institute, Massachusetts Institute of Technology, Cambridge, MA 02139, USA}
\affiliation{Black Hole Initiative, Harvard University, Cambridge, MA 02138, USA}
\author{Lindley A. Winslow}
\email{lindley.winslow@mpp.mpg.de}
\affiliation{Laboratory for Nuclear Science, Massachusetts Institute of Technology, Cambridge, MA 02139, USA}
\affiliation{Max-Planck-Institut f\"ur Physik, 85748 Garching bei M\"unchen, Germany}
\date{\today}
\maketitle

\section{Overview}
\label{sec:overview}
This Supplemental Material develops the Letter's proposed measurement of the
Fulling--Davies--Unruh effect. The observable is the ratio of excitation and
de-excitation responses of a collective degree of freedom in an accelerated
charge bunch coupled to a Rindler wave packet. For eternal acceleration and a
monochromatic proper-time mode, the Kubo--Martin--Schwinger (KMS) condition fixes this ratio.
In the laboratory, the acceleration lasts for a finite proper time, the Rindler
wave packet must be reconstructed from positive-Minkowski-frequency modes, and
the desired signal must be distinguished from the much larger coherent Li\'enard-Wiechert field. We give an
initial assessment of these requirements; a full feasibility study must also
establish the accelerator and receiver performance at a specific facility.

The protocol alternates lower- and upper-state preparations of a transverse,
electric-dipole-active collective degree of freedom while keeping the
longitudinal bunch trajectory and receiver configuration fixed.

The two preparations probe absorption from and emission into the same co-moving wave packet mode (a ``Rindler'' mode). In the inertial laboratory, this packet is represented by paired
positive-Minkowski-frequency wave packets. The receiver measures their
occupations and phase-sensitive covariance after removal of the coherent field.
These moments reconstruct the two ordered Rindler responses, whose ratio is
compared with the detailed-balance prediction for the measured acceleration
history and accepted wave packet.

Before frequency conversion, the receiver selects the polarization perpendicular
to the plane containing each observation direction and the acceleration axis.
The ideal Li\'enard field lies in that plane, so this selection
removes it while retaining a component of the transverse transition field.
The remaining coherent field is predicted from independent beam diagnostics and
control measurements. Where needed, it is displaced optically before photon
counting. Its residual mean and covariance are propagated through the same
reconstruction as the science signal. Mean corrections carry their own
uncertainty, and fluctuations about the corrected mean contribute to the
statistical covariance.

Three levels of the measurement must remain distinct:
\begin{description}
\item[Source response] The absorption and emission responses
$P_{\rm abs}^{R}$ and $P_{\rm em}^{R}$ into the selected field sector. They
coincide with single-transition probabilities at leading order.
\item[Field moments] The partner occupations $N_{\mathrm I}$ and
$N_{\mathrm{II}}$, their anomalous covariance $M_{\mathrm I,\mathrm{II}}$, and
the reconstructed ordered counts $n_\pm$ of the Rindler wave packet.
\item[Receiver outputs] Converted photon counts and homodyne quadratures,
related to the field moments by a calibrated transfer map.
\end{description}

We first formulate detailed balance for a finite-duration Rindler wave packet
and describe its Minkowski partners. We then relate the packet to laboratory
frequencies, specify the bunch response, and explain how calibrated measurements
reconstruct it. The remaining sections compare classical-field predictions,
quantify uncertainties, describe frequency and acceleration tests, and assess
the experimental benchmark.

\section{Detailed balance for a finite-duration Rindler wave packet}
\label{sec:finite-detailed-balance}

\subsection{Reciprocal transition response}

Let $\facc>0$ denote ordinary frequency, rather than angular frequency,
conjugate to the proper time $\tau$ of the accelerating bunch. Consider an ideal
reciprocal two-level degree of freedom with level spacing $h\facc$. If that
degree of freedom is prepared in its lower state, it can be excited by absorbing
a Rindler quantum. If it is prepared in its upper state, it can de-excite by
emitting a Rindler quantum. We adopt the sign convention of the Letter,
\begin{align}
 S_\tau(+\facc)&\ \longleftrightarrow\
 \text{bunch excitation / Rindler absorption},\nonumber\\
 S_\tau(-\facc)&\ \longleftrightarrow\
 \text{bunch de-excitation / Rindler emission}.
 \label{eq:sign-convention}
\end{align}
We call the $+\facc$ response the \emph{weak branch}: it measures absorption
from the thermal bath. The stronger $-\facc$ response includes spontaneous
emission as well as stimulated emission.

For a thermal bosonic mode with mean occupation $n$, the upward and downward
responses are proportional to $n$ and $n+1$. Their difference is fixed by the
canonical commutator, while their ratio is fixed by temperature:
\begin{equation}
 \frac{S_\tau(+\facc)}{S_\tau(-\facc)}
 =\frac{n}{n+1}
 =\exp\!\left[-\frac{h\facc}{k_B T}\right].
 \label{eq:ideal-detailed-balance}
\end{equation}
For the Minkowski vacuum sampled along a uniformly accelerated trajectory,
$T=\TU=\hbar a/(2\pi c k_B)$. It is convenient to introduce
\begin{equation}
 \fU\equiv\frac{k_B\TU}{h}=\frac{a}{4\pi^2c}.
 \label{eq:unruh-frequency}
\end{equation}
The ideal Unruh prediction is therefore
\begin{equation}
 \ln\!\frac{S_\tau(-\facc)}{S_\tau(+\facc)}
 =\frac{\facc}{\fU}.
 \label{eq:ideal-unruh-slope}
\end{equation}
Equations~\eqref{eq:ideal-detailed-balance} and
\eqref{eq:ideal-unruh-slope} define the central observable. The experiment
compares two reciprocal transition responses made with the same source matrix
element and the same field wave packet. The common matrix element cancels in
the ratio, while the quantum $+1$ does not. This is the same ordered-noise
distinction used in quantum-noise spectroscopy and motional-sideband-asymmetry
measurements \cite{Clerk10,Aspelmeyer14,Weinstein14}.

\subsection{Experiment-defined finite-duration wave packet}

The ideal discussion above refers to a monochromatic Rindler mode extending for
all proper time. A finite accelerator can neither prepare nor measure such a
mode exactly. The acceleration history and source-interaction profile determine
a target proper-time waveform. The receiver accepts a limited range of angles,
polarizations, and frequencies; its impulse response specifies how a short input
pulse is transmitted. These properties determine the waveform that can be
reconstructed. After canonical normalization, its annihilation operator is
\begin{equation}
 \delta\hat b_f
 =\int_0^\infty\dd\mu(\lambda)
 \left[\alpha_f(\lambda)\hat a_\lambda
       +\beta_f(\lambda)\hat a_\lambda^\dagger\right].
 \label{eq:finite-mode}
\end{equation}

Here $\lambda$ labels the accepted laboratory frequencies, directions, and
optical paths, with the local polarization fixed by the selection described
below.

  The term proportional to $\beta_f$ expresses
the negative-Minkowski-frequency component of the Rindler waveform through
creation operators of positive-energy laboratory modes.

The wave packet is canonically normalized when
\begin{equation}
 [\delta\hat b_f,\delta\hat b_f^\dagger]=1,
 \qquad
 \int\dd\mu(\lambda)
 \left(|\alpha_f(\lambda)|^2-|\beta_f(\lambda)|^2\right)=1.
 \label{eq:finite-mode-normalization}
\end{equation}
In the Minkowski vacuum, the two ordered counts are
\begin{equation}
 n_{+,0}\equiv\langle\delta\hat b_f^\dagger\delta\hat b_f\rangle
 =\|\beta_f\|^2,
 \qquad
 n_{-,0}\equiv\langle\delta\hat b_f\delta\hat b_f^\dagger\rangle
 =\|\alpha_f\|^2=n_{+,0}+1.
 \label{eq:finite-ordered-counts}
\end{equation}
The exact detailed-balance prediction for this particular wave packet is
\begin{equation}
 R_f\equiv\frac{n_{+,0}}{n_{-,0}}
 =\frac{\|\beta_f\|^2}{\|\alpha_f\|^2}.
 \label{eq:finite-ratio}
\end{equation}
If the packet is narrow in boost frequency, this reduces to the monochromatic
expression used in the Letter,
\begin{equation}
 R_f\longrightarrow\exp\!\left(-\frac{\facc}{\fU}\right).
 \label{eq:narrowband-kms}
\end{equation}

An intermediate representation is useful when the packet is expanded in
orthonormal right-Rindler modes. Let $g_f(\omega)$ be a normalized profile in
boost angular frequency,
$\int_0^\infty\dd\omega\,|g_f(\omega)|^2=1$. Its source-off occupation is
\begin{equation}
 n_{+,0}
 =\int_0^\infty\dd\omega\,
 |g_f(\omega)|^2
 \frac{1}{\exp(2\pi c\omega/a)-1}.
 \label{eq:boost-profile-occupation}
\end{equation}
Thus a finite-duration packet averages the Planck spectrum occupation over boost
frequency. The ratio $n_{+,0}/(n_{+,0}+1)$ equals the monochromatic exponential
at one nominal frequency only when the profile is sufficiently narrow. The
Letter uses Eq.~\eqref{eq:narrowband-kms} to display the benchmark and its
scaling. The final experimental fit must instead evaluate
Eq.~\eqref{eq:finite-ratio} using the same acceleration window and calibrated
receiver response used to construct the data wave packet.

Finite-duration packets centered at nearby values of $\facc$ need not be
orthogonal. This has two distinct consequences. First, the statistical
estimators are correlated. If $\underline n_{+,i}$ denotes the estimate for packet
$i$, the analysis must account for
\begin{equation}
 \Sigma^{(f)}_{ij}
 \equiv
 \operatorname{Cov}\!\left(\underline n_{+,i},\underline n_{+,j}\right)
 \label{eq:cross-frequency-covariance}
\end{equation}
rather than treating neighboring frequency points as independent. Second, a
test injection in packet $j$ can reveal an error in the mean response assigned
to packet $i$, relative to the response predicted from their known overlap.
This excess response is receiver cross-talk and enters the leakage and bias
budget below.

\subsection{Reference classes and preparation-resolved response}

A receiver that reconstructs $\delta\hat b_f$ in empty Minkowski space already
finds the occupation $n_{+,0}=\|\beta_f\|^2$. This is a property of the
Minkowski vacuum under the chosen Rindler-mode decomposition and must not be
confused with radiation emitted by the bunch. Conversely, the field change
caused by a bunch transition is conditional on the source preparation. Unruh
and Wald showed that a transition described in the accelerated frame as
absorption of a Rindler quantum can leave a positive-energy Minkowski partner
excitation in the field \cite{UnrWal84}.

The experiment therefore uses three physically distinct baselines:
\begin{enumerate}
 \item {A receiver-only or bunch-absent class measures detector dark
 counts, unwanted light produced by the frequency conversion pump, and the receiver's own source-off
 vacuum response.}
 \item An accelerated, transition-detuned reference class preserves the
 accelerator environment and coherent Li\'enard field while suppressing the
 selected reciprocal bunch transition.
 \item Lower-state and upper-state classes differ from that reference through
 the prepared source response of interest.
\end{enumerate}
Let $n_{+,0}$ be the wave-packet occupation inferred in the detuned reference,
and let $n_{+,g}$ and $n_{+,e}$ be the corresponding occupations after
lower- and upper-state preparation. These lowercase $n$'s are mean occupations,
not probabilities. At the leading order retained here, a transition into the selected sector
changes its Rindler occupation by one quantum. The change in the expectation value then equals the mean transition
probability per accepted shot:
\begin{equation}
 P_{\rm abs}^{R}=n_{+,0}-n_{+,g},
 \qquad
 P_{\rm em}^{R}=n_{+,e}-n_{+,0}.
 \label{eq:preparation-differences}
\end{equation}
This equality holds at the leading order used in the \emph{source map}: the
rule describing how the prepared bunch changes the field state during the
interaction, derived in \hyperref[app:moment-derivation]{the partner-moment appendix}. Beyond this order,
the occupation differences are integrated responses rather than, in general,
transition probabilities. The corresponding transition rates would in that case be inferred
by varying the interaction duration or coupling strength.

The final bunch state need not be measured on each shot. Such a measurement
could provide an independent route to detailed balance, but we expect it to be
more difficult than the field readout proposed here. The initial populations
must still be prepared or independently measured. When pure preparation is
unavailable, a joint fit can use two or more known mixtures with different
lower- and upper-state populations.

This formulation does not require an eternally accelerated detector in a
stationary state to produce a nonzero unconditional far-zone flux. In idealized
stationary models, radiation-like detector terms can cancel against
vacuum--source interference in the renormalized stress tensor
\cite{HuRaval00,Iso13}. The present observable is different: it is the
finite-time change in one field wave packet, compared across reciprocal source
preparations while its definition is held fixed.

\section{Minkowski partner decomposition and Rindler-mode reconstruction}
\label{sec:partners}

\subsection{Canonical partner decomposition}

The Bogoliubov transformation between the accelerated and inertial descriptions
mixes annihilation and creation operators. In the paired representation used
here, suitable orthonormal laboratory wave packets organize this mixing into
pairs, each containing a positive-Minkowski-frequency packet in sector I and
one in sector II. Both partners belong to the selected polarization. This
representation applies to the field transformation; receiver loss is included
separately through its environmental inputs.

Let $r_k\geq0$ specify the creation--annihilation mixing in pair $k$, and let
$c_{f,k}$ be its amplitude in the selected Rindler packet. We write
$\delta\hat a\equiv\hat a-\langle\hat a\rangle$ for an operator with its coherent mean
removed. We fix the phases of a common partner basis before constructing the packets,
so that the coefficients within each pair are real and nonnegative:
\begin{equation}
 \delta\hat b_f
 =\sum_k c_{f,k}\left(
 \cosh r_k\,\delta\hat a_{\mathrm I,k}
 +\sinh r_k\,\delta\hat a_{\mathrm{II},k}^\dagger\right),
 \qquad
 \sum_k|c_{f,k}|^2=1.
 \label{eq:multimode-bogoliubov}
\end{equation}
Each expression in parentheses has unit commutator because
$\cosh^2r_k-\sinh^2r_k=1$. The coefficients $c_{f,k}$ may still be complex;
they specify how the pairs combine into the Rindler waveform selected by the
bunch trajectory, interaction profile, and receiver acceptance. The phase
choice is a convention for this basis, not an assumption that the receiver
introduces no phase shifts. The same convention must be used in the calculated
waveforms, calibrated response, and measured covariances, and held fixed across
the source preparations. When different packets are compared, their relative
phases and overlaps must likewise be evaluated in consistent conventions.

The Letter retains the pair that makes the leading contribution in this
chosen decomposition and neglects the remaining pairs. This \emph{one-pair
approximation} concerns how many laboratory waveforms must be measured; it does
not discard the creation-operator term. Its accuracy must be checked through
the omitted contributions to the occupation and response ratio, as described
below. With one pair retained,
\begin{equation}
 \delta\hat b_f
 =u_f\delta\hat a_{\mathrm I}
 +v_f\delta\hat a_{\mathrm{II}}^\dagger,
 \qquad |u_f|^2-|v_f|^2=1.
 \label{eq:single-pair}
\end{equation}
The Roman numerals label positive-Minkowski-frequency partners, not the signs
of Rindler frequency. In the thermal tail, $|u_f|\simeq1$ and $|v_f|\ll1$,
so partner I carries the larger coefficient. Nevertheless, its annihilation
operator gives zero on the Minkowski vacuum, leaving only the partner-II
creation term:
\begin{equation}
 \delta\hat b_f|0_M\rangle
 =v_f\delta\hat a_{\mathrm{II}}^\dagger|0_M\rangle,
 \qquad
 \delta\hat b_f^\dagger|0_M\rangle
 =u_f^*\delta\hat a_{\mathrm I}^\dagger|0_M\rangle.
 \label{eq:transition-dictionary}
\end{equation}
Thus the weak $+\facc$ absorption transition is accompanied in the laboratory
field by a partner-II excitation, whereas the strong $-\facc$ emission
transition is accompanied by a partner-I excitation.

\subsection{Reconstruction from partner moments}

The receiver should reconstruct the complete Rindler wave packet. Define the centered partner occupations
\begin{equation}
 N_J\equiv
 \langle\delta\hat a_J^\dagger\delta\hat a_J\rangle,
 \qquad J=\mathrm I,\mathrm{II},
 \label{eq:partner-occupations}
\end{equation}
and the phase-sensitive cross correlation
\begin{equation}
 M_{\mathrm I,\mathrm{II}}
 \equiv\langle\delta\hat a_{\mathrm I}
                   \delta\hat a_{\mathrm{II}}\rangle.
 \label{eq:anomalous-covariance}
\end{equation}
The quantity $M_{\mathrm I,\mathrm{II}}$ is conventionally called an
\emph{anomalous covariance} because it correlates two annihilation operators
rather than an annihilation and a creation operator. It records the phase-sensitive correlation needed to combine the two packets
into the selected Bogoliubov mode.

In the one-pair approximation the ordered Rindler counts are
\begin{align}
 n_\pm={}&|u_f|^2\left(N_{\mathrm I}+\frac12\right)
 +|v_f|^2\left(N_{\mathrm{II}}+\frac12\right)
 +2\Ree\!\left(u_f^*v_fM_{\mathrm I,\mathrm{II}}^*\right)
 \mp\frac12.
 \label{eq:reconstruction}
\end{align}
The final $\mp1/2$ terms are a symmetric way of writing normal and antinormal
ordering. Their difference gives
\begin{equation}
 n_--n_+
 =\langle[\delta\hat b_f,\delta\hat b_f^\dagger]\rangle=1.
 \label{eq:canonical-difference}
\end{equation}
Photon counting determines $N_{\mathrm I}$ and $N_{\mathrm{II}}$. Dual homodyne
measurements with a common phase reference determine the real and imaginary
parts of $M_{\mathrm I,\mathrm{II}}$. All three are required in
Eq.~\eqref{eq:reconstruction}. Separate partner counts provide a
detailed-balance cross-check once the counting efficiencies and source
populations are calibrated. They do not by themselves reconstruct the quantum ordered
Rindler response, because they omit the phase-sensitive correlation.

\subsection{How many partner pairs must be measured?}

The one-pair formula used in the Letter is adequate only if a single pair
captures nearly all of the selected Rindler wave packet. This statement can be
made quantitative. Keeping only the first $K$ pairs in
Eq.~\eqref{eq:multimode-bogoliubov} gives
\begin{equation}
 \delta\hat b_f^{(K)}
 =\sum_{k=1}^{K}c_{f,k}\left(
 \cosh r_k\,\delta\hat a_{\mathrm I,k}
 +\sinh r_k\,\delta\hat a_{\mathrm{II},k}^\dagger\right).
 \label{eq:truncated-partner-expansion}
\end{equation}
Its commutator is
\begin{equation}
 [\delta\hat b_f^{(K)},\delta\hat b_f^{(K)\dagger}]
 =C_K,
 \qquad
 C_K\equiv\sum_{k=1}^{K}|c_{f,k}|^2.
 \label{eq:captured-commutator-weight}
\end{equation}
We refer to $C_K$ as the \emph{captured commutator weight}. The condition $C_K=1$ means that the
retained partner pairs reproduce the full unit commutator of the target wave
packet, while $1-C_K$ is the part omitted by the receiver model.
For a single target whose annihilation and creation components lie in orthogonal
laboratory sectors, those components can define one exact pair. Additional pairs
may be required for a family of packets or for the basis realizable by the
receiver.

Before specifying the hardware, one must calculate this decomposition from the
measured or simulated trajectory and interaction profile. The calculation should
report $C_K$ as more pairs are included, the omitted contributions to $n_+$ and
$R_f$, and the effect of unequal losses in the retained pairs. If one pair does
not recover the target to the required accuracy, the receiver must measure
additional partner wave packets or the target must be redesigned.

\section{Mapping Rindler Modes to Laboratory Modes}
\label{sec:kinematics}

\subsection{Acceleration and finite-duration scales}

For an electron subject to an approximately constant longitudinal electric
field $E$, the proper acceleration and rapidity satisfy
\begin{equation}
 a=\frac{eE}{m_e},
 \qquad
 \eta=\frac{a\tau}{c},
 \qquad
 \Delta\tau=\frac{c}{a}\Delta\eta.
 \label{eq:trajectory-scales}
\end{equation}
A record of duration $\Delta\tau$ cannot sharply resolve proper-time frequencies
far below
\begin{equation}
 \fminacc\simeq\frac{1}{\Delta\tau}.
 \label{eq:minimum-frequency}
\end{equation}
The order-unity meaning of ``minimum'' depends on the selected window, but the
scaling is unavoidable. Combining Eqs.~\eqref{eq:unruh-frequency} and
\eqref{eq:minimum-frequency} gives
\begin{equation}
 \frac{\fminacc}{\fU}\simeq\frac{4\pi^2}{\Delta\eta}.
 \label{eq:finite-duration-ratio}
\end{equation}
This identity explains an initially counterintuitive feature of the proposal.
Increasing the wakefield gradient raises the Unruh temperature, but it also
shortens the proper-time duration of a fixed rapidity interval and raises the
infrared cutoff by the same factor. At fixed $\Delta\eta$, the thermal
suppression at a fixed $\kappa\equiv\facc/\fminacc$ is therefore unchanged. A
realistic wakefield experiment operates in the Boltzmann tail and extracts the
slope from several modes rather than observing the peak of the Planck spectrum.

The following trajectory provides a useful reference for estimating receiver
and exposure requirements. Selecting an operating point will require a joint
optimization of accelerator performance, bunch preparation, and detector
capabilities:
\begin{equation}
 E_*=1~\mathrm{GV/m},
 \qquad
 \eta_i=1,
 \qquad
 \eta_f=8,
 \qquad
 \Delta\eta=7.
 \label{eq:reference-trajectory}
\end{equation}
It gives
\begin{equation}
 \Delta\tau_*=12~\mathrm{ps},
 \qquad
 \fU=15~\mathrm{GHz},
 \qquad
 \fminacc=84~\mathrm{GHz}.
 \label{eq:reference-time-frequency-scales}
\end{equation}
The fit benchmark uses $\kappa_*=1.2$, or
\begin{equation}
 f_{\rm acc,*}=100~\mathrm{GHz},
 \qquad
 \exp(-f_{\rm acc,*}/\fU)=1.2\times10^{-3}.
 \label{eq:reference-tail}
\end{equation}

\begin{table}[t]
\centering
\caption{Representative fixed-$\Delta\eta=7$ kinematic scales. Source strength
and receiver efficiency are not included. Increasing $E$ shifts the absolute
proper-time and laboratory-frequency scales, but leaves
$\fminacc/\fU=4\pi^2/7$ unchanged.}
\label{tab:alternative-regimes}
\small
\setlength{\tabcolsep}{7pt}
\renewcommand{\arraystretch}{1.18}
\begin{tabular}{@{}ccccc@{}}
\toprule
$E$ & $\Delta\tau$ & $\fU$ & $\fminacc$ & $1.2\fminacc$\\
\midrule
$0.225~\mathrm{GV/m}$ & $53~\mathrm{ps}$ & $3.3~\mathrm{GHz}$ & $19~\mathrm{GHz}$ & $23~\mathrm{GHz}$\\
$1~\mathrm{GV/m}$ & $12~\mathrm{ps}$ & $15~\mathrm{GHz}$ & $84~\mathrm{GHz}$ & $100~\mathrm{GHz}$\\
$10~\mathrm{GV/m}$ & $1.2~\mathrm{ps}$ & $150~\mathrm{GHz}$ & $840~\mathrm{GHz}$ & $1.0~\mathrm{THz}$\\
\bottomrule
\end{tabular}
\end{table}

\subsection{Retarded time and laboratory frequency}

Radiation emitted at rapidity $\eta$ and observed at polar angle $\theta$
arrives at retarded laboratory time
\begin{equation}
 u(\eta,\theta)
 =\frac{c}{a}\left[
 \sinh\eta-\cos\theta(\cosh\eta-1)
 \right].
 \label{eq:retarded-time}
\end{equation}
It is useful to define
\begin{equation}
 D(\eta,\theta)
 \equiv\cosh\eta-\cos\theta\sinh\eta.
 \label{eq:D-definition}
\end{equation}
Then
\begin{equation}
 \frac{\dd u}{\dd\eta}=\frac{c}{a}D,
 \qquad
 \frac{\dd\tau}{\dd u}=\frac{1}{D}.
 \label{eq:retarded-time-jacobian}
\end{equation}
A proper-time phase $\exp(\mp i2\pi\facc\tau)$ therefore appears to the inertial
laboratory with instantaneous sideband-offset magnitude
\begin{equation}
 \delta\nu_{\rm lab}(\eta,\theta)
 =\facc\frac{\dd\tau}{\dd u}
 =\frac{\facc}{D(\eta,\theta)}
 =\frac{\facc}{\gamma(1-\beta\cos\theta)}.
 \label{eq:sideband-offset}
\end{equation}
Here $\gamma=\cosh\eta$ and $\beta=\tanh\eta$. The quantity
$\delta\nu_{\rm lab}$ is a positive offset magnitude. The opposite Rindler
signs refer to conjugate proper-time phases: when expressed in retarded time
$u$, their phases rotate in opposite directions. They do not signify
negative-energy laboratory photons. Both converted outputs have positive
laboratory frequency.

A frequency converter may route either partner above or below a pump. To avoid
building an implementation-dependent choice into the physics, we write
\begin{equation}
 \nu_{J,\rm out}
 =\nu_{c,J}+s_J\bar\nu_{J,\rm lab}>0,
 \qquad s_J=\pm1,
 \qquad J=\mathrm I,\mathrm{II},
 \label{eq:output-routing}
\end{equation}
where
\begin{equation}
 \bar\nu_{J,\rm lab}
 =\int_0^\infty\dd\nu\,\nu|\psi_{J,f}(\nu)|^2
 \label{eq:mode-weighted-frequency}
\end{equation}

is the exact mean laboratory frequency before conversion. Here
$|\psi_{J,f}(\nu)|^2$ is the normalized marginal spectrum after integrating
over the accepted directions in the selected polarization. For the energy
budget we use the ray estimate derived below,
$\bar\nu_{\mathrm{II},\rm lab}^{\rm est}(\kappa_*)=26~\mathrm{THz}$.
The calibrated optical output remains near $300~\mathrm{THz}$; its photon
energy is set by the conversion process.

\subsection{Angular collection}

The classical Li\'enard radiation has a null along the beam axis. Its normalized instantaneous angular profile is

\begin{equation}
 p(\eta,\theta)
 =\frac{3}{8\pi\gamma^6}
 \frac{\sin^2\theta}{(1-\beta\cos\theta)^5},
 \qquad
 \int p\,\dd\Omega=1.
 \label{eq:normalized-angular-profile}
\end{equation}
For an axially symmetric cone ${\cal C}:0\leq\theta\leq\theta_{\max}$, define
\begin{align}
 F_{\cal C}(\eta)
 &=2\pi\int_0^{\theta_{\max}}\dd\theta\,\sin\theta\,p(\eta,\theta),
 \label{eq:instantaneous-angular-fraction}\\
 \overline F_{\cal C}
 &=\frac{\displaystyle
   \int_{\eta_i}^{\eta_f}\dd\eta\,\cosh\eta\,F_{\cal C}(\eta)}
  {\displaystyle
   \int_{\eta_i}^{\eta_f}\dd\eta\,\cosh\eta}.
 \label{eq:energy-weighted-angular-fraction}
\end{align}
The factor $\cosh\eta$ appears because
$\dd t=(c/a)\cosh\eta\,\dd\eta$ and the total Li\'enard power is constant for
constant longitudinal proper acceleration. A measured nonuniform trajectory is
handled by replacing this weight with the reconstructed radiated-energy weight.

\begin{table}[t]
\centering
\caption{Fractions of the pulse energy collected within a forward cone for the
reference $\eta:1\to8$ trajectory. The classical field and selected azimuthal
signal are normalized to their respective all-angle emitted energies. The
signal column uses the ray weights in Eq.~\eqref{eq:selected-ray-weights};
it does not include the loss from selecting one polarization or subsequent
receiver losses.}
\label{tab:angular-fractions}
\small
\setlength{\tabcolsep}{12pt}
\renewcommand{\arraystretch}{1.16}
\begin{tabular}{@{}ccc@{}}
\toprule
$\theta_{\max}$ & Classical field $\overline F_{\cal C}$ & Selected signal\\
\midrule
$10~\mathrm{mrad}$ & $0.9415$ & $0.9481$\\
$20~\mathrm{mrad}$ & $0.9711$ & $0.9744$\\
$50~\mathrm{mrad}$ & $0.9889$ & $0.9902$\\
$100~\mathrm{mrad}$ & $0.9949$ & $0.9955$\\
$200~\mathrm{mrad}$ & $0.9979$ & $0.9981$\\
$0.5~\mathrm{rad}$ & $0.9996$ & $0.9996$\\
\bottomrule
\end{tabular}
\end{table}

The $0.2~\mathrm{rad}$ aperture used in the Letter therefore contains about
$99.79\%$ of the pulse-integrated Li\'enard radiation in this idealized
trajectory, despite the exact zero on axis.

For the signal, choose the transition dipole along $x$, perpendicular to the
mean acceleration along $z$. At observation angle $(\theta,\phi)$,
$\bm e_\theta$ lies in the meridional plane containing the observation direction
and the $z$ axis, while $\bm e_\phi$ is perpendicular to that plane. The
classical Li\'enard field is polarized along $\bm e_\theta$. The receiver retains
$\bm e_\phi$, giving
\begin{equation}
 \Pi_\phi\bm E_{\rm cl}=0,\qquad
 \Pi_\phi=\bm e_\phi\bm e_\phi^{\mathsf T}
 \label{eq:polarization-selection}
\end{equation}
for the ideal current along the beam axis. This requires a direction-dependent
polarization basis, implemented in separate angular channels or by calibrated
vector optics. Relative phases between opposite azimuths are retained when the
transmitted field is combined. The rejected field is available as a beam monitor.

In the ray approximation, Eq.~(12) of the Letter reads
$\dd P_{\rm sig.}/\dd\Omega=(3P'_x/8\pi\gamma D^3)\sin^2\phi$,
where $P'_x$ is the constant rest-frame dipole power. Integrating over azimuth
with $\dd\Omega=\sin\theta\,\dd\theta\,\dd\phi$, using
$\dd t/\dd\eta=c\gamma/a$ for the source laboratory time, and dividing
by the photon energy $h\facc/D$ gives
\begin{equation}
 \frac{\dd N_{x,\phi}}{\dd\eta\,\dd\theta}\propto
 \frac{3\sin\theta}{8D^2},\qquad
 \frac{\dd E_{x,\phi}}{\dd\eta\,\dd\theta}\propto
 \frac{3\sin\theta}{8D^3}.
 \label{eq:selected-ray-weights}
\end{equation}
The transverse field need not vanish per unit solid angle on axis; the density
per $\dd\theta$ still vanishes there through its angular Jacobian.

The energy budget below compares the classical pulse with the weak signal,
so we need an estimate of the laboratory energy per signal photon. We call
$\nu_0$ the mean laboratory frequency in this ray model when
$\facc=\fminacc$, and obtain it by averaging the Doppler-shifted frequency
with the photon-number weight in Eq.~\eqref{eq:selected-ray-weights}:
\begin{equation}
 \mathcal D_{\cal C}\equiv
 \frac{\displaystyle\int_{\eta_i}^{\eta_f}\dd\eta
 \int_0^{\theta_{\max}}\dd\theta\,\sin\theta\,D^{-3}}
 {\displaystyle\int_{\eta_i}^{\eta_f}\dd\eta
 \int_0^{\theta_{\max}}\dd\theta\,\sin\theta\,D^{-2}},
 \qquad \nu_0=\fminacc\mathcal D_{\cal C}.
 \label{eq:characteristic-lab-frequency}
\end{equation}
The integrals cover the full reference rapidity interval and collection
cone. At the reference trajectory and aperture,
\begin{equation}
 \mathcal D_{\cal C}=2.62\times10^2,\qquad
 \nu_0=22~\mathrm{THz},\qquad
 \bar\nu_{\mathrm{II},\rm lab}^{\rm est}(\kappa)=\kappa\nu_0.
 \label{eq:reference-ray-frequency}
\end{equation}
At $\facc=\kappa\fminacc$, every ray frequency is multiplied by $\kappa$,
so the same average gives $\bar\nu_{\mathrm{II},\rm lab}^{\rm est}=\kappa\nu_0$.
This is the characteristic energy used in the order-of-magnitude budget; the
exact packet mean remains defined by Eq.~\eqref{eq:mode-weighted-frequency}.
Averaging the energies of the accepted photons does not impose another
collection loss on their number.

\subsection{Construction and interpretation of Fig.~3}

Figure 3 of the Letter places two different physical quantities on the same
angle--frequency axes. Panel (a) is an ideal support representation for the selected transverse
transition field at one definite Rindler frequency, chosen for illustration at
\begin{equation}
 \facc=\fminacc=84~\mathrm{GHz}.
\end{equation}
Writing $D=D(\eta,\theta)$, its pulse-integrated density is represented as
\begin{equation}
 \rho_{\rm II}(\theta,\nu)
 \propto
 \int_{\eta_i}^{\eta_f}\dd\eta\,  \frac{3\sin\theta}{8D^3}
 \delta\!\left[
 \log_{10}\nu-
 \log_{10}\!\left(\frac{\facc}{D}\right)
 \right].
 \label{eq:illustrative-partner-support}
\end{equation}

Panel (b) is the Fourier spectrum of the complete real Li\'enard
pulse before polarization selection.
The receiver and
partner decomposition are formulated in terms of positive-Minkowski-frequency
laboratory modes, so only $\nu>0$ is displayed. For a real waveform, the
negative-frequency part is the complex conjugate of the positive-frequency part
and carries no independent power information. The plotted density is
\begin{equation}
 \rho_{\rm cl}(\theta,\nu)
 \propto
 \sin\theta\,(\ln10)\nu
 \left|
 \int\dd u\,
 \frac{\sin\theta}{D[\eta(u),\theta]^3}
 \e^{-i2\pi\nu u}
 \right|^2,
 \qquad \nu>0.
 \label{eq:illustrative-classical-spectrum}
\end{equation}
Both panels use the density $\dd E/(\dd\theta\,\dd\log_{10}\nu)$, the same
trajectory and aperture, and the same axes and color scale. Each is normalized
to its own maximum, so the figure compares where the two spectra lie, not their
absolute quantum-to-classical power ratio.

For this illustration the rapidity interval has unit weight: no taper, detector
response, or image-space smoothing is applied. The wave packet used in the
actual likelihood is instead constructed from the measured acceleration
envelope and the calibrated channel impulse response. Changing that window can
alter three independently testable quantities: the exact ordered ratio $R_f$ of
Eq.~\eqref{eq:finite-ratio}; the captured commutator weight $C_K$ of
Eq.~\eqref{eq:captured-commutator-weight}; and the covariance
$\Sigma^{(f)}_{ij}$ between neighboring wave-packet estimators in
Eq.~\eqref{eq:cross-frequency-covariance}. This statistical covariance differs from an error in the \emph{mean} response
to an injected packet, after allowing for its known overlap with the target;
the latter enters the leakage and bias budget.

It may be possible to shape the wakefield so that the interaction approximates
a smoother temporal window. A taper can reduce Fourier sidelobes, but it can
also reduce the captured signal or increase overlap with an unwanted waveform.
No rejection benefit is credited until the complete source-to-receiver
calculation has been repeated and the improvement has been verified by
end-to-end injections.

Finally, the source duration and the arrival duration are not interchangeable.
At the reference point,
\begin{equation}
 \Delta t_{\rm src}=2.5~\mathrm{ns},
 \qquad
 \Delta u_{\cal C}=51~\mathrm{ps},
 \qquad
 \frac{\Delta t_{\rm src}}{\Delta u_{\cal C}}=49.6.
 \label{eq:time-coordinate-distinction}
\end{equation}
Receiver-side pulse powers must be converted to pulse energies with the arrival
window $\Delta u_{\cal C}$, not with the much longer source-coordinate
duration.

\section{Collective bunch response}
\label{sec:source}

\subsection{Coherent displacement and connected response}

The bunch produces a large, reproducible classical field simply because charge
is accelerated. That coherent displacement is not the quantum response whose
detailed balance we wish to test. To separate the two, write the collective
dipole operator as
\begin{equation}
 \delta\hat{\bm D}
 =\hat{\bm D}-\langle\hat{\bm D}\rangle.
 \label{eq:connected-dipole}
\end{equation}

The coherent current determines the classical Li\'enard waveform. The transverse
dipole fluctuations determine the source's ability to exchange a quantum with
the selected Rindler packet. For the selected transition, we use the correlation of the
$x$ component of the dipole fluctuation.
Their frequency-dependent
response is described by the \emph{connected dipole structure factor}, the
Fourier transform of the dipole-fluctuation correlation:
\begin{equation}
 S_{D_xD_x}^{>}(\omega)
 =\int\dd\Delta\tau\,\e^{i\omega\Delta\tau}
 \left\langle\delta\hat D_x(\tau+\Delta\tau)\delta\hat D_x(\tau)\right\rangle.
 \label{eq:connected-dipole-structure-factor}
\end{equation}
A large bunch energy or rms energy spread does not by itself determine the
occupation or one-quantum response of this particular collective coordinate.
The coherent first moment and the connected fluctuation spectrum must be kept
separate.

\subsection{Bunch collective mode model}

We model one connected, dipole-active collective coordinate by
\begin{equation}
 \hat H_{\rm int}(\tau)
 =\hbar g_f
 \left(
 \hat\sigma_+\delta\hat b_f
 +\hat\sigma_-\delta\hat b_f^\dagger
 \right).
 \label{eq:source-interaction}
\end{equation}
Here $\hat\sigma_+$ and $\hat\sigma_-$ raise and lower the chosen bunch
transition, and $g_f$ is its coupling to the field packet. The same matrix
element governs both directions. The chosen transition is transverse and may have zero mean dipole. Identifying
and preparing that collective oscillation is a target for future bunch modeling, and ultimately
a measurement of its dipole-active fluctuations. Steering or shaping
the bunch may help prepare a suitable mode.

For a narrow proper-time band, we use the oscillator-strength
normalization

\begin{equation}
 \Gamma_0^{\rm band}(\facc)
 =\frac{e^2(2\pi\facc)^2}{2\pi\epsilon_0m_ec^3}\,N_e,
 \qquad N_e=\frac{Q_b}{e}.
 \label{eq:source-rate-envelope}
\end{equation}
This expression uses the dipole oscillator-strength sum of the $N_e$ electrons
as a response envelope. The \emph{oscillator strength} of a transition is a
dimensionless measure of its electric-dipole coupling: it combines the squared
transition dipole moment with the transition frequency so that, when all
allowed transitions are summed, the total strength is fixed by the number of
charged particles. Only a fraction of that total strength will generally lie
in the desired collective coordinate and frequency band.

We separate this coupling strength from the probability of preparing either
source state. The resulting \emph{population-free} interaction strength over the
acceleration interval is
\begin{equation}
 \epsilon_b^2
 =\Gamma_0^{\rm band}\Delta\tau\,{\cal R}_{\rm src}.
 \label{eq:epsilon-b}
\end{equation}

The dimensionless factor ${\cal R}_{\rm src}$ describes how much of this
response enters the selected azimuthal-polarization packet. It includes the
participating directional oscillator strength and source--packet overlap.
Transmission through the filter and subsequent readout losses enter the
receiver efficiencies separately.

For pure lower- and upper-state preparations,
\begin{equation}
 P_{\rm abs}^{R}=\epsilon_b^2 n_{+,0},
 \qquad
 P_{\rm em}^{R}=\epsilon_b^2(n_{+,0}+1).
 \label{eq:source-probabilities}
\end{equation}
Consequently,
\begin{equation}
 \frac{P_{\rm abs}^{R}}{P_{\rm em}^{R}}=R_f,
 \qquad
 P_{\rm em}^{R}-P_{\rm abs}^{R}=\epsilon_b^2.
 \label{eq:source-ratio-difference}
\end{equation}
The first equality is the detailed-balance test for the selected wave packet.
The second isolates the part fixed by its unit commutator. A
source-independent classical drive would instead give equal stimulated upward and
downward responses and has no corresponding $+1$.

\subsection{Microscopic interpretation of \texorpdfstring{${\cal R}_{\rm src}$}{Rsrc}}

For bunch states $|a\rangle$ and $|b\rangle$ with energies $E_a$ and $E_b$,
let $\bm d_{ab}$ be the transition dipole matrix element and $v_{\rm int}$ a
characteristic internal velocity in the bunch rest frame. The signed oscillator
strength and its sum rule are
\begin{equation}
 \mathfrak f_{ab}
 =\frac{2m_e(E_b-E_a)}{3\hbar^2e^2}
  |\bm d_{ab}|^2,
 \qquad
 \sum_b\mathfrak f_{ab}
 =N_e+{\cal O}(v_{\rm int}^2/c^2).
 \label{eq:oscillator-strength-sum-rule}
\end{equation}
For a ground-like state with negligible downward dipole oscillator strength, the sum rule supplies an approximate upper envelope for the total dipole response. \cite{Aucar95}.

Our chosen transition couples through the $x$ component of the dipole.
It is therefore useful to specify the strength in that direction:
\begin{equation}
 f^x_{ab}=\frac{2m_e(E_b-E_a)}{\hbar^2e^2}|d^x_{ab}|^2,
 \qquad \sum_b f^x_{ab}=N_e,
 \label{eq:directional-oscillator-strength}
\end{equation}
The sum rule has the same nonrelativistic domain as
Eq.~\eqref{eq:oscillator-strength-sum-rule}. That equation uses the squared
length of the dipole vector divided by three, whereas $f^x_{ab}$ uses its
$x$ component without this division. Thus an exclusively $x$-directed
transition has $\mathfrak f_{ab}=f^x_{ab}/3$. This factor changes the
normalization convention, not the physical transition strength.

We next determine how much of that transition response remains when the
receiver accepts only azimuthal polarization. The transition is driven by
the electric field along the dipole direction. Let $\hat E_x(\tau)$ be that
component measured in the bunch's instantaneous rest frame, with the $x$
axis transported without rotation along the trajectory. Its vacuum correlation
between two proper times is
\begin{equation*}
 C_x(s)\equiv
 \langle0_M|\hat E_x(\tau)\hat E_x(\tau')|0_M\rangle,
 \qquad s=\tau-\tau'.
\end{equation*}
Uniform acceleration makes this correlation depend only on $s$. Its Fourier
components at $\pm\facc$ determine the two reciprocal transition rates.
For the electromagnetic vacuum,
$C_x(s)=C_0\operatorname{csch}^4y$, with
$C_0=\hbar(a/c)^4/(16\pi^2\epsilon_0c^3)$ and
$y=a(s-i0)/(2c)$ \cite{ZhuYuLu06}. Here
$\operatorname{csch}y=1/\sinh y$, and $i0$ denotes the infinitesimal imaginary
time displacement specifying the vacuum correlation.

Decompose the field into the same azimuthal and meridional plane-wave
polarizations used by the receiver. Their contributions to the $x$-field
correlation, denoted $C_{x,\phi}$ and $C_{x,\theta}$, are
\begin{equation}
 C_{x,\phi}(s)=C_0\left(\frac34\operatorname{csch}^4y+
                         \frac12\operatorname{csch}^2y\right),
 \qquad C_{x,\theta}=C_x-C_{x,\phi}.
 \label{eq:selected-electric-correlation}
\end{equation}
Taking their Fourier transforms at the transition frequency and dividing the
selected contribution by the full one gives the fraction of either reciprocal
stationary rate retained in azimuthal polarization:
\begin{equation}
 \chi_\phi(\facc)=\frac{3(2\pi\facc)^2}{4[(2\pi\facc)^2+(a/c)^2]}.
 \label{eq:polarization-rate-fraction}
\end{equation}
At the reference frequency this fraction is $0.40$. It multiplies absorption
and emission equally, so their detailed-balance ratio is unchanged. The
geometrical $3/4$ fraction of the instantaneous ray pattern is recovered only
when $2\pi\facc\gg a/c$.

To relate this result to the source normalization, let $\Gamma_{x,\rm vac}$
be the spontaneous-emission rate of the same dipole at rest. Let
$\Gamma_{0,x}$ and $\Gamma_{0,\phi}$ be the accelerated rate coefficients
for the full transverse field and the selected polarization, respectively,
before multiplication by $n_{+,0}$ or $n_{+,0}+1$. Then
\begin{equation*}
 \Gamma_{x,\rm vac}=\frac{(2\pi\facc)^3|d_{eg}^x|^2}
 {3\pi\epsilon_0\hbar c^3},\qquad
 \Gamma_{0,x}=\Gamma_{x,\rm vac}
 \left[1+\left(\frac{a}{2\pi c\facc}\right)^2\right],\qquad
 \Gamma_{0,\phi}=\frac34\Gamma_{x,\rm vac}.
\end{equation*}
These expressions describe stationary acceleration with all directions
included. The fraction of the selected radiation represented by the finite
accepted packet enters through the overlap defined next.

To make the wave-packet matching explicit, let $h_\lambda(\xi)$ be the
normalized positive-frequency laboratory waveform emitted by one quantum of
the collective source mode $\lambda$, and let $\psi_f(\xi)$ be the normalized
laboratory waveform selected for the Rindler measurement. The variable $\xi$
collects laboratory frequency, direction, and polarization. Define
\begin{equation}
 {\cal W}_{\lambda f}
 =\left|
 \int\dd\mu(\xi)\,
 \psi_f^*(\xi)h_\lambda(\xi)
 \right|^2.
 \label{eq:source-wavepacket-overlap}
\end{equation}
This number lies between zero and one in the normalized positive-frequency
field inner product. It is the fraction of the one-quantum source waveform
captured by the selected receiver wave packet: it equals one for identical
waveforms and vanishes for orthogonal waveforms.

The source factor is obtained by adding the contributions of the collective
modes in the chosen frequency band. A mode $\lambda$ contributes its
directional strength $f^x_\lambda$ multiplied by the fraction
${\cal W}_{\lambda,\phi f}$ of its azimuthally polarized waveform in the
accepted packet:
\begin{equation}
 {\cal R}_{\rm src}
 =\frac{1}{4N_e}\sum_{\lambda\in{\rm band}}
 f^x_\lambda{\cal W}_{\lambda,\phi f}
 =\frac14\zeta_x{\cal W}_{\phi f},
 \label{eq:microscopic-rsrc}
\end{equation}
Here $\zeta_x=N_e^{-1}\sum_{\lambda\in{\rm band}}f^x_\lambda$ is the fraction
of directional oscillator strength in the band. The quantity
${\cal W}_{\phi f}$ is the average of ${\cal W}_{\lambda,\phi f}$, with
weights proportional to $f^x_\lambda$. Both are at most one for a ground-like state in this band approximation.

The factor $1/4=(1/3)(3/4)$ has two origins. The first factor converts the
directional strength to the normalization of
Eq.~\eqref{eq:source-rate-envelope}; the second is the selected rate relative
to $\Gamma_{x,\rm vac}$, just derived. Thus the polarization-dependent
accelerated response is already included in ${\cal R}_{\rm src}$.

For a particular accelerator, Eq.~\eqref{eq:microscopic-rsrc} requires a model
of the bunch's small collective oscillations. A normal-mode calculation finds
those oscillations directly. A Vlasov calculation evolves the beam's continuous
phase-space density under its self-consistent fields; a particle-in-cell
calculation represents it by numerical particles coupled to fields on a grid.
The dipole-active fluctuations must then be quantized to determine transition
strengths, linewidths, populations, and emitted waveforms. Their overlap with
the selected packet gives ${\cal R}_{\rm src}$. Driving the bunch with a known
field and measuring the response in
Eq.~\eqref{eq:connected-dipole-structure-factor} provides an empirical
calibration. Numerical values below use the 2022 CODATA constants \cite{NISTsigma}.

The reference calculation assumes
\begin{equation}
 {\cal R}_{\rm src}=0.16,\qquad \zeta_x{\cal W}_{\phi f}=0.64.
 \label{eq:selected-source-target}
\end{equation}
Thus the required directional strength and accepted-waveform overlap have
product $0.64$. This is a source target to be checked by the calculations and
measurements above. Keeping ${\cal R}_{\rm src}$ explicit allows the exposure
estimate to be updated when the participating transition is characterized.

\subsection{Reference source probabilities and finite coupling effects}

At the reference point
\begin{equation}
 E=1~\mathrm{GV/m},
 \quad Q_b=0.65~\mathrm{nC},
 \quad \Delta\eta=7,
 \quad \kappa=1.2,
 \quad {\cal R}_{\rm src}=0.16,
 \label{eq:reference-source-inputs}
\end{equation}
we obtain
\begin{equation}
 \epsilon_b^2=0.058,\qquad
 P_{\rm abs}^{R}=6.7\times10^{-5},\qquad
 P_{\rm em}^{R}=5.8\times10^{-2}.
 \label{eq:reference-source-results}
\end{equation}

These probabilities use the narrowband approximation to
$n_{+,0}=\|\beta_f\|^2$; an illustrative finite-window comparison is given
below Eq.~\eqref{eq:moment-change-table}.

These estimates retain the leading term in $\epsilon_b^2$. Here
\emph{finite-coupling corrections} mean departures from this approximation when
the interaction is sufficiently strong or lasts sufficiently long that the
source populations change appreciably.

For illustration, let $t$ be the proper-time interaction duration and
$\Gamma_{0,\rm tot}=\Gamma_{0,\phi}+\Gamma_{0,\theta}$ the common rate scale
including both polarizations. Transitions into modes not collected by the
receiver still affect the source populations. Consider a two-state model with
constant reciprocal rates
\[
 \Gamma_\uparrow=\Gamma_{0,\rm tot}n_{+,0},
 \qquad
 \Gamma_\downarrow=\Gamma_{0,\rm tot}(n_{+,0}+1).
\]
It gives
\begin{align}
 P_{g\to e}^{\rm tot}(t)
 &=\frac{n_{+,0}}{2n_{+,0}+1}
 [1-\e^{-\Gamma_{0,\rm tot}(2n_{+,0}+1)t}],\nonumber\\
 P_{e\to g}^{\rm tot}(t)
 &=\frac{n_{+,0}+1}{2n_{+,0}+1}
 [1-\e^{-\Gamma_{0,\rm tot}(2n_{+,0}+1)t}].
 \label{eq:finite-coupling-two-state}
\end{align}
The ratio remains $n_{+,0}/(n_{+,0}+1)$, but both probabilities fall below
the linear estimate as the populations evolve. The selected occupation changes
are $-\chi_\phi{\cal W}_{\phi f}P_{g\to e}^{\rm tot}$ and
$+\chi_\phi{\cal W}_{\phi f}P_{e\to g}^{\rm tot}$, so this common selection
factor also cancels in their ratio. With $\zeta_x{\cal W}_{\phi f}=0.64$ and
$\zeta_x\leq1$, the total source coupling lies in the range
\begin{equation}
 \epsilon_{\rm tot}^2\equiv\Gamma_{0,\rm tot}\Delta\tau
 =\frac{\epsilon_b^2}{\chi_\phi{\cal W}_{\phi f}}
 \in[0.14,0.23].
 \label{eq:total-coupling-range}
\end{equation}
At the reference point the reduction is approximately $6.9$--$10.5\%$.
Here the population evolution depends only on the present populations, not
their earlier history. The actual response need not satisfy that Markov
approximation \cite{Moustos18}; a duration or coupling scan can test it rather
than presuming it. We retain the leading-order benchmark until the source
dynamics have been established.

\section{Phase-preserving receiver and calibration}
\label{sec:receiver}

\subsection{Receiver requirements}

The Letter's ``$\sim$THz detector array'' collects the radiation; adding the
powers recorded by independent detectors would not recover the desired
phase-sensitive observable. The Rindler packet may span laboratory frequencies
from the millimeter-wave band through the terahertz and mid-infrared. Its
relative phase must be preserved across this range and across the accepted
angles and polarizations.

For implementation, the collected field can be divided into \emph{receiver
channels}: separately calibrated portions specified by frequency range,
direction, polarization, and optical path. We use \emph{channel} only in this
hardware sense. A partner wave packet may span several channels, and a channel
need not correspond to one detector element.

The conversion pump is the strong field that drives translation into the
optical band. A second reference field, the homodyne local oscillator, is
combined with the converted light to measure its field quadratures. These
fields perform different functions even when derived from a common source.
The reference architecture proceeds in six steps:

\begin{enumerate}
 \item Collect the radiation while preserving its polarization and relative phase.
 Before conversion, retain the local $\bm e_\phi$ component and direct the
 rejected $\bm e_\theta$ component to a monitor of the classical field.
 \item Measure, for each retained channel, the amplitude attenuation and phase
 delay introduced by the collection optics, cables or waveguides, converters,
 and readout electronics.
 \item Use phase-locked frequency conversion to translate the partner-I and
 partner-II waveforms into optical modes that can be measured with a common
 phase reference.
 \item Measure each selected partner's occupation by photon counting. If a large
 coherent field remains in that mode, cancel it optically before the counter
 rather than merely subtracting its mean from the recorded counts. Calibrate
 this cancellation along with the rest of the counting chain.
 \item In a phase-sensitive configuration, mix each converted field with a
 strong phase-locked reference and repeat the measurement at several relative
 phases. These balanced-homodyne measurements determine the two phase components
 of the field, called quadratures, and hence the complex partner covariance.
 \item Correct the measured moments with the calibrated transfer map and combine
 them according to Eq.~\eqref{eq:reconstruction}.
\end{enumerate}

Experiments have separately selected and sorted temporal waveforms, measured
two-mode correlations, characterized coherent terahertz fields, and converted
microwave signals to optical frequencies
\cite{Ansari17,Garikapati23,Serino23,Serino25,Babichev04,Eichler11,
panFLASHTHz,liebermeisterTHz,Sahu22,Multani26}. Sum-frequency conversion has preserved squeezed-vacuum quantum coherences from $1550$ to $532$~nm \cite{Vollmer14}. These results support individual
steps, not an existing receiver for the complete wakefield wave packet. The integration of these capabilities into a single experiment represents a significant technical challenge.

On-chip single-photon circuits have demonstrated excitation-light suppression exceeding $95$~dB \cite{Elshaari17}. This is a precedent for filtering spectrally separated pump light, not for the same-band classical-background rejection required here.

A direct active decoder implementing
$\delta\hat b_f=u_f\delta\hat a_{\mathrm I}
+v_f\delta\hat a_{\mathrm{II}}^\dagger$ would provide a useful independent
cross-check once the two-mode receiver has been calibrated and tested. Such a decoder
would be a phase-sensitive parametric device: it mixes the partner-I input with
the creation operator of the positive-frequency partner-II input. The dagger
does not represent a propagating negative-energy wave. Because an active
decoder necessarily adds its own vacuum output and combines the diagonal and
cross-correlation terms before they can be checked separately, it is not the
primary science readout. Joint tomography leaves those contributions available
for independent consistency tests.

\subsection{Partner waveforms along the measured trajectory}

Retaining the single partner pair of Eq.~\eqref{eq:single-pair}, we write the
temporal weights used to select its two waveforms as follows. The functions
$z_{\mathrm I}$ and $z_{\mathrm{II}}$ specify the amplitude and phase assigned
to the field at each arrival time $u$; their proper-time phases are evaluated
along the measured trajectory:
\begin{align}
 z_{\mathrm I}(u;\facc)
 &= {\cal A}_{\mathrm I}{\cal M}(\eta)W(\eta)
 \exp\!\left[-i2\pi\facc\frac{c}{a}\eta(u)\right],\nonumber\\
 z_{\mathrm{II}}(u;\facc)
 &= {\cal A}_{\mathrm{II}}{\cal M}(\eta)W(\eta)
 \exp\!\left[+i2\pi\facc\frac{c}{a}\eta(u)\right].
 \label{eq:partner-kernels}
\end{align}
The exponential factors are the two conjugate proper-time phase histories
mapped into laboratory retarded time. The complex constants ${\cal A}_{J}$ set
the normalization and phase convention. The amplitude factor ${\cal M}$ accounts for propagation, collection, and
projection of the transverse source field into the accepted polarization.
These amplitudes, as well as the displayed phases, are needed to define the
normalized laboratory partner waveforms. The window $W$
contains the finite acceleration envelope and the calibrated temporal response
used to select the partner wave packet. In the Letter's order-of-magnitude estimate it is
taken to be a common real function over the fit band; in the final analysis it
must be measured rather than assumed.

The field coefficients can be calculated by following the coupling along
the measured trajectory. Let $\chi(\tau)$ specify how that coupling is
switched on and off in proper time. Before restricting the collection
aperture, the selected-polarization coefficients have the form
\begin{align}
 \alpha_f^\phi(\nu,\theta,\phi)&\propto-\sqrt\nu\sin\phi
 \int\dd\tau\,\chi(\tau)D(\tau,\theta)
 \e^{i2\pi\facc\tau-i2\pi\nu u(\tau,\theta)},\nonumber\\
 \beta_f^\phi(\nu,\theta,\phi)&\propto-\sqrt\nu\sin\phi
 \int\dd\tau\,\chi(\tau)D(\tau,\theta)
 \e^{i2\pi\facc\tau+i2\pi\nu u(\tau,\theta)}.
 \label{eq:selected-source-coefficients}
\end{align}
The factor $-\sin\phi$ projects the $x$-directed dipole into the azimuthal
polarization, and $D(\tau,\theta)$ transforms the field to the source rest
frame. The exponentials combine the transition's proper-time phase with the
laboratory wave's phase at the source. Their opposite signs in the laboratory
frequency distinguish the annihilation and creation coefficients.

Here the field measure is $\dd^3k$, with $\nu=c|\bm k|/(2\pi)$;
$\sqrt\nu$ is the electric-field normalization in that measure. The aperture
and measured polarization response select the components that reach the
receiver. Finally, rescale the coefficients so that their squared-norm
difference is one, as in Eq.~\eqref{eq:finite-mode-normalization}.
The same coefficients then determine the finite-packet ratio in
Eq.~\eqref{eq:finite-ratio}.

The distinction between the two ordered responses is established by the source
preparation and the ordering-sensitive hardware before a classical data record
is formed. Lower- and upper-state shots couple to $\delta\hat b_f$ and
$\delta\hat b_f^\dagger$, respectively, and the receiver maps the associated
partner moments to calibrated outputs. Software estimates the moments of those
physical output modes; it does not create operator ordering by applying two
filters to one already classical waveform.

\subsection{Pure-loss limit and general transfer map}

The polarization filter also attenuates the signal it transmits. Denote
its power transmission for partner $J$ by $T_{\phi,J}$ and the efficiency
of the remaining receiver by $\eta_{J,0}$. The total is then
$\eta_J=T_{\phi,J}\eta_{J,0}$. We take $T_{\phi,J}=0.90$ in the reference
calculation. Any further loss needed to achieve the proposed rejection must
be included in $\eta_J$ when estimating the required exposure.

The simplest useful receiver model is a lossy but otherwise ideal conversion of
each positive-frequency partner. We use the superscript $(m)$, for
``measured,'' to label the output mode whose moments are sampled by the
receiver. Quantities without this superscript refer to the incident partner
mode, before the receiver losses. The two are related by
\begin{equation}
 \delta\hat a_J^{(m)}
 =\sqrt{\eta_J}\,\e^{i\varphi_J}\delta\hat a_J
 +\sqrt{1-\eta_J}\,\hat v_J,
 \qquad J=\mathrm I,\mathrm{II}.
 \label{eq:pure-loss-map}
\end{equation}
Here $\hat v_J$ is the environmental input associated with loss. For vacuum
environmental modes,
\begin{equation}
 N_J^{(m)}=\eta_JN_J,
 \qquad
 M_{\mathrm I,\mathrm{II}}^{(m)}
 =\sqrt{\eta_{\mathrm I}\eta_{\mathrm{II}}}
 \e^{i(\varphi_{\mathrm I}+\varphi_{\mathrm{II}})}
 M_{\mathrm I,\mathrm{II}}.
 \label{eq:pure-loss-moments}
\end{equation}
Dividing by the calibrated efficiencies and removing the measured phase shifts
recovers the input moments:
\begin{equation}
 N_J=\frac{N_J^{(m)}}{\eta_J},
 \qquad
 M_{\mathrm I,\mathrm{II}}
 =\frac{\e^{-i(\varphi_{\mathrm I}+\varphi_{\mathrm{II}})}}
 {\sqrt{\eta_{\mathrm I}\eta_{\mathrm{II}}}}
 M_{\mathrm I,\mathrm{II}}^{(m)}.
 \label{eq:pure-loss-inversion}
\end{equation}
Correcting the mean response does not recover information lost to the
environment. The uncertainty is divided by the same efficiency as the measured
moment. For fixed calibrated losses and phases, define
$\eta_{\rm tom}=\sqrt{\eta_{\mathrm I}\eta_{\mathrm{II}}}$. For the phase
component of the anomalous covariance that enters the response, the
vacuum-noise estimate in Eq.~\eqref{eq:covariance-uncertainty} gives
\begin{equation}
 \sigma\!\left(\Delta M_{\mathrm I,\mathrm{II}}\right)
 =\frac{\sigma\!\left(\Delta M_{\mathrm I,\mathrm{II}}^{(m)}\right)}
 {\eta_{\rm tom}}
 \simeq\frac{1}{\eta_{\rm tom}\sqrt{2N_{\rm eff}}}.
 \label{eq:input-covariance-uncertainty}
\end{equation}
Here $N_{\rm eff}$ is the effective sample size of the source--reference
comparison, defined below. This loss-amplified readout uncertainty enters the
signal-to-noise ratio in Eq.~\eqref{eq:absorption-snr} and the total propagated
statistical uncertainty $\sigma_{\rm stat}(\Delta n_+)$ in
Eq.~\eqref{eq:full-statistical-variance}. Uncertainty in the
calibrated efficiencies and phases must also be propagated; a mean error left
by that calibration is treated separately as a bias or a bounded offset.

A real receiver can mix the two inputs, change their amplitudes and phases,
and add background fields. For the two partner inputs retained here, these
effects on their mean amplitudes are written as

\begin{equation}
 \bm\alpha_m
 =S_{\rm rx}\bm\alpha_{\rm in}+\bm d_{\rm rx},
 \qquad
 \bm\alpha=(\alpha_{\mathrm I},\alpha_{\mathrm{II}}^*)^{\mathsf T},
 \label{eq:first-moment-receiver-map}
\end{equation}
with
\begin{equation}
 S_{\rm rx}=
 \begin{pmatrix}
 g_{\mathrm I} & \ell_{\mathrm I\leftarrow\mathrm{II}}\\
 \ell_{\mathrm{II}\leftarrow\mathrm I} & g_{\mathrm{II}}
 \end{pmatrix}.
 \label{eq:first-moment-matrix}
\end{equation}
The diagonal entries describe the desired gain and phase of each partner path;
the off-diagonal entries describe unintended mixing between the two paths. At
the second-moment level, the same calibration is summarized by an affine map,
meaning a linear transformation plus an additive offset:
\begin{equation}
 \bm m_{\rm meas}=R_{\rm rx}\bm m_{\rm in}+\bm b_{\rm rx},
 \qquad
 \bm m=(N_{\mathrm I},N_{\mathrm{II}},
 \Ree M_{\mathrm I,\mathrm{II}},
 \Ima M_{\mathrm I,\mathrm{II}}).
 \label{eq:receiver-affine-map}
\end{equation}
With the bunch absent, send a known field through the source region and
measure it after collection, polarization selection, and readout. Repeat the
test for the accepted and rejected polarizations so that desired transmission
and unwanted leakage are determined separately.
With the bunch
present, calibration fields are introduced downstream so that they do not drive
the source. Injecting each partner alone and then both with a varied relative
phase determines the gains, mixing, and phase response in $S_{\rm rx}$ and
$R_{\rm rx}$. Vacuum inputs, loads at known temperatures, and runs with the
conversion pump off determine offsets and added noise.

To recover the input moments, one subtracts $\bm b_{\rm rx}$ from the
measurements and applies $R_{\rm rx}^{-1}$. This inversion becomes unreliable
if distinct inputs produce nearly identical outputs: small measurement or
calibration errors can then produce large errors in the inferred input. Only
wave packets for which this amplification stays within the error budget are
used in the measurement.

If calibration shows that other input modes contribute appreciably, their
amplitudes and correlations must be included in the measured response. The
two-partner matrices above apply only when those additional contributions can
be neglected within the error budget.

Recovering the moments is not sufficient unless the calibration also preserves
the bosonic commutation relations. Collect the two operators into
$\hat{\bm a}=(\delta\hat a_{\mathrm I},\delta\hat a_{\mathrm{II}}^\dagger)^{\mathsf T}$ and
define
\begin{equation}
 J=\begin{pmatrix}1&0\\0&-1\end{pmatrix}.
\end{equation}
The opposite signs record the commutators of an annihilation operator and a
creation operator. Let $J_{\rm add}$ describe the independently calibrated commutator contribution
of environmental fields entering through loss and noise ports. The receiver
must preserve the commutation relations, a check we call \emph{canonical
closure}:
\begin{equation}
 S_{\rm rx}J S_{\rm rx}^\dagger+J_{\rm add}=J_m,
 \label{eq:receiver-canonical-closure}
\end{equation}
where $J_m$ is the output commutator matrix. Equivalently, the target wave
packet and its calibrated companion output must satisfy the corresponding
symplectic consistency relation. The reconstructed target is accepted only if
$[\delta\hat b_f^{\rm(rec)},(\delta\hat b_f^{\rm(rec)})^\dagger]=1$ within the
independent calibration uncertainty. This check should be performed on calibration
data, not inferred from the science difference $n_--n_+$.

\subsection{End-to-end rejection of unwanted modes}

An optical filter's \emph{extinction} describes how strongly it suppresses
unwanted light. Strong suppression by individual components does not guarantee
that a bright classical waveform is absent from the final reconstructed mode.
That must be tested by sending known unwanted inputs through the complete
receiver and applying the same analysis used for the signal.

Inject one test waveform $q$ at a time downstream of the source, with all tests
normalized to the same input power. For a waveform whose ideal target response
is zero, any reconstructed target amplitude measures leakage. After the usual
offset subtraction and receiver correction, let $\bm w_f$ combine the inferred
partner amplitudes into the target. Define
\begin{equation}
 \ell_f(q)
 =\bm w_fS_{\rm rx}^{-1}
 \left[\bm\alpha_m(q)-\bm d_{\rm rx}\right],
 \qquad
 L_f(q)=|\ell_f(q)|^2.
 \label{eq:end-to-end-mode-leakage}
\end{equation}
The number $L_f(q)$ is the fraction of the test-waveform power that the full
analysis mistakenly reconstructs as target-mode power.

One unwanted input is the conjugate sideband: a field on the other side of
the conversion reference that can reach the same output band as the desired
signal. Testing this response is often called an \emph{image-response} test.
Let ${\cal Q}_{\rm conj}$ be the collection of such test waveforms and
${\cal Q}_{\rm L}$ the classical Li\'enard waveforms, including the
perturbations expected from beam fluctuations. The largest measured power
leakages in the two tests are
\begin{equation}
 L_{\rm img}=\max_{q\in{\cal Q}_{\rm conj}}L_f(q),\qquad
 L_{\rm tom}=\max_{q\in{\cal Q}_{\rm L}}L_f(q).
 \label{eq:image-and-total-leakage}
\end{equation}
Thus $L_{\rm tom}$ measures leakage of the Li\'enard field through the
complete receiver and reconstruction, relative to the response to the desired
signal.

For one channel, let $T_\phi$ and $T_\theta$ be the filter's power
transmissions for the desired and unwanted polarizations. Their ratio is
$\rho_{\rm pol}=T_\theta/T_\phi$. After this filter, the remaining receiver
can provide further mode rejection. Denote that additional, signal-normalized
power ratio by $L_{\rm rest}$. Our targets are
\begin{equation}
 L_{\rm tom}=\rho_{\rm pol}L_{\rm rest},\qquad
 \rho_{\rm pol}=2\times10^{-10},\qquad L_{\rm rest}=10^{-6},\qquad
 L_{\rm tom}=2\times10^{-16}.
 \label{eq:leakage-target}
\end{equation}
The product describes the complete rejection only when $L_{\rm rest}$ is
measured on the field that actually leaves the filter.

The polarization filter must be aligned with the polarization we wish to
retain. A small angular error $\delta\psi$ allows additional unwanted light
to pass, increasing the power-leakage ratio from $\rho_{\rm pol}$ for the
correctly aligned filter to approximately
$\rho_{\rm pol}+(\delta\psi)^2$. 
The filter's intrinsic leakage and the contribution from misalignment must
together remain below the required limit.

\subsection{Environmental occupation and usable bandwidth}

Ambient photons in the retained
laboratory channels can add spurious occupation to the reconstructed wave packet. If
the positive-frequency inputs have occupation $n_{\rm env}(\nu)$, then
\begin{equation}
 \langle\delta\hat b_f^\dagger\delta\hat b_f\rangle_{\rm env}
 =\|\beta_f\|^2
 +\int\dd\nu\,
 \left(|\alpha_f|^2+|\beta_f|^2\right)n_{\rm env}(\nu),
 \label{eq:environmental-occupation}
\end{equation}
up to correlations between channels.

For a channel ${\cal C}_j$, define its unnormalized weight in the occupation
integral by
\begin{equation}
 w_j
 =\int_{{\cal C}_j}\dd\mu(\lambda)\,
 \left(|\alpha_f(\lambda)|^2+|\beta_f(\lambda)|^2\right),
 \qquad
 \bar w_j=\frac{w_j}{1+2n_{+,0}},
 \qquad
 \sum_j\bar w_j=1.
 \label{eq:normalized-channel-weight}
\end{equation}
The normalized weight $\bar w_j$ describes how strongly channel $j$
contributes to the packet's occupation response. Its thermal contribution is
$w_j n_{{\rm env},j}$, where $n_{{\rm env},j}$ is the environmental occupation
averaged within that channel with the weights
$|\alpha_f|^2+|\beta_f|^2$ in Eq.~\eqref{eq:normalized-channel-weight}. Thus both the weight
and the thermal occupation matter. A mid-infrared mean output frequency does
not by itself make low-frequency input radiation negligible.

For a thermal load at $4~\mathrm K$,
\begin{equation}
 n_{\rm env}(0.3~\mathrm{THz})\simeq2.8\times10^{-2},
 \qquad
 n_{\rm env}(1~\mathrm{THz})\simeq6.2\times10^{-6}.
 \label{eq:thermal-occupation-examples}
\end{equation}
At the lower frequency, the room-temperature occupation is substantial.
Each input therefore needs a measured thermal background. A \emph{thermal
termination} is an absorbing load placed at an input port; its temperature sets
the radiation entering through that port. Such loads, shielding, and explicit
models of the remaining radiation provide ways to control or characterize the
background. If a channel's contribution cannot be reduced or calibrated to the
required accuracy, it must be excluded and $R_f$, $C_K$, and the statistical
sensitivity recomputed.

\subsection{Minkowski mode photon counting}

Photon counting measures $N_{\mathrm I}$ and $N_{\mathrm{II}}$ for the
Rindler reconstruction. The two counts can also be inspected separately to
check the source response and estimate the optical signal power.
For partner $J$, factor the
probability that one collected partner photon produces one registered optical
count as
\begin{equation}
 \eta_{J,\rm count}^{(N)}  =T_{\phi,J}\,\eta_{J,\rm tm}\,
  \eta_{J,\rm conv}^{\rm ph}\,
  \eta_{J,\rm out}\,
  \eta_{J,d}.
 \label{eq:factorized-counting-efficiency}
\end{equation}

In order, the factors are the filter transmission, the fraction retained
when selecting the partner waveform, the probability of converting an input
photon into the optical mode while preserving phase, the output-path
transmission, and the detector's counting efficiency. These receiver losses
are separate from the source's coupling to the accepted packet, which is
already included in ${\cal R}_{\rm src}$.
Each receiver channel is calibrated separately;
its contribution to the partner wave packet is then included with the measured
weight. 
In the reference design, the remaining counting chain has efficiency $0.020$.
Including the filter transmission gives
\begin{equation}
 \eta_{\rm count}^{(N)}=0.90\times0.020=0.018
 \label{eq:selected-counting-efficiency}
\end{equation}
as the required counting probability for a photon in the
selected input packet.

\section{Preparation-conditioned tomography and statistical sensitivity}
\label{sec:estimator}

\subsection{Source-induced partner moments}

We retain the leading nonzero change in the field moments, of order
$\epsilon_b^2$, where $\epsilon_b$ is the dimensionless interaction amplitude
specified by Eq.~\eqref{eq:epsilon-b}. Higher-order terms in the source map
are neglected; their importance is tested by varying the interaction duration
or coupling. With the partner phases chosen so that $u_f$ and $v_f$ are real,
as in \hyperref[app:moment-derivation]{the partner-moment appendix}, the
lower- and upper-state preparations give
\begin{equation}
\begin{array}{c|ccc|c}
\text{preparation}
&\Delta N_{\mathrm I}
&\Delta N_{\mathrm{II}}
&\Delta M_{\mathrm I,\mathrm{II}}
&\Delta n_+\\
\hline
|g\rangle
&0
&\epsilon_b^2|v_f|^2
&-\epsilon_b^2u_fv_f/2
&-\epsilon_b^2|v_f|^2\\
|e\rangle
&\epsilon_b^2|u_f|^2
&0
&-\epsilon_b^2u_fv_f/2
&+\epsilon_b^2|u_f|^2
\end{array}
\label{eq:moment-change-table}
\end{equation}
The final column is not obtained by counting either positive-frequency partner
alone. It follows by substituting all three measured moment changes into
Eq.~\eqref{eq:reconstruction}. Equation~\eqref{eq:moment-change-table} uses the finite packet's coefficients
$u_f,v_f$ in the single-pair approximation, whereas the numerical estimates
below also use the narrowband approximation.\footnote{For the reference trajectory and collection cone,
evaluating Eq.~\eqref{eq:selected-source-coefficients} with the unit-peak
envelope $\chi(\eta)=\cos^{7/2}[\pi(\eta-4.5)/7]$ on
$1\leq\eta\leq8$, zero otherwise, $\zeta_x=1$, and ideal matching
over the full collected frequency support gives
$P_{\rm abs}^{R}\simeq7.4\times10^{-5}$ at $\kappa=1.2$,
about $10\%$ above the benchmark.
This favorable example supports the order-of-magnitude power and classical contamination estimates,
but the measured interaction window and receiver response must enter
the detailed-balance prediction and sensitivity analysis.}

For the lower-state preparation, the partner-II occupation contributes
$|v_f|^2\Delta N_{\mathrm{II}}=\epsilon_b^2|v_f|^4$ to $\Delta n_+$, while
the anomalous covariance contributes
$2\Ree(u_f^*v_f\Delta M_{\mathrm I,\mathrm{II}}^*)
=-\epsilon_b^2|u_f|^2|v_f|^2$. Their sum is
$-\epsilon_b^2|v_f|^2$. In the thermal tail,
$|u_f|^2/|v_f|^2\simeq870$ at the reference point, so the phase-sensitive
covariance term is much larger than the weighted partner-II number term and is
essential to the negative change in the reconstructed Rindler occupation. For
the upper-state preparation, the positive change is dominated by the partner-I
occupation.

Both preparations must nevertheless be analyzed with the same definition of
the two partner waveforms and the same coefficients $u_f,v_f$ in
Eq.~\eqref{eq:reconstruction}. For each data class, the independently measured
receiver offsets are subtracted, the calibrated gains and phases are used to
infer $N_{\mathrm I}$, $N_{\mathrm{II}}$, and $M_{\mathrm I,\mathrm{II}}$,
and all three moments are combined by that equation. Only then are the source
and reference occupations compared. Calibration parameters may be updated to
track measured drift, but the correction procedure is common to all classes
and is not adjusted to favor either response. Thus the comparison changes the
bunch preparation, not the field observable or its normalization.

At the reference point,
\begin{equation}
 u_*=\sqrt{1+n_{+,0}},
 \qquad
 v_*=\sqrt{n_{+,0}},
 \qquad
 u_*v_*\simeq0.034,
 \label{eq:reference-uv}
\end{equation}
and
\begin{equation}
 \Delta M_{\mathrm I,\mathrm{II},*}
 =-9.9\times10^{-4}.
 \label{eq:reference-anomalous-change}
\end{equation}
The corresponding reconstructed occupations are
\begin{equation}
 n_{+,0}=1.15\times10^{-3},
 \quad
 n_{+,g}=1.08\times10^{-3},
 \quad
 n_{+,e}=5.9\times10^{-2}.
 \label{eq:reference-full-mode-occupations}
\end{equation}

\subsection{Differential effective sample size}

Loss reduces the measured anomalous covariance by the effective two-mode
quadrature efficiency
\begin{equation}
 \eta_{\rm tom}\equiv\sqrt{\eta_{\mathrm I}\eta_{\mathrm{II}}}.
 \label{eq:tomography-efficiency}
\end{equation}
The weak response is a difference of two estimates, one from lower-state shots
and one from reference shots, so uncertainty in both contributes. Let
$N_g^{\rm eff}$ and $N_0^{\rm eff}$ count the equivalent independent records
after allocation among the four homodyne phases, event selection, and correction
for correlations between shots. For independent source and reference estimates with the same single-record
variance, the effective sample size of their difference is
\begin{equation}
 N_{\rm eff}
 =\left[
 (N_g^{\rm eff})^{-1}+(N_0^{\rm eff})^{-1}
 \right]^{-1}.
 \label{eq:differential-exposure}
\end{equation}
A raw accelerator shot count cannot be substituted for
$N_{\rm eff}$.

For quadratures normalized so that the vacuum variance is $1/2$,
\begin{equation}
 \sigma\!\left(\Delta M_{\mathrm I,\mathrm{II}}^{(m)}\right)
 \simeq\frac{1}{\sqrt{2N_{\rm eff}}}.
 \label{eq:covariance-uncertainty}
\end{equation}

When the uncertainty is dominated by measuring the anomalous covariance
against vacuum fluctuations, Eqs.~\eqref{eq:moment-change-table} and
\eqref{eq:covariance-uncertainty} give the weak-response significance

\begin{equation}
 {\rm SNR}_{\rm abs}
 \simeq
 \eta_{\rm tom}\epsilon_b^2
 \sqrt{n_{+,0}(n_{+,0}+1)}
 \sqrt{\frac{N_{\rm eff}}{2}}.
 \label{eq:absorption-snr}
\end{equation}

\begin{figure}[t]
\centering
\includegraphics[width=0.79\linewidth]{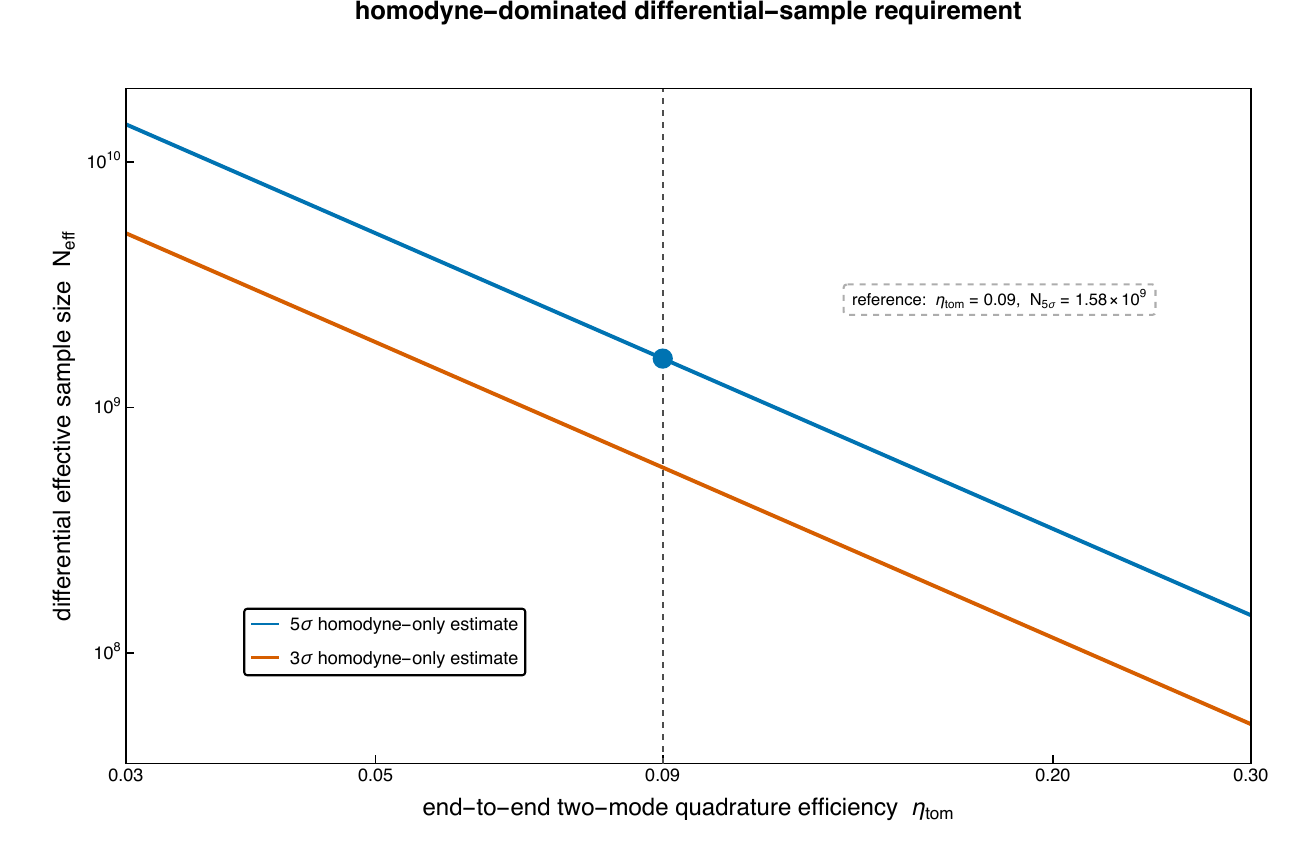}
\caption{Homodyne-only sample requirement for the selected weak response, with
${\cal R}_{\rm src}=0.16$. The curves give the differential effective sample
sizes for nominal $3\sigma$ and $5\sigma$ measurements of the anomalous
covariance. Halving the total efficiency $\eta_{\rm tom}$ requires four times
the exposure. At $\eta_{\rm tom}=0.09$, the $5\sigma$ value is
$1.6\times10^9$. Both preparation classes contribute to this differential
sample size; photon counting, calibration, and other controls add to the
complete measurement cost.}
\label{fig:shot-requirement}
\end{figure}

The reference calculation includes the filter transmission in the total
quadrature efficiency:
\begin{equation}
 \eta_{\rm tom}=0.90\times0.10=0.09,\qquad
 N_{\rm eff}=3.0\times10^9
 \label{eq:selected-homodyne-benchmark}
\end{equation}
The remaining receiver contributes the factor $0.10$. These values give
${\rm SNR}_{\rm abs}=6.9$ in the vacuum-limited approximation above.
Figure~\ref{fig:shot-requirement} shows how this homodyne-only estimate changes
with efficiency. The complete occupation reconstruction also requires photon
counts and calibration, whose uncertainties must be included.

For the counting configuration, take a separate differential effective sample
size $N_{\rm count}=1.5\times10^9$. Let $b_J$ be the residual mean
photon number in partner $J$ after optical background suppression, expressed
at the receiver input by correcting for the counting efficiency. The example
assumes $b_J\leq6\times10^{-4}$. These are photons still present in the
measured mode, not a count remaining after numerical background subtraction.
For coherent or weak thermal light, the count-variance estimate is
\begin{equation}
 \operatorname{Var}(\Delta\underline N_J)
 \lesssim\frac{b_J/\eta_{J,\rm count}^{(N)}+b_J^2}
 {N_{\rm count}}.
 \label{eq:counting-variance-benchmark}
\end{equation}
The first term accounts for random photon arrivals and the uncertainty
amplification when correcting for loss; the second allows for the additional
intensity fluctuations of thermal light. For partner II, use the larger
source-on occupation $b_{\mathrm{II}}+P_{\rm abs}^{R}$.

We propagate the count and homodyne uncertainties through the same occupation
reconstruction, using Eq.~\eqref{eq:full-statistical-variance}. Including an
independent calibration standard uncertainty of $0.03P_{\rm abs}^{R}$ gives
\begin{equation}
 \sigma_{\rm tot}(\Delta n_+)\simeq 1.1\times10^{-5},\qquad
 \frac{P_{\rm abs}^{R}}{\sigma_{\rm tot}}\simeq 6.1.
 \label{eq:full-readout-reference}
\end{equation}
A further mean error bounded by $0.03P_{\rm abs}^{R}$ is treated separately
from this standard uncertainty. Allowing it to reduce the inferred signal
gives a significance of $5.9$. The illustrative finite-coupling reduction in
Eq.~\eqref{eq:total-coupling-range} lowers this to $5.3$--$5.5$, with the
same noise and absolute bias bounds. These values require the stated photon
loads and calibration performance; they are not implied by the polarizer's
extinction alone.

The tighter classical-background budget below applies to the classical part
of this total error. That part must meet its own bound; the $3\%$ aggregate
bias allowance cannot also be assigned to it while claiming a subpercent
classical contribution. If the aggregate bias is reduced to $0.5\%$, the
leading significance becomes $6.1$ at the same variance.

The effective sample sizes also determine the acquisition cost. Equal
lower-state and reference allocations have
$N_g^{\rm eff}=N_0^{\rm eff}=2N_{\rm eff}$, giving
\begin{equation}
 N_{g+0}^{\rm eff}=4N_{\rm eff}=1.2\times10^{10}.
 \label{eq:balanced-record-count}
\end{equation}
Including the separate counting configuration requires
$4(N_{\rm eff}+N_{\rm count})=1.8\times10^{10}$ effective
records. At the target usable rate of $100$~Hz, this is about $5.7$ years of
ideal live time, before upper-state preparations, scans, and calibration runs.
The conversion to running time assumes the same source response, receiver
efficiency, and effective-record yield on each shot. Reusing reference data or
improving efficiency can reduce this cost only as supported by measured
correlations and stability.

\subsection{Effective weak Minkowski mode power}

The preparation-resolved partner counts are
\begin{equation}
 N_{\mathrm{II},\rm det}^{(g)}
 =\eta_{\mathrm{II},\rm count}^{(N)}P_{\rm abs}^{R},
 \qquad
 N_{\mathrm I,\rm det}^{(e)}
 =\eta_{\mathrm I,\rm count}^{(N)}P_{\rm em}^{R}.
 \label{eq:partner-detected-counts}
\end{equation}
For the weak branch, the power during the arrival window is
\begin{equation}
 \overline P_{\mathrm{II},\rm det}^{\rm arrival}
 =\frac{N_{\mathrm{II},\rm det}^{(g)}
        h\nu_{\mathrm{II},\rm out}}
       {\Delta u_{\cal C}}.
 \label{eq:weak-arrival-power}
\end{equation}

The reference weak response gives $1.2\times10^{-6}$ registered photons per
shot. Their energy, divided by the arrival window, is equivalent to an optical
power of $4.7\times10^{-15}$ W. This arrival-window power scales as

\begin{align}
 \overline P_{\mathrm{II},\rm det}^{\rm arrival}(\kappa,E)
 &\simeq 4.7\times10^{-15}~\mathrm W
 \left(\frac{Q_b}{0.65~\mathrm{nC}}\right)
 \left(\frac{{\cal R}_{\rm src}}{0.16}\right)\nonumber\\
 &\quad\times
 \left(\frac{\eta_{\rm count}^{(N)}}{0.018}\right)
 \left(\frac{E}{1~\mathrm{GV/m}}\right)^2
 \left(\frac{\kappa}{1.2}\right)^2
 \frac{\nu_{\mathrm{II},\rm out}}{300~\mathrm{THz}}\nonumber\\
 &\quad\times
 \exp\!\left[-\frac{4\pi^2}{\Delta\eta}(\kappa-\kappa_*)\right].
 \label{eq:weak-power-scaling}
\end{align}
The exact narrowband factor replaces the final exponential by
$(\e^{x_*}-1)/(\e^{4\pi^2\kappa/\Delta\eta}-1)$, with
$x_*=f_{\rm acc,*}/\fU$. This is an auxiliary output-power estimate for the
counting chain, not the complete ordered response. The latter is reconstructed
from all three moments in Eq.~\eqref{eq:reconstruction}.

\subsection{Direct weak-spectrum estimator}
\label{subsec:direct-spectrum}

The weak partner-II branch also furnishes a direct estimate of the occupation
of the selected finite-duration Rindler wave packet. After source--reference subtraction, define
\begin{equation}
 \underline n_{\rm dir}(\facc)
 =\frac{
 \Delta N_{\mathrm{II},\rm det}^{(g)}(\facc)
 -\Delta{\cal B}_{\mathrm{II}}(\facc)
 }{
 \eta_{\mathrm{II},\rm count}^{(N)}(\facc)\,
 \epsilon_b^2(\facc)
 }.
 \label{eq:direct-spectrum-estimator}
\end{equation}
Here $\Delta{\cal B}_{\mathrm{II}}$ is the residual partner-II bias after
subtracting the classical first moment and correcting for receiver baselines. In the
narrowband stationary limit,
$\underline n_{\rm dir}\rightarrow n_U(\facc)$; for the actual experiment
it estimates $\|\beta_f\|^2$.

This estimator uses only the weak partner count and does not determine the
anomalous covariance. It also needs absolute calibration of the bunch response,
preparation population, and full counting chain at each frequency. The ordered
ratio replaces much of this requirement by relative calibration of the two
branches and remains the primary observable. The direct spectrum provides an
independent check of the wave-packet model and its frequency and acceleration
dependence.

\subsection{Multifrequency thermometry and acceleration scaling}

\begin{figure}[t]
\centering
\includegraphics[width=0.6\linewidth]{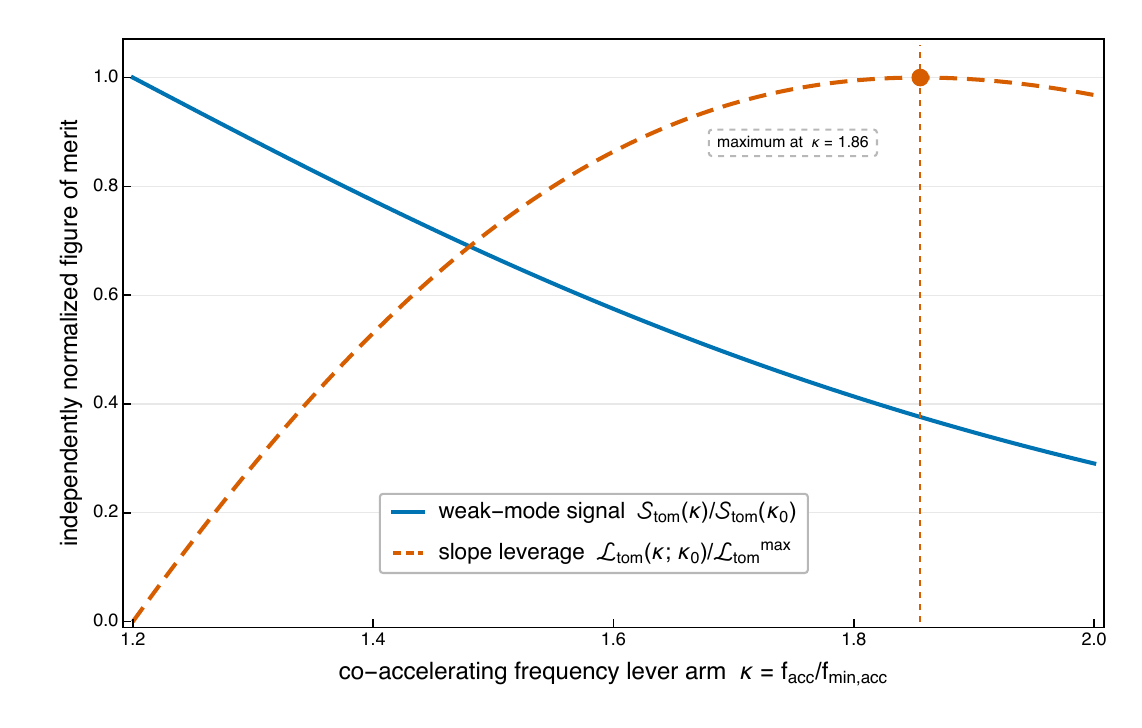}
\par\vspace{1.2ex}
\includegraphics[width=0.6\linewidth]{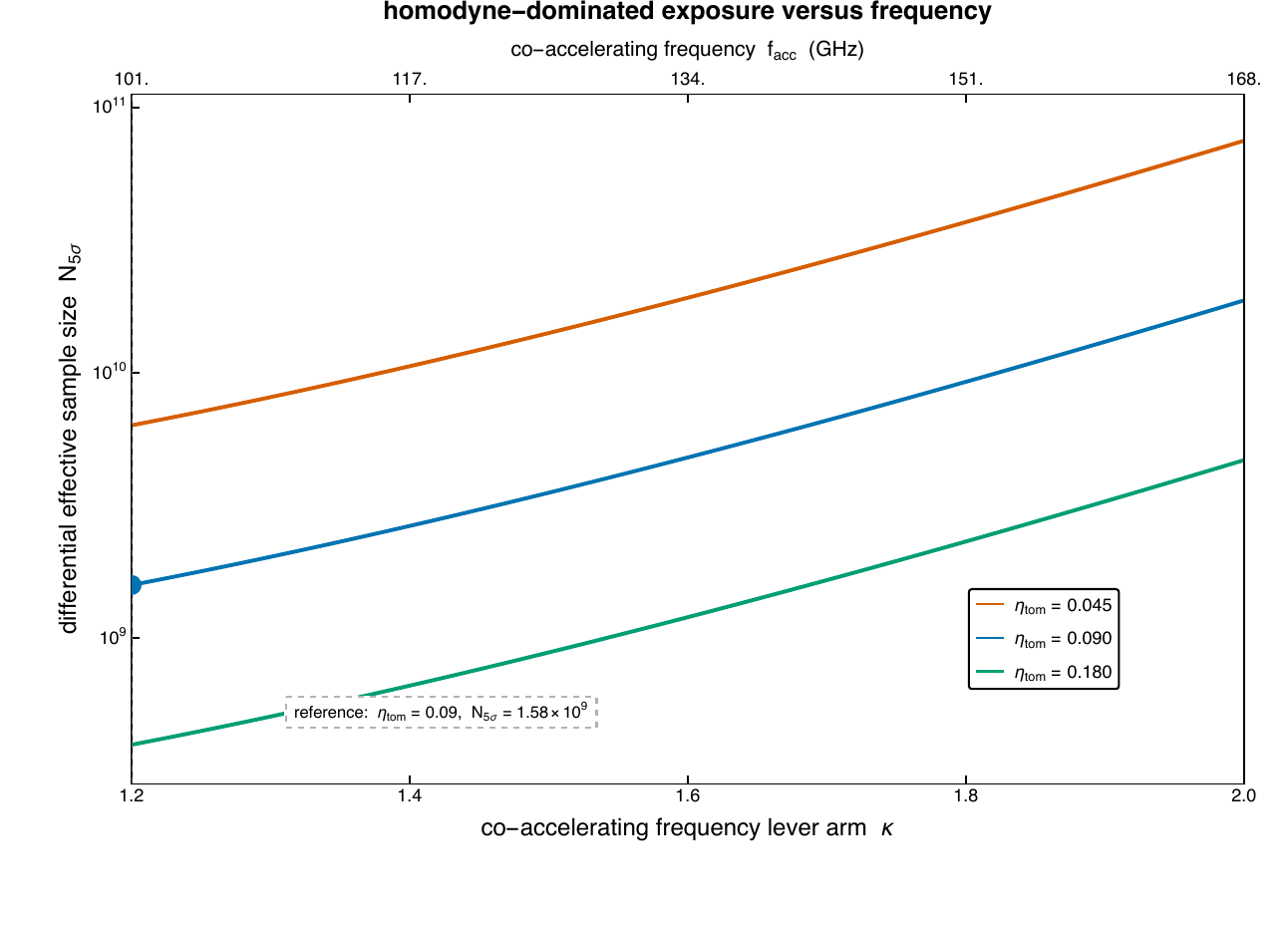}
\caption{Choosing frequencies for a temperature measurement. The upper panel shows
the weak-mode signal ${\cal S}_{\rm tom}(\kappa)$
[Eq.~\eqref{eq:kappa-signal}, solid blue], divided by its value at
$\kappa_0=1.2$, and the two-point slope leverage
${\cal L}_{\rm tom}(\kappa;\kappa_0)$
[Eq.~\eqref{eq:kappa-leverage}, dashed orange], divided by its maximum
${\cal L}_{\rm tom}^{\max}$ over $1.2\leq\kappa\leq2.0$.
The leverage peaks near $\kappa=1.86$. The lower panel shows
$N_{5\sigma}(\kappa,\eta_{\rm tom})$
[Eq.~\eqref{eq:frequency-dependent-exposure}] for
$\eta_{\rm tom}=0.045$, $0.090$, and $0.180$, including the $90\%$
polarizer transmission in each scenario. The curves are homodyne-only exposure
estimates.
Together they guide selection of a second frequency and allocation of running
time. These benchmark curves hold source and receiver factors fixed; the final
optimization must include their measured frequency dependence, captured
commutator weight, and correlations between packet estimates.}
\label{fig:frequency-design}
\end{figure}

A single ratio gives a temperature estimate; several frequencies test whether
one temperature describes the response across a band. Using
$\kappa\equiv\facc/\fminacc$, and holding the source and receiver factors
fixed, the narrowband signal controlling the statistical sensitivity has the
frequency dependence
\begin{equation}
 {\cal S}_{\rm tom}(\kappa)
 \propto\epsilon_b^2(\kappa)
 \sqrt{n_+(\kappa)[n_+(\kappa)+1]}
 \propto\frac{\kappa^2}{2\sinh(2\pi^2\kappa/7)}.
 \label{eq:kappa-signal}
\end{equation}
Within the frequency range considered here, lower frequencies give the largest
weak response because the thermal occupation falls exponentially with
$\facc$. This decrease outweighs the quadratic increase of the source factor
$\epsilon_b^2(\kappa)$. A combined measurement across well-separated frequencies, however, constrains the
temperature more strongly. With equal independent exposures at
$\kappa_0=1.2$ and $\kappa>\kappa_0$, the two measurement uncertainties add
in quadrature, while their separation increases the sensitivity to a slope.
Their combined value for thermometry is described by
\begin{equation}
 {\cal L}_{\rm tom}(\kappa;\kappa_0)
 =\frac{\kappa-\kappa_0}
 {\sqrt{{\cal S}_{\rm tom}(\kappa_0)^{-2}
       +{\cal S}_{\rm tom}(\kappa)^{-2}}}.
 \label{eq:kappa-leverage}
\end{equation}
This slope-leverage measure balances frequency separation against the falling
signal. The upper panel of Fig.~\ref{fig:frequency-design} plots
${\cal S}_{\rm tom}(\kappa)/{\cal S}_{\rm tom}(\kappa_0)$ and
${\cal L}_{\rm tom}(\kappa;\kappa_0)/{\cal L}_{\rm tom}^{\max}$, where
${\cal L}_{\rm tom}^{\max}$ is the maximum of
${\cal L}_{\rm tom}(\kappa;\kappa_0)$ over the plotted interval
$1.2\leq\kappa\leq2.0$, with $\kappa_0=1.2$. Leverage peaks near
$\kappa\simeq 1.86$, motivating a first scan at $1.2$, $1.4$, $1.6$, and
$1.8$. The lowest points resolve the weak response most efficiently; the upper
point constrains the slope; the intermediate points test curvature and
frequency-dependent calibration.

The same tradeoff can be stated in the experimentally more direct language of
exposure. From Eq.~\eqref{eq:absorption-snr}, the differential effective sample
size required for a nominal $5\sigma$ measurement is
\begin{align}
 N_{5\sigma}(\kappa,\eta_{\rm tom})
 &=2\left[
 \frac{5}{\eta_{\rm tom}\,\epsilon_b^2(\kappa)
 \sqrt{n_+(\kappa)[n_+(\kappa)+1]}}
 \right]^2\nonumber\\
 &=1.6\times10^9
 \left(\frac{0.09}{\eta_{\rm tom}}\right)^2
 \left(\frac{1.2}{\kappa}\right)^4
 \left[\frac{\sinh(2\pi^2\kappa/7)}{\sinh(2\pi^2\times1.2/7)}\right]^2.
 \label{eq:frequency-dependent-exposure}
\end{align}

For the reference source and $\eta_{\rm tom}=0.09$, the homodyne-only
$5\sigma$ sample requirements at $\kappa=1.2,1.4,1.6$, and $1.8$ are
approximately $1.6\times10^9$, $2.6\times10^9$, $4.8\times10^9$, and
$9.2\times10^9$, respectively. At $\kappa=2$ the requirement rises to
$1.9\times10^{10}$. Wider frequency separation therefore improves the slope
measurement at the cost of longer acquisition. These values are evaluated
with the exact exponent $4\pi^2\kappa/7$ and rounded only at the end.

For each frequency and acceleration, the final fit uses the Bogoliubov
coefficient functions calculated for the measured trajectory and calibrated
receiver response to evaluate
Eq.~\eqref{eq:finite-ratio}, including correlations between estimates. It does
not assume the monochromatic formula. Repeating the measurement at different
accelerations then tests the predicted dependence on $a$, with geometry and
converter settings included in the prediction for each accepted packet.

\section{Classical backgrounds and quantum discrimination}
\label{sec:classical}

\subsection{The Li\'enard-Wiechert field as a coherent first moment}

For the background estimate, we use the radiation of the bunch's total
charge moving coherently along the prescribed trajectory along the beam axis. This
provides the conservative reference field before polarization rejection. Its
far-zone electric field is, up to an overall sign,
\begin{equation}
 \bm E_{\rm cl}^{\rm rad}(u,R_{\rm obs},\theta)
 =Z_{\rm cl}(\eta,\theta)\bm e_\theta,\qquad
 Z_{\rm cl}(\eta,\theta)=\frac{Q_ba}{4\pi\epsilon_0c^2R_{\rm obs}}
 \frac{\sin\theta}{D^3}.
 \label{eq:classical-waveform}
\end{equation}
Here $R_{\rm obs}$ is the distance to the observer. Integrating the radiated
power over solid angle and then over the source laboratory duration gives
\begin{equation}
 P_{\rm cl}=\frac{Q_b^2a^2}{6\pi\epsilon_0c^3},\qquad
 E_{\rm cl}^{\rm shot}=P_{\rm cl}\frac ca
 (\sinh\eta_f-\sinh\eta_i).
 \label{eq:absolute-classical-energy}
\end{equation}
The acceleration $a=eE/m_e$ is proper acceleration. At the reference
trajectory these expressions give $P_{\rm cl}=2.9\times10^6$ W and
$E_{\rm cl}^{\rm shot}=7.4\times10^{-3}$ J. We neglect finite size effects of the bunch on the coherence of the radiation.

To model the field accepted by the receiver, let $\bm T_{\rm rx}$ describe how the collection optics, polarization filter,
and remaining receiver transmit each field component, including its amplitude
and phase. Let $\boldsymbol\psi_{J,\phi f}$ be the normalized vector waveform
of the selected partner $J$. Its classical amplitude is the projection
\begin{equation}
 \alpha_J^{\rm cl}=\int\dd\mu(\lambda)\,
 \boldsymbol\psi_{J,\phi f}^{\dagger}(\lambda)
 \bm T_{\rm rx}\bm E_{\rm cl}^{(+)}(\lambda),
 \qquad J=\mathrm I,\mathrm{II}.
 \label{eq:classical-partner-amplitudes}
\end{equation}
For the ideal trajectory along the beam axis, polarization selection makes both
amplitudes zero. Imperfect polarization transport or transverse radiation
can leave different residual amplitudes in the two partners. These residuals
must be characterized along with the desired source response.

The remaining reproducible field is removed by correcting its complex mean
amplitude; when needed to limit count noise, that cancellation may performed
optically before photon detection. If every shot were to repeat the same trajectory
and receiver response,

\begin{equation}
 \delta\alpha_J^{\rm cl}
 \equiv\alpha_J^{\rm cl}
 -\langle\alpha_J^{\rm cl}\rangle=0,
 \label{eq:perfect-classical-subtraction}
\end{equation}
and no connected classical covariance would remain. In a real accelerator the
charge, current profile, pointing, timing, plasma state, transport, and detector
gain fluctuate. Those fluctuations create the residual background that must be
measured and controlled.

\subsection{What aspects of the signal could a classical field reproduce?}
\label{subsec:classical-mimicability}
Removal of the deterministic first moment does not, by itself, make the remaining
field covariance intrinsically quantum. The two Minkowski partners are distinct
positive-frequency wave packets, so a real classical field may have unrelated
complex amplitudes in the two packets; reality relates each amplitude to its own
negative-frequency conjugate, not to the amplitude of the other partner. More
generally, a zero-mean classical stochastic field may be assigned any
positive-semidefinite two-mode covariance. It may therefore reproduce a
Planck-shaped partner-II spectrum, an arbitrary mean partner-count ratio, or an
anomalous covariance satisfying the classical bound $|M_{\mathrm I,\mathrm{II}}|^2\leq N_{\mathrm I}N_{\mathrm{II}}$. In
semiclassical photodetection, such classical mixtures of coherent fields are
described by a nonnegative Glauber--Sudarshan $P$ distribution
\cite{Glauber63,Clerk10}.

The proposed test makes a narrower comparison. As shown in the Letter's Appendix, a real classical field that is
independent of the source preparation drives the same reciprocal transition
equally in both directions, after accounting for populations and the common
matrix element. It has no commutator contribution corresponding to the quantum
$+1$. A Planck-shaped spectrum or a partner covariance within this classical bound is therefore a
consistency check, not proof of a quantum field. The test combines the
transition-rate asymmetry, the predicted wave-packet ratio, the measured
anomalous covariance, and an independently calibrated commutator.

\subsection{Equal reciprocal responses to a classical field}

The source-side distinction between classical and quantum fields can be seen
without making any assumption about the outgoing modal powers. Consider the
same reciprocal two-level coordinate, now driven by a real classical field:
\begin{equation}
 \hat H_{\rm int}(\tau)
 =-\chi(\tau)\hat{\bm d}(\tau)\cdot\bm E_{\rm cl}(\tau),
 \label{eq:classical-antenna-interaction}
\end{equation}
where
\begin{equation}
 \hat{\bm d}(\tau)
 =\bm d_{eg}\hat\sigma_+\e^{+i2\pi\facc\tau}
 +\bm d_{eg}^*\hat\sigma_-\e^{-i2\pi\facc\tau}.
 \label{eq:dipole-operator}
\end{equation}
Here $\chi(\tau)$ specifies when the interaction is active, and
$\bm d_{eg}$ is the transition dipole. For real $\chi$, first-order
perturbation theory gives
\begin{align}
 P_{g\to e}^{\rm cl}
 &=\frac{1}{\hbar^2}
 \left|\bm d_{eg}\cdot
 \int\dd\tau\,\chi(\tau)\bm E_{\rm cl}(\tau)
 \e^{+i2\pi\facc\tau}\right|^2,
 \label{eq:classical-excitation}\\
 P_{e\to g}^{\rm cl}
 &=\frac{1}{\hbar^2}
 \left|\bm d_{eg}^*\cdot
 \int\dd\tau\,\chi(\tau)\bm E_{\rm cl}(\tau)
 \e^{-i2\pi\facc\tau}\right|^2.
 \label{eq:classical-deexcitation}
\end{align}
Reality of the field implies that the two Fourier components are complex
conjugates. Hence
\begin{equation}
 P_{g\to e}^{\rm cl}=P_{e\to g}^{\rm cl}.
 \label{eq:classical-reciprocal-null}
\end{equation}
This is the equality of the stimulated Einstein $B$ coefficients. A classical
wave can drive both transitions, but it drives them symmetrically. It has no
operator commutator and no intrinsic $n$ versus $n+1$ distinction.

The same conclusion holds for a source-independent real classical stochastic
field. If
\begin{equation}
 C_{ij}^{\rm cl}(\tau,\tau')
 =\mathbb E\!\left[E_{{\rm cl},i}(\tau)E_{{\rm cl},j}(\tau')\right],
 \label{eq:classical-covariance}
\end{equation}
then commuting classical variables satisfy
$C_{ij}^{\rm cl}(\tau,\tau')=C_{ji}^{\rm cl}(\tau',\tau)$. Relabeling the
integration variables in the two ensemble-averaged transition probabilities
again gives equal upward and downward responses.

The equal-rate prediction concerns the reciprocal source transition, not
arbitrary outgoing field correlations. A true quantum detailed balance measurement must therefore check
all of the following:
\begin{enumerate}
 \item reciprocal lower- and upper-state transitions with unequal rates;
 \item the canonical difference $n_--n_+=1$;
 \item an anomalous covariance consistent with the intended Bogoliubov mode;
 \item the acceleration-dependent finite-duration wave-packet ratio $R_f$;
 \item a controlled classical source drive that obeys the reciprocal
 equality in Eq.~\eqref{eq:classical-reciprocal-null}. Downstream injections
 test the receiver response separately.
\end{enumerate}
A gain or background field that changes with the preparation setting could
mimic an asymmetry. Reversing the preparations, exchanging readout paths, and
using known test injections check for this dependence.

\subsection{Estimating the classical residual from shot-resolved data}

Beam and receiver variations make the Li\'enard waveform differ from shot to shot.
An individual science record does not identify each fluctuation as classical or
quantum. The classical residual is instead characterized in transition-detuned
and other control runs that suppress the selected bunch transition while
retaining the accelerator and receiver backgrounds.

Let $\lambda_i$ denote measured beam and receiver properties that affect the
background, such as charge, timing, or gain. A predictor for the classical
partner amplitudes is fitted to independent control data. To first order about
the reference operating point,
\begin{equation}
 \delta\alpha_J^{\rm cl}
 =\sum_i
 \frac{\partial\alpha_J^{\rm cl}}{\partial\lambda_i}
 \delta\lambda_i,
 \qquad J=\mathrm I,\mathrm{II},
 \label{eq:nuisance-linearization}
\end{equation}
with covariance
\begin{equation}
 C_{JK}^{\rm cl}
 =\sum_{i\ell}
 \frac{\partial\alpha_J^{\rm cl}}{\partial\lambda_i}
 \Sigma_{i\ell}
 \frac{\partial\alpha_K^{\rm cl*}}{\partial\lambda_\ell},
 \qquad
 \Sigma_{i\ell}=\langle\delta\lambda_i\delta\lambda_\ell\rangle.
 \label{eq:nuisance-covariance}
\end{equation}
These equations show how measured fluctuations and their correlations change
the predicted field, but we do not restrict the final predictor to be linear.

Fluctuations in charge, timing, or acceleration along the beam axis change the
size or timing of the classical Li\'enard field without changing its polarization.
An ideal filter therefore rejects them on each shot: their derivatives in
Eq.~\eqref{eq:nuisance-linearization}, after projection into the accepted
mode, vanish. A real filter transmits a small fraction of this field and
its fluctuations. Transverse motion or polarization misalignment can instead
produce a field directly in the accepted polarization; its measured covariance
must therefore be included in the error budget.

After fitting the classical-background model to control measurements and
calibrating the receiver, both are tested on a separate set of control
measurements. The difference between the observed and predicted field
determines the residual mean and shot-to-shot fluctuations. The same
procedure is applied to the anomalous covariance. Receiver-only, pump-off,
vacuum-load, and thermal-load records are used to separate receiver and
environmental contributions from the beam-dependent residual. The resulting
mean and covariance are then propagated through the same receiver inverse and
the same $n_+$ reconstruction used for the source data.

Let $\sigma_A$ be the fractional amplitude error remaining after prediction
from the beam diagnostics, measured in this independent sample. This is the
\emph{post-conditioned} error, not the raw shot-to-shot beam variation. The
source-level energy dynamic range is
\begin{equation}
 D_{\rm cl}
 \equiv\frac{E_{\rm cl}^{\rm shot}}{E_{\rm weak}^{\rm shot}}.
 \label{eq:dynamic-range}
\end{equation}

To put the classical and weak responses on a common energy scale, we use
the ray estimate $h\kappa\nu_0$ for the mean energy of a weak-branch photon.
Thus $E_{\rm weak}^{\rm shot}\simeq P_{\rm abs}^{R}h\kappa\nu_0$, giving
\begin{equation}
 E_{\rm weak,*}^{\rm shot}\simeq 1.2\times10^{-24}~\mathrm J,
 \qquad D_{\rm cl,*}\simeq 6.3\times10^{21}.
 \label{eq:reference-energy-budget}
\end{equation}
The numerator retains the full classical reference energy. Its reduction by
collection, polarization selection, and mode rejection is described by the
calibrated transfer to the reconstructed packet, not by another factor applied
to the numerator.

For any distinct physical or instrumental mechanism $r$, define its
energy-equivalent shift in the reconstructed Rindler wave packet, normalized
to the raw classical energy,
\begin{equation}
 \Lambda^{(r)}
 \equiv\frac{\Delta E_{f,r}^{\rm shot}}
 {E_{\rm cl}^{\rm shot}}.
 \label{eq:operational-lambda-component}
\end{equation}
The classification concerns the shift of the \emph{reconstructed observable},
not just the field amplitude: a zero-mean fluctuating field can still add a
nonzero occupation. After reference and background corrections, we classify
residual shifts as follows:

\begin{itemize}
 \item Subtract a reliably measured offset from the reported response.
 $\Lambda^{\rm bias}$ describes only a mean shift left after that correction,
 not the size of the background already removed. The uncertainty of the
 subtraction remains part of the corrected moment covariance.
 \item {An unresolved offset whose magnitude can be bounded contributes
 to the nonnegative allowance $\Lambda^{\rm other}$. The underlying shift
 may have either sign; this allowance bounds its absolute size and is therefore
 added to other terms in the error budget.}
 \item Fluctuations about the independently corrected mean contribute to the
 statistical uncertainty and receive $N_{\rm eff}^{-1/2}$ suppression.
 Uncertainty in the mean correction must still be retained separately.
\end{itemize}

For example, let $\underline q_{\rm raw}$ be a response estimated from the
science records, and let $\underline b$ be the background offset estimated
from control data. The corrected estimate is
$\underline q_{\rm corr}=\underline q_{\rm raw}-\underline b$, with variance
\begin{equation}
 \operatorname{Var}(\underline q_{\rm corr})=
 \operatorname{Var}(\underline q_{\rm raw})+
 \operatorname{Var}(\underline b)-2\operatorname{Cov}
 (\underline q_{\rm raw},\underline b).
 \label{eq:correction-variance}
\end{equation}
The correction therefore carries its own sampling uncertainty, including any
correlation with the science estimate. Collecting more science shots does not
reduce the uncertainty of a fixed calibration sample. A possible change in the
background between calibration and science conditions is bounded separately.
Similarly, a zero-mean fluctuating field can have nonzero mean power. That
power must agree between the compared classes, or its difference must be
corrected; averaging then reduces the uncertainty of the corrected estimate.

Receiver gain or phase drift enters $\Lambda^{\rm bias}$ when repeated
calibration resolves a mean shift, or $\Lambda^{\rm other}$ when only a bound
is available. The same rule applies to preparation errors remaining after
population calibration: a resolved preparation-dependent shift is a bias; an
unresolved bounded shift is included in $\Lambda^{\rm other}$.

Unwanted-mode leakage and errors in the classical background predictor contribute to
$\Lambda^{\rm bias}$ or $\Lambda^{\rm other}$ through their residual mean
shifts. The estimate below uses $\sigma_A$ and the complete $L_{\rm tom}$ to
quantify sampling uncertainty from the leaked Li\'enard field. Fluctuations
already in the accepted polarization are included in the measured covariance
but are not reduced by this polarization-rejection estimate. Thermal
occupation, detector offsets, dark counts, and converter fluorescence are
measured in separate background runs; any remaining mean shift is assigned to
the same bias or bounded-uncertainty terms. Their counting and readout
fluctuations also enter the measured statistical covariance, and additional error sources will contribute to the final error analysis for a given facility.

To distinguish the total statistical uncertainty from the particular
classical-residual term, let $\Delta\underline{\bm m}_{\rm in}$ be the
estimated source--reference difference of the four input moments in
Eq.~\eqref{eq:receiver-affine-map}, written as a column vector. For fixed
calibrated mode coefficients, Eq.~\eqref{eq:reconstruction} gives
\begin{align}
 \sigma_{\rm stat}^2(\Delta n_+)
 &=\bm t_f\,\operatorname{Cov}
 \!\left(\Delta\underline{\bm m}_{\rm in}\right)\bm t_f^{\mathsf T},
 \label{eq:full-statistical-variance}\\
 \bm t_f
 &=\left(|u_f|^2,\,|v_f|^2,\,
 2\Ree(u_f^*v_f),\,2\Ima(u_f^*v_f)\right).
 \nonumber
\end{align}
The covariance matrix includes the sampling uncertainties of the counting and
homodyne measurements, their correlations, and the fluctuating backgrounds
remaining after correction. With a fixed receiver matrix it is obtained from
the measured covariance by applying $R_{\rm rx}^{-1}$ on the left and its
transpose on the right. Statistical uncertainties of calibration data and of
the mode coefficients must also be propagated when these are estimated rather
than fixed. The symbol $\sigma_{\Xi}^{\rm stat}$ below describes only the
classical-residual contribution, expressed relative to the weak signal. The total readout uncertainty must account for any additional sources of error, but this gives a preliminary assessment.

The classical contribution is most directly compared with the weak signal
in occupation units. Let $\sigma_{\rm cl}[\Delta\underline n_+(f)]$ be the
standard uncertainty in the corrected occupation difference attributable to
the residual classical Li\'enard field. Relative to the expected weak response,
it is
\begin{equation}
 \sigma_{\Xi}^{\rm stat}(f)=
 \frac{\sigma_{\rm cl}[\Delta\underline n_+(f)]}{P_{\rm abs}^{R}(f)}.
 \label{eq:classical-relative-uncertainty}
\end{equation}
The following energy-budget expression is its order-of-magnitude estimate. All terms refer to the same selected response
and the same calibrated mode reconstruction.

With $z$ the chosen statistical confidence multiplier, the Letter summarizes
the classical-residual budget as
\begin{equation}
 \Lambda_z^{\rm eff}
 =|\Lambda^{\rm bias}|
 +z\frac{\sigma_A^2L_{\rm tom}}{\sqrt{N_{\rm eff}}}
 +\Lambda^{\rm other},
 \label{eq:effective-leakage}
\end{equation}
and
\begin{equation}
 \Xi_z^{\rm tot}
 =D_{\rm cl}\Lambda_z^{\rm eff}
 =\Xi^{\rm bias}+z\,\sigma_{\Xi}^{\rm stat},
 \label{eq:total-contamination}
\end{equation}
where
\begin{equation}
 \Xi^{\rm bias}
 =D_{\rm cl}
 \left(|\Lambda^{\rm bias}|+\Lambda^{\rm other}\right),
 \qquad
 \sigma_{\Xi}^{\rm stat}
 \simeq D_{\rm cl}
 \frac{\sigma_A^2L_{\rm tom}}
 {\sqrt{N_{\rm eff}}}.
 \label{eq:contamination-components}
\end{equation}
At the reference point,
\begin{equation}
 \begin{gathered}
 D_{\rm cl}\simeq 6.3\times10^{21},\qquad \sigma_A=10^{-3},\\
 L_{\rm tom}=2\times10^{-16},\qquad N_{\rm eff}=3\times10^9,
 \end{gathered}
 \label{eq:background-benchmark}
\end{equation}
so
\begin{equation}
 \sigma_{\Xi}^{\rm stat}\simeq 2.3\times10^{-5}.
 \label{eq:background-result}
\end{equation}

This value estimates how precisely the residual contribution from the Li\'enard field
can be determined after the mean correction. It assumes that the fourth moments
of the residual field, which control the variance of quadratic field estimates,
are characterized by the adopted model and effective sample size. The photon
load at the counters and the full counting, homodyne, and calibration variance
in Eq.~\eqref{eq:full-statistical-variance} must still be checked separately.

We allocate at most one percent of the weak response to the combined
classical uncertainty and residual-offset allowance:
\begin{equation}
 \Xi_z^{\rm tot}\lesssim10^{-2}.
 \label{eq:science-contamination-requirement}
\end{equation}

With the ray-energy estimate, the largest standard error from the residual
Li\'enard field is $4.5\times10^{-4}$ over $1.2\leq\kappa\leq2$, attained at $\kappa=2$.
We require the remaining classical contributions, including residual offsets
and any correction uncertainty not already counted, to fit within $0.005$
of the weak response. Adding five times the largest modeled standard error gives
\begin{equation}
 \Xi_5^{\rm tot}\lesssim0.005+
 5\max_{1.2\leq\kappa\leq2}\sigma_{\Xi}^{\rm stat}
 \simeq 0.0073<0.01.
 \label{eq:classical-band-allowance}
\end{equation}
The $0.005$ is a requirement on the controls, not a measured performance.
The factor of ``five'' sets the statistical allowance for this background estimate. A confidence statement for
the entire frequency band also requires the correlations between its estimates.
For example, at $\kappa=2$, Eq.~\eqref{eq:absorption-snr} gives a
homodyne-limited signal-to-noise ratio
$\mathrm{SNR}_{\rm abs}\simeq2.0$ at the reference efficiency and exposure,
even when the residual classical Li\'enard background is negligible.
The classical bound therefore does not imply one-percent total precision or
a discovery at every frequency. The full test also uses the complete readout
uncertainty and the reciprocal-ratio and acceleration measurements.

The suppression used in this budget must be demonstrated with known classical
test fields passed through the complete receiver and analyzed with the same
calibrated reconstruction as the science data.

\begin{figure}[p]
\centering
\includegraphics[width=0.80\linewidth]{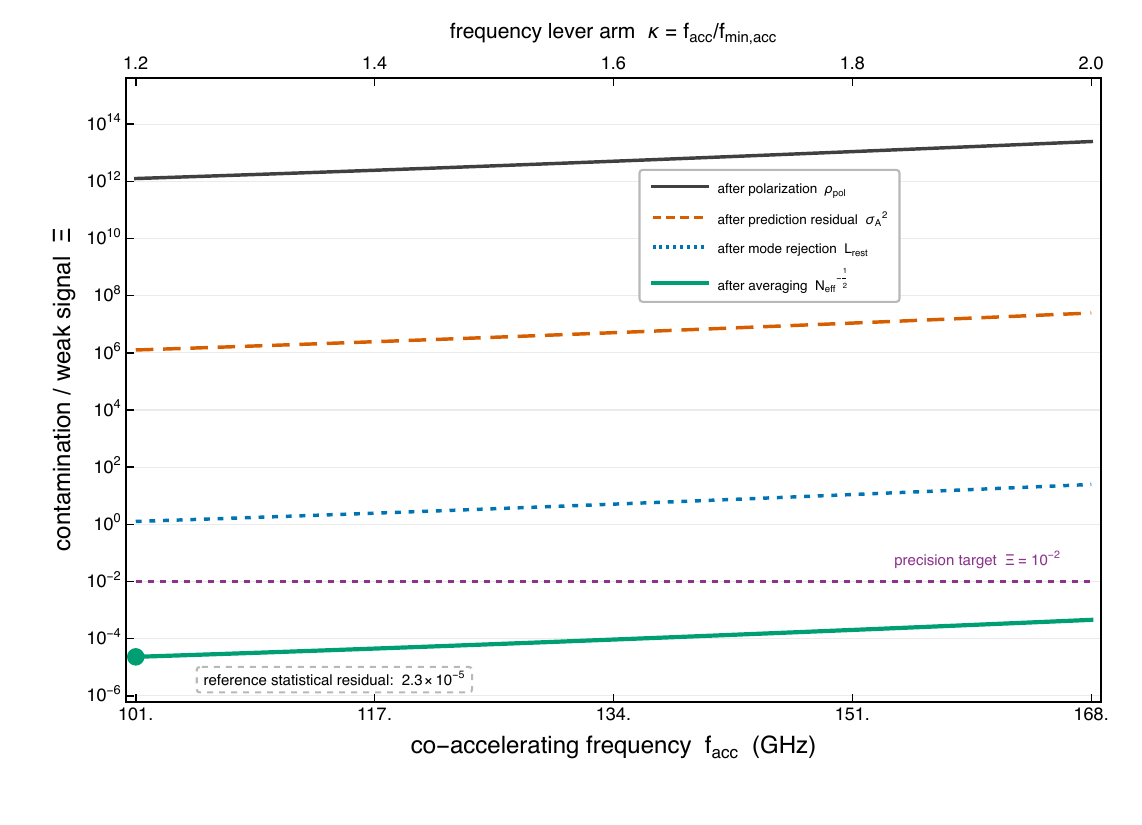}
\par\vspace{1.2ex}
\includegraphics[width=0.74\linewidth]{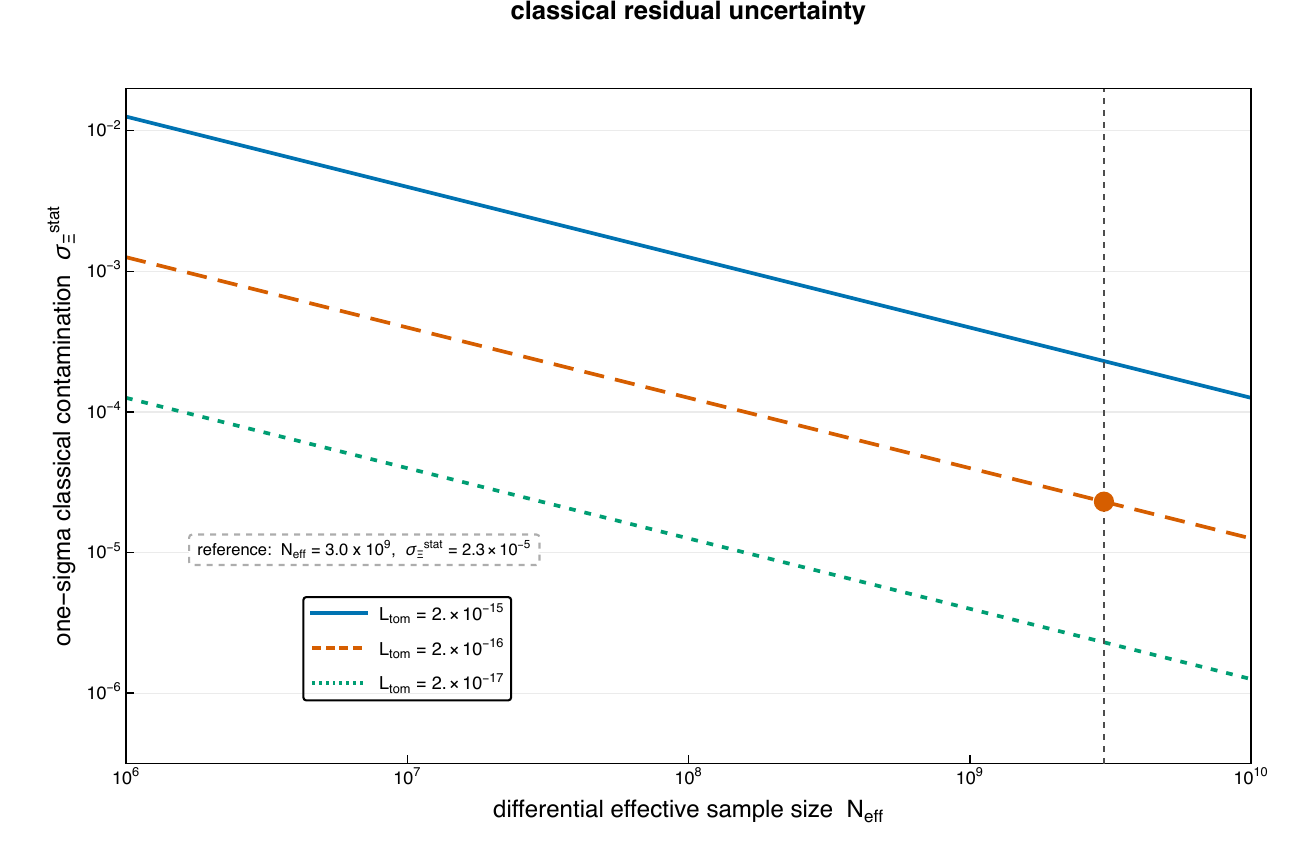}
\caption{Classical-background rejection and exposure for the selected response.
The upper panel begins with the field admitted by polarization selection,
$D_{\rm cl}\rho_{\rm pol}$. Successive curves include the residual prediction
factor $\sigma_A^2$, further mode rejection $L_{\rm rest}$, and averaging
by $N_{\rm eff}^{-1/2}$. The reference values are
$N_{\rm eff}=3\times10^9$ and complete $L_{\rm tom}=2\times10^{-16}$.
Only the final curve is a statistical uncertainty. The lower panel shows that
uncertainty for complete leakages $2\times10^{-15}$, $2\times10^{-16}$,
and $2\times10^{-17}$ at $\kappa=1.2$. Both panels use the photon-energy
estimate $h\kappa\nu_0$, with $\nu_0=22$ THz. The $1\%$ line is the total
classical allowance: residual offsets, other classical uncertainties, and
the chosen statistical multiplier must also be included as in
Eq.~\eqref{eq:classical-band-allowance}.}
\label{fig:classical-control}
\end{figure}

Figure~\ref{fig:classical-control} starts with the background that passes
polarization selection. The upper panel shows the successive reductions
associated with field prediction, further mode rejection, and statistical
averaging. The lower panel compares three complete $L_{\rm tom}$ values as
the exposure is increased. The order of factors in this budget is not a
prescription for the hardware: polarization rejection still precedes
conversion and detection. Residual mean shifts and other statistical
contributions are assessed separately.

\subsection{Control measurements}

The following controls distinguish different causes of an apparent signal:
\begin{itemize}
 \item \textbf{Receiver-only and conversion-pump-off runs.} With no
 bunch input, compare the operating receiver with the same readout while the
 conversion pump is blocked. The homodyne local oscillator, when used, is kept
 on. This comparison helps distinguish pump-dependent light and noise from
 detector dark counts and electronic offsets. Because switching off the pump
 also removes the desired conversion, pump-on background measurements are also necessary.
 \item \textbf{Transition-detuned accelerated runs} use the same beam trajectory
 and receiver settings as the source data, but shift or disable the selected
 collective transition. They measure the Li\'enard field and its fluctuations in
 the closest available signal-null configuration.
 \item \textbf{Deliberately mismatched waveforms} use a different chirp, phase,
 or packet profile. The response is compared with that predicted from its
 overlap with the target. An excess reveals unwanted acceptance by the
 receiver or analysis, including leakage of the bright classical field.
 \item \textbf{Polarization and acceleration controls} inject accepted
 and rejected polarizations separately, rotate the filter by known small
 angles, and vary transverse pointing. These measurements test leakage into
 the selected field. An unaccelerated comparison tests the predicted dependence
 on acceleration.
 \item \textbf{Preparation-label and receiver-path reversals} exchange the
 physical preparations assigned to the acquisition labels or swap the partner
 inputs between the two readout paths. A source asymmetry follows the physical
 preparation; a fixed readout imbalance follows the hardware.
 \item \textbf{Known classical injections} add repeatable or randomized waveforms of known
 amplitude to test the receiver's gain and rejection. A separate drive applied
 to the source can test whether it responds equally in the two reciprocal
 directions.
\end{itemize}

Alternating the preparations avoids measuring one class systematically earlier
than another, when slow drift could imitate a source difference. Receiver
calibration and background prediction must use data independent of the sample
used to quote the residual uncertainty. Hiding the weak result until the
controls are complete---a blinded analysis---may prove useful.

\section{Status of the experimental benchmark}
\label{sec:status}

The field model predicts the ordered $n$ versus $n+1$ response, the ratio in
Eq.~\eqref{eq:finite-ratio}, and the positive-frequency partner representation.
Whether a facility can measure them depends on the bunch response, the captured
fraction of the packet, phase-preserving receiver efficiency, background
rejection, and long-term stability. The benchmark specifies targets for these
quantities, not demonstrated performance.

No existing apparatus combines all of the required operations. The benchmark is
therefore assessed component by component, while explicitly distinguishing a
demonstrated operation from its required realization over the complete
angle--frequency support of the selected wave packet.
Tables~\ref{tab:source-mode-evidence}--\ref{tab:control-exposure-evidence}
summarize that distinction.

The experiment must capture the selected packet with a manageable number of
partner pairs, achieve the source response ${\cal R}_{\rm src}=0.16$, and
retain a total quadrature efficiency of $9\%$ while meeting the rejection
targets. Calibration must remain reliable over the acquisition. The counting
budget also requires the actual residual photon occupations assumed in
Eq.~\eqref{eq:counting-variance-benchmark}.

Three measurements or calculations should precede a long science campaign:
\begin{enumerate}
 \item Calculate the accepted field coefficients for a realistic trajectory,
 including the finite interaction and polarization selection. Construct their
 laboratory partner pairs, and show quantitatively how rapidly the
 captured commutator weight $C_K$ approaches unity. Repeat the reconstruction
 with realistic differential losses.
 \item Measure, or calculate with a validated bunch model, the connected dipole
 response, the relevant source populations, and the overlap between the emitted
 one-quantum field and the selected Rindler wave packet.
 \item Send the measured Li\'enard waveform and other relevant unwanted waveforms
 through the complete receiver. Test backgrounds in the accepted polarization
 separately, and demonstrate the required
 reconstruction, conjugate-sideband rejection, rejection of neighboring and
 mismatched wave packets, and stability over the retained bandwidth.
\end{enumerate}
Failure of any one test would identify which subsystem or target wave packet
must be redesigned before the benchmark can be regarded as experimentally
credible.

\begin{table}[htbp]
\centering
\caption{Experimental precedents and theoretical inputs for the desired, component figures of merit in the benchmark proposal.}
\label{tab:source-mode-evidence}
\small
\setlength{\tabcolsep}{6pt}
\renewcommand{\arraystretch}{1.16}
\begin{tabular}{@{}lll@{}}
\toprule
\statusname{\textbf{Figure of merit}} & \statustarget{\textbf{Active target}} &
\statusdetail{\textbf{Evidence and remaining requirement}}\\
\midrule
\statusname{Wakefield and charge} & \statustarget{$E=1~\mathrm{GV/m}$; $Q_b=0.65~\mathrm{nC}$} &
\statusdetail{\precedentgap{Gradients above $1~\mathrm{GV/m}$, meter-scale acceleration, and nC-class drive bunches are established; FACET-II has reported a high-gradient stage with a $1.6~\mathrm{nC}$ drive bunch \cite{Doche17,AWAKE18,natureZhang,facet2}. Emittance preservation in a plasma stage has also been demonstrated \cite{Lindstrom24}.}{The charge scale is available, but preparation or control of a comparably charged, high-quality participating collective mode has not been demonstrated.}}\\
\midrule
\statusname{Collective bunch response} & \statustarget{${\cal R}_{\rm src}=0.16$} &
\statusdetail{\precedentgap{Free-electron energy-loss heralding of nonclassical light has been demonstrated in a dielectric waveguide \cite{Arend25}; the oscillator-strength sum rule supplies a separate theoretical normalization \cite{Aucar95}.}{The accelerated-bunch structure factor, reciprocal-state populations, and collected radiative yield have not been measured. This is a principal source-side uncertainty.}}\\
\midrule
\statusname{Partner-wave-packet reconstruction} & \statustarget{Measure both waveforms in each retained pair; bound the omitted contributions
to $n_+$ and $R_f$ below the required accuracy.} &
\statusdetail{\precedentgap{Quantum-pulse-gate tomography \cite{Ansari17}, programmable single-photon temporal-mode sorting \cite{Serino23,Serino25}, and spatiotemporal sorting \cite{Garikapati23} have been demonstrated. An experimental preprint also reports single-photon-level temporal-mode selection in thin-film lithium niobate \cite{Babel26}.}{The demonstrated sorters do not yet span the required angle--frequency range. The number of partner pairs needed to recover the target commutator weight, and their end-to-end separation, remain to be established \cite{Garikapati23,Zhang19Mode}.}}\\
\bottomrule
\end{tabular}
\end{table}

\begin{table}[htbp]
\centering
\caption{Receiver evidence for the active benchmark. The quoted targets refer to
the complete phase-preserving chirped wave packet rather than to any one narrowband
component.}
\label{tab:receiver-evidence}
\small
\setlength{\tabcolsep}{6pt}
\renewcommand{\arraystretch}{1.16}
\begin{tabular}{@{}lll@{}}
\toprule
\statusname{\textbf{Figure of merit}} & \statustarget{\textbf{Active target}} &
\statusdetail{\textbf{Evidence and remaining requirement}}\\
\midrule
\statusname{Photon conversion and counting} & \statustarget{$\eta_{\rm count}^{(N)}=0.018$} &
\statusdetail{\precedentgap{Upconversion photon counting has reached $37\%$ overall efficiency at $3~\mu\mathrm m$ \cite{Huang21MIR}; an experimental preprint reports $(1.98\pm0.09)\%$ full-system efficiency near $4$~THz \cite{Vishnuradhan26}. Long-wave-infrared upconversion detection spans $9.69$--$12.38~\mu\mathrm m$ \cite{Liu25LWIR}, while optical photodetection reaches $(98.0\pm0.5)\%$ at $1550$~nm \cite{Reddy20}.}{The cited component results do not establish $\eta_{\rm count}^{(N)}=0.018$
for the full partner waveform across the required frequency and angular range.
The remaining step is to convert its separately collected components, preserve
their relative phases, and recombine them into the same optical mode before
detection, while retaining the quoted overall efficiency.}}\\
\midrule
\statusname{Homodyne covariance readout} & \statustarget{$\eta_{\rm tom}=0.09$} &
\statusdetail{\precedentgap{Dual-mode optical homodyne tomography \cite{Babichev04} and high common-mode rejection \cite{jinBHD} are established. Quantum-noise-limited balanced homodyne detection reaches a $15.3$~GHz 3-dB bandwidth \cite{Tasker24}.}{The relative phases of all receiver channels contributing to the two partner packets must remain calibrated throughout the required interleaved acquisition.}}\\
\bottomrule
\end{tabular}
\end{table}

\begin{table}[htbp]
\centering
\caption{Calibration, rejection, and exposure evidence for the active benchmark.}
\label{tab:control-exposure-evidence}
\small
\setlength{\tabcolsep}{6pt}
\renewcommand{\arraystretch}{1.16}
\begin{tabular}{@{}lll@{}}
\toprule
\statusname{\textbf{Figure of merit}} & \statustarget{\textbf{Active target}} &
\statusdetail{\textbf{Evidence and remaining requirement}}\\
\midrule
\statusname{Thermal and vacuum calibration} & \statustarget{Near-vacuum input verified in each receiver channel} &
\statusdetail{\precedentgap{Subcycle and electro-optic measurements of vacuum fluctuations and correlations have been demonstrated \cite{Riek15,BeneaChelmus19,Leitenstorfer22}. Subcycle squeezing has also been measured \cite{Riek17}; subcycle tomography has been proposed \cite{Hubenschmid24}.}{Some accepted frequencies have appreciable room-temperature occupation. Their contribution must be reduced with cooling and shielding or included in a calibrated model of the incoming radiation.}}\\
\midrule
\statusname{Polarization selection} & \statustarget{$\rho_{\rm pol}=2\times10^{-10}$; $T_\phi=0.90$} &
\statusdetail{\precedentgap{THz polarimetry reports about $90$ dB extinction at $0.7$ THz and over $90\%$ pass transmission; static optical polarimetry reports a crossed/parallel ratio of $2\times10^{-10}$ \cite{Park24Polarizer,Fang10Polarimeter}.}{The blocked THz signal is at the measurement noise level. For this receiver,
rejection and signal transmission must be measured across the accepted band
and for the local polarization in each angular channel.}}\\
\midrule
\statusname{Classical rejection} & \statustarget{$L_{\rm rest}=10^{-6}$; $L_{\rm tom}=2\times10^{-16}$; $\sigma_A=10^{-3}$} &
\statusdetail{\precedentgap{A near-infrared photonic lantern, which directs spatial field modes into
different fiber ports, reports at least $60$ dB isolation for selected ports
with fundamental-mode input and at most $0.49$ dB excess loss
\cite{BecerraDeana25Lantern}.}{The strongest rejection is limited by the measurement noise and was measured
on the two-ended structure before it was cut. The proposed combined rejection
must be tested on the actual field leaving the filter; spatial filtering
already used there cannot be credited again downstream.}}\\
\midrule
\statusname{Statistical exposure} & \statustarget{$N_{\rm eff}=3\times10^9$; $N_{\rm count}=1.5\times10^9$} &
\statusdetail{\precedentgap{FACET-II specifies operation at rates up to $30~\mathrm{Hz}$ \cite{facet2}. An experimental preprint reports $100$~Hz laser-wakefield operation with $47$~MeV, $15$~pC bunches \cite{Smartsev26}.}{The example requires $1.8\times10^{10}$ effective records across the balanced
source and reference samples: about $5.7$ years at a usable $100$~Hz, before
other preparations and scans. Combining this usable rate with the benchmark charge, energy, and collective-state
preparation remains a development target. Losses, correlations, and downtime increase
the required running time.}}\\
\bottomrule
\end{tabular}
\end{table}

\clearpage

\section{End-to-end analysis sequence}
\label{sec:analysis-protocol}

The proposed analysis proceeds as follows.
\begin{enumerate}
 \item \textbf{Record the source and apparatus state on every shot.} Measure the
 bunch charge and current profile, acceleration history, pointing, timing,
 plasma diagnostics, preparation label, and the receiver settings needed to
 interpret that shot.
 \item \textbf{Construct the target and measured wave packets.} Use the measured
 trajectory and source-interaction profile to define the target proper-time
 waveform. Use the measured angular aperture, polarization selection, frequency range,
 and channel responses to determine which part of that waveform can be
 reconstructed.
 \item \textbf{Check how completely the receiver captures the target.} Perform
 the canonical partner decomposition, increase the number of retained pairs
 until the captured commutator weight $C_K$ is sufficiently close to unity, and
 propagate the omitted weight and differential-loss sensitivity as explicit
 uncertainties.
 \item \textbf{Calibrate the receiver.} Send each partner waveform through the
 receiver separately, then send both while varying their relative phase.
 Repeat with rejected-polarization, conjugate-sideband, and mismatched
 waveforms to measure unwanted responses. Known thermal inputs and no-signal
 runs determine added noise and offsets; conversion-pump-off runs help locate
 their origin. Record which optics each test traverses and check the
 commutator relation as well as the amplitude and phase response.
 \item \textbf{Fix the analysis choices before evaluating the final sample.}
 Specify the transfer-matrix inversion, wave-packet weights, nuisance predictor,
 event-selection criteria, and control thresholds using calibration and
 development data. A fully blinded implementation is optional, but the sample
 used to quote the final result must remain independent of these choices.
 \item \textbf{Interleave the source and control preparations.} Alternate
 transition-detuned reference, lower-state, upper-state, unaccelerated,
 polarization-control, pump-off, and reversal shots so that slow drift cannot
 become correlated with one preparation label.
 \item \textbf{Characterize the field remaining after polarization selection.}
 Predict it from the beam diagnostics and rejected bright output. Cancel any
 large coherent amplitude optically before counting. Use independent runs
 with the transition suppressed to determine the residual mean field, photon
 occupations, and covariance, including backgrounds in the accepted polarization.
 \item \textbf{Determine the effective statistical exposure.} Measure
 correlations between successive residual records in each class and compute
 $N_{\rm eff}$ for homodyne and $N_{\rm count}$ for counting.
 One way to check the uncertainty is to divide the data
 into blocks of consecutive shots, repeatedly draw whole blocks (allowing repeated draws of the same block) to form new data sets, and repeat the estimate on each data set.
 The spread of the resulting estimates measures the sampling uncertainty while
 retaining correlations within a block. Choose blocks longer than the measured
 correlation time, and check that increasing their length does not
 substantially change the uncertainty.
 \item \textbf{Reconstruct the partner moments and the Rindler response.} Infer
 $N_{\mathrm I}$, $N_{\mathrm{II}}$, and
 $M_{\mathrm I,\mathrm{II}}$ from photon counting and phase-sensitive
 measurements, and combine them according to Eq.~\eqref{eq:reconstruction}.
 \item \textbf{Form the reciprocal source responses.} Use
 Eq.~\eqref{eq:preparation-differences} together with the measured source
 populations to infer $P_{\rm abs}^{R}$ and $P_{\rm em}^{R}$. Report their
 difference and ratio, the anomalous covariance, the direct weak-spectrum
 estimate, and the auxiliary partner counts.
 \item \textbf{Test the frequency and acceleration dependence.} Compare the
 reciprocal responses at the same interaction duration and packet definition.
 Across several $\kappa$ values, fit their ratio to $R_f$ using the measured
 correlations between packet estimates. A duration or coupling scan provides
 a check on finite-coupling corrections. Repeat the comparison at more than
 one independently reconstructed acceleration.
 \item \textbf{Interpret the result after the control tests are satisfied.}
 Use the complete statistical covariance and bias bounds to assess the
 reciprocal response, and check the classical allowance in
 Eq.~\eqref{eq:classical-band-allowance} separately. The commutator calibration,
 mismatched-waveform and classical-injection tests, preparation reversals,
 and acceleration dependence must support the same interpretation. In a
 blinded analysis, reveal the science result only after these checks.
\end{enumerate}

The experiment compares reciprocal bunch transitions through changes in the
same Rindler wave packet, reconstructed with inertial apparatus. Measurements
at several frequencies and accelerations test the predicted KMS dependence
rather than agreement at one point. The rate difference and ratio, partner
covariance, calibrated commutator, and control results must all support the same
interpretation.

\appendix

\section{Derivation of the preparation-induced partner moments}
\label{app:moment-derivation}

The map below describes the field in the observed polarization and packet,
with the unobserved field modes averaged over. At order $\epsilon_b^2$,
the orthogonal modes initially in vacuum contribute no additional term to
this selected-field map. They still affect the source's population evolution
at higher order, as discussed in Eq.~\eqref{eq:total-coupling-range}.

For completeness, we derive the entries in
Eq.~\eqref{eq:moment-change-table}. Choose phases so that $u_f$ and $v_f$ are
real and write
\begin{equation}
 \delta\hat b_f=u_f\delta\hat a_{\mathrm I}
 +v_f\delta\hat a_{\mathrm{II}}^\dagger.
\end{equation}
On the Minkowski vacuum $|0_{\mathrm I}0_{\mathrm{II}}\rangle$,
\begin{equation}
 \delta\hat b_f^\dagger\delta\hat b_f|0\rangle
 =v_f^2|0\rangle
 +u_fv_f|1_{\mathrm I}1_{\mathrm{II}}\rangle.
 \label{eq:bf-number-on-vacuum}
\end{equation}
Let $\delta\hat\rho_g$ denote the field-state change after a lower-state bunch
interacts, with its final state unobserved. To leading nontrivial order,
\begin{align}
 \delta\hat\rho_g
 =\epsilon_b^2\Big[&
 v_f^2|0_{\mathrm I}1_{\mathrm{II}}\rangle
       \langle0_{\mathrm I}1_{\mathrm{II}}|
 -v_f^2|0\rangle\langle0|\nonumber\\
 &-\frac{u_fv_f}{2}
 \left(|1_{\mathrm I}1_{\mathrm{II}}\rangle\langle0|
 +\mathrm{H.c.}\right)\Big].
 \label{eq:lower-state-field-map}
\end{align}
Here $\mathrm{H.c.}$ denotes the Hermitian conjugate. The resulting moment
changes are
\begin{equation}
 \Delta N_{\mathrm I}=0,
 \qquad
 \Delta N_{\mathrm{II}}=\epsilon_b^2v_f^2,
 \qquad
 \Delta M_{\mathrm I,\mathrm{II}}
 =-\frac{\epsilon_b^2u_fv_f}{2}.
\end{equation}
Substituting these values into Eq.~\eqref{eq:reconstruction} yields
$\Delta n_+=-\epsilon_b^2v_f^2$. Exchanging
$\delta\hat b_f\leftrightarrow\delta\hat b_f^\dagger$ gives the upper-state row and
$\Delta n_+=+\epsilon_b^2u_f^2$. The derivation shows explicitly how the partner occupations and their
phase-sensitive covariance combine into the Rindler response.

\section{Reference scaling formulas}
\label{app:benchmark-formulas}

For $\Delta\eta=7$, define
\begin{equation}
 x(\kappa)=\frac{4\pi^2\kappa}{\Delta\eta}\simeq 5.64\kappa,
\end{equation}
At fixed rapidity endpoints the population-free interaction strength scales as
\begin{equation}
 \epsilon_b^2(\kappa,E)
 \simeq 0.058
 \left(\frac{Q_b}{0.65~\mathrm{nC}}\right)
 \left(\frac{{\cal R}_{\rm src}}{0.16}\right)
 \left(\frac{E}{1~\mathrm{GV/m}}\right)
 \left(\frac{\kappa}{1.2}\right)^2.
 \label{eq:epsilon-scaling}
\end{equation}
In the narrowband stationary approximation,
\begin{equation}
 n_+(\kappa)=\frac{1}{\e^{x(\kappa)}-1},
 \qquad
 \sqrt{n_+(n_++1)}
 =\frac{1}{2\sinh[x(\kappa)/2]}.
 \label{eq:narrowband-thermal-functions}
\end{equation}
The reciprocal source probabilities are
\begin{equation}
 P_{\rm abs}^{R}=\epsilon_b^2n_+,
 \qquad
 P_{\rm em}^{R}=\epsilon_b^2(n_++1),
 \label{eq:benchmark-probabilities}
\end{equation}
and the anomalous moment change is
\begin{equation}
 \Delta M_{\mathrm I,\mathrm{II}}
 =-\frac{\epsilon_b^2}{2}\sqrt{n_+(n_++1)}.
 \label{eq:benchmark-anomalous}
\end{equation}
The exact finite-window calculation replaces $n_+$ and $n_++1$ by
$\|\beta_f\|^2$ and $\|\alpha_f\|^2$.

The statistical classical contamination scales as
\begin{align}
 \sigma_{\Xi}^{\rm stat}
 &\simeq 2.3\times10^{-5}
 \left(\frac{Q_b}{0.65~\mathrm{nC}}\right)
 \left(\frac{0.16}{{\cal R}_{\rm src}}\right)
 \left(\frac{1~\mathrm{GV/m}}{E}\right)\nonumber\\
 &\quad\times
 \left(\frac{1.2}{\kappa}\right)^3
 \frac{\e^{x(\kappa)}-1}{\e^{x(1.2)}-1}
 \left(\frac{\sigma_A}{10^{-3}}\right)^2
 \left(\frac{L_{\rm tom}}{2\times10^{-16}}\right)
 \left(\frac{3\times10^9}{N_{\rm eff}}\right)^{1/2}.
 \label{eq:contamination-scaling}
\end{align}

The factors can be traced to three quantities. At fixed rapidity endpoints
and aperture, the classical pulse energy scales as $Q_b^2E$, the weak
transition probability as $Q_b{\cal R}_{\rm src}E\kappa^2n_+$, and the
ray-estimated energy per photon as $E\kappa$. Their combination gives the
$E^{-1}\kappa^{-3}$ dependence above. Using an exact packet mean in place of
$\kappa\nu_0$ multiplies the estimate by
$\kappa\nu_0/\bar\nu_{\mathrm{II},\rm lab}$, without any further photon-energy
factor. This expression describes the statistical uncertainty of the corrected Li\'enard background. Residual mean offsets and fixed calibration
uncertainties remain separate parts of the error budget.

\section{Notation}
\label{app:symbol-guide}

\begin{description}
\item[$\facc$] Ordinary frequency conjugate to accelerating-frame proper time.
\item[$\fU$] Unruh frequency, $a/(4\pi^2c)$.
\item[$\fminacc$] Order-of-magnitude finite-duration frequency scale,
$1/\Delta\tau$.
\item[$\delta\hat b_f$] Annihilation operator of the canonically normalized
finite-duration Rindler wave packet.
\item[$\mathrm I,\mathrm{II}$] Positive-Minkowski-frequency partner sectors;
not signs of Rindler frequency.
\item[$N_{\mathrm I},N_{\mathrm{II}}$] Centered partner occupations.
\item[$M_{\mathrm I,\mathrm{II}}$] Phase-sensitive anomalous covariance between
the partners.
\item[$n_+,n_-$] Normally and antinormally ordered counts of the reconstructed
Rindler mode.
\item[$R_f$] Exact ordered ratio of the experiment-defined finite-duration
Rindler wave packet,
$\|\beta_f\|^2/\|\alpha_f\|^2$.
\item[$C_K$] Captured commutator weight of the first $K$ canonical partner
pairs; $1-C_K$ is the portion of the target unit commutator omitted by the
retained receiver basis.
\item[$\Sigma^{(f)}_{ij}$] Statistical covariance between the reconstructed
occupations of two overlapping finite-duration wave packets.
\item[$\underline n_{\rm dir}$] Direct weak-spectrum estimator of the selected wave-packet
occupation $\|\beta_f\|^2$.
\item[$P_{\rm abs}^{R},P_{\rm em}^{R}$] Finite-duration absorption and emission responses in the selected field
packet; equal to single-transition probabilities at leading order.
\item[${\cal R}_{\rm src}$] Source response in the selected azimuthal-polarization packet, including
the participating directional oscillator strength and waveform overlap, but
not preparation populations or receiver transmission losses.
\item[$\epsilon_b^2$] Population-free one-quantum interaction strength over
the finite acceleration interval.
\item[$\eta_{J,\rm count}^{(N)}$] End-to-end photon-counting probability for
partner $J$.
\item[$\eta_{\rm tom}$] Effective two-mode quadrature efficiency.
\item[$(m)$] Superscript labeling the measured receiver-output
mode or its moments, before correcting for receiver loss and phase shifts.
\item[$\sigma_{\rm stat}(\Delta n_+)$] Total statistical
uncertainty of the reconstructed source--reference occupation difference,
obtained by propagating the moment covariance.
\item[$N_{\rm eff}$] Differential effective homodyne sample size.
\item[$N_{\rm count}$] Effective sample size of the source--reference difference in the counting
configuration.
\item[$b_J$] Residual mean photons in partner $J$ after optical suppression, expressed
at the receiver input by correcting for measured loss.
\item[$T_{\phi,J}$] Filter power transmission in the accepted local polarization.
\item[$\rho_{\rm pol}$] Rejected-to-accepted polarization power-transmission ratio.
\item[$L_{\rm rest}$] Additional signal-normalized power leakage of the unwanted field through
the receiver after the polarization filter.
\item[$\nu_0$] Mean laboratory frequency estimated from the ray distribution at
$\facc=\fminacc$; the corresponding estimate at $\facc=\kappa\fminacc$
is $\kappa\nu_0$.
\item[$L_{\rm img}$] Maximum conjugate-sideband leakage after receiver
inversion; ``image response'' is the conventional mixer term for the unwanted
conjugate frequency component.
\item[$L_{\rm tom}$] Li\'enard-field power leakage through the complete selected-mode readout,
relative to its signal response; equals $\rho_{\rm pol}L_{\rm rest}$ only
when their measured responses compose in this way.
\item[$D_{\rm cl}$] Raw source-level classical-to-weak energy ratio.
\item[$\Lambda^{\rm bias}$] Signed residual mean offset,
expressed as an energy-equivalent fraction of the raw classical energy.
\item[$\Lambda^{\rm other}$] Nonnegative bound on the
absolute size of unresolved residual offsets, whose signs need not be positive.
\item[$\sigma_{\Xi}^{\rm stat}$] Statistical uncertainty of
the calibrated classical-residual contribution, expressed relative to the weak
signal; not the total counting and homodyne uncertainty.
\item[$\Xi_z^{\rm tot}$] Total classical contamination in units of the weak
complete Rindler-wave-packet signal.
\end{description}

\raggedright
\bibliographystyle{apsrev4-2-author-truncate}
\bibliography{all}